%% file: main.tex
\documentclass[modern]{lsstdescnote_customized}
\usepackage{lsstdesc_macros}
\usepackage{xcolor}
\usepackage{soul}
\newcommand{\themekey}[1]{\textbf{\textcolor{Ink}{#1}}}
\newcommand{\themebullet}{\noindent\textcolor{ThemeBlue}{\textbf{·}}\hspace{4pt}}
\definecolor{DESCred}{rgb}{0.63,0.00,0.20}
\hypersetup{colorlinks=true,breaklinks=true,
  citecolor=DESCred,filecolor=DESCred,linkcolor=DESCred,urlcolor=DESCred}
  
\usepackage[dvipsnames,table]{xcolor}
\usepackage[most]{tcolorbox}
\tcbuselibrary{skins,breakable}

\usepackage{booktabs}   % nicer horizontal rules
\usepackage{tabularx}   % auto-width table with X columns
\usepackage{enumitem}   % customizable lists
\usepackage{array}      % extended column definitions
\usepackage{ragged2e}   % better ragged-right in X columns

\usepackage{hyperref}
\usepackage[acronym]{glossaries}
\makenoidxglossaries
\input{methods_and_challenges}

\definecolor{ThemeBlue}{HTML}{2563EB}      % Modern blue
\definecolor{AccentBlue}{HTML}{DBEAFE}     % Light blue background
\definecolor{ThemeMint}{HTML}{10B981}      % Fresh green
\definecolor{AccentMint}{HTML}{D1FAE5}     % Light green background
\definecolor{DESCRed}{HTML}{a71437}
\definecolor{ThemeRed}{HTML}{c24b60}
\definecolor{Ink}{HTML}{1F2937}
\colorlet{InkMuted}{black!60}

\tcbset{
  before skip=12pt, after skip=12pt,
  boxsep=0pt, left=12pt, right=12pt, top=10pt, bottom=10pt,
  breakable, enhanced,
}

\newtcolorbox{ThemeBoxA}[1][]{
  colback=white,
  colframe=ThemeRed,
  boxrule=1pt,
  leftrule=0pt,
  borderline west={3pt}{0pt}{DESCRed},
  fonttitle=\sffamily\bfseries, 
  coltitle=DESCRed,
  title={#1},
  arc=2pt,
}

\newif\ifshowinstructions
\showinstructionstrue

\shorttitle{AI/ML for DESC}
\shortauthors{LSST~DESC}

\begin{document}

\title{Opportunities in AI/ML for the Rubin LSST\\Dark Energy Science Collaboration}
\input{authors}

\begin{abstract}
\small
The Vera C.\ Rubin Observatory's Legacy Survey of Space and Time (LSST) will produce unprecedented volumes of heterogeneous astronomical data---images, catalogs, and alerts---that challenge traditional analysis pipelines. The LSST Dark Energy Science Collaboration (DESC) aims to derive robust constraints on dark energy and dark matter from these data, requiring methods that are statistically powerful, scalable, and operationally reliable. Artificial intelligence and machine learning (AI/ML) are already embedded across DESC science workflows, from photometric redshifts and transient classification to weak lensing inference and cosmological simulations. Yet their utility for precision cosmology hinges on trustworthy uncertainty quantification, robustness to covariate shift and model misspecification, and reproducible integration within scientific pipelines. This white paper surveys the current landscape of AI/ML across DESC's primary cosmological probes and cross-cutting analyses, revealing that the same core methodologies and fundamental challenges recur across disparate science cases. Since progress on these cross-cutting challenges would benefit multiple probes simultaneously, we identify key methodological research priorities, including Bayesian inference at scale, physics-informed methods, validation frameworks, and active learning for discovery. With an eye on emerging techniques, we also explore the potential of the latest foundation model methodologies and LLM-driven agentic AI systems to reshape DESC workflows, provided their deployment is coupled with rigorous evaluation and governance. Finally, we discuss critical software, computing, data infrastructure, and human capital requirements for the successful deployment of these new methodologies, and consider associated risks and opportunities for broader coordination with external actors. Taken together, DESC's combination of community-accessible data, demanding scientific requirements, and mature simulation infrastructure makes the collaboration an excellent testbed for developing and validating robust AI/ML practices for fundamental physics.
\end{abstract}

\maketitle
\newpage
\input{contributors}

\newpage
{
\setlength{\parskip}{0pt}
\makeatletter
\renewcommand{\l@section}[2]{%
  \ifnum \c@tocdepth >\z@
    \addpenalty\@secpenalty
    \addvspace{2pt plus 1pt}%
    \setlength\@tempdima{1.5em}%
    \begingroup
      \parindent \z@ \rightskip \@pnumwidth
      \parfillskip -\@pnumwidth
      \leavevmode \bfseries
      \advance\leftskip\@tempdima
      \hskip -\leftskip
      #1\nobreak\hfil \nobreak\hb@xt@\@pnumwidth{\hss #2}\par
    \endgroup
  \fi}
\renewcommand{\l@subsection}[2]{%
  \ifnum \c@tocdepth >\@ne
    \addpenalty\@secpenalty
    \addvspace{1pt}%
    \setlength\@tempdima{2.3em}%
    \begingroup
      \parindent \z@ \rightskip \@pnumwidth
      \parfillskip -\@pnumwidth
      \leavevmode
      \advance\leftskip\@tempdima
      \hskip -\leftskip
      #1\nobreak\hfil \nobreak\hb@xt@\@pnumwidth{\hss #2}\par
    \endgroup
  \fi}
\renewcommand{\l@subsubsection}[2]{%
  \ifnum \c@tocdepth >2
    \addvspace{0.5pt}%
    \setlength\@tempdima{3.8em}%
    \begingroup
      \parindent \z@ \rightskip \@pnumwidth
      \parfillskip -\@pnumwidth
      \leavevmode
      \advance\leftskip\@tempdima
      \hskip -\leftskip
      #1\nobreak\hfil \nobreak\hb@xt@\@pnumwidth{\hss #2}\par
    \endgroup
  \fi}
\makeatother
\tableofcontents
}

\input{sections/executive_summary}
\input{sections/introduction}
\input{sections/sec3_ai_in_desc}
\input{sections/sec4_methodology}
\input{sections/sec5_emerging}
\input{sections/sec6_infrastructure}
\input{sections/sec7_coordination}

\input{sections/sec8_risks_challenges}
\input{sections/conclusion}

\appendix

\section{Index of AI/ML Methodologies and Challenges}

\printnoidxglossary[type=methods,title={AI/ML Methods Index}]

\printnoidxglossary[type=challenges,title={Cross-cutting Challenges Index}]

\section{Glossary of Acronyms}
\printnoidxglossary[type=\acronymtype, title={Acronyms and Abbreviations}]

\section{Acknowledgments}
\input{desc-standard-ack}

\newpage
\bibliography{refs}

\end{document}

%% file: methods_and_challenges.tex
\newglossary[mlg]{methods}{mls}{mln}{AI/ML Methods}
\newglossary[clg]{challenges}{cls}{cln}{Cross-cutting Challenges}

\newcommand{\meth}[1]{\gls{#1}}
\newcommand{\challenge}[1]{\gls{#1}}

\definecolor{aiMethBayes}{HTML}{263654} % Rhino (User Anchor)
\definecolor{aiMethGen}{HTML}{5D4C76}   % Muted Purple
\definecolor{aiMethDeep}{HTML}{4A7A8C}  % Muted Cyan
\definecolor{aiMethPhys}{HTML}{5C7A63}  % Muted Green
\definecolor{aiMethRepr}{HTML}{3E8E7E}  % Muted Teal
\definecolor{aiMethVis}{HTML}{5B7C99}   % Steel Blue

\newcommand{\aimethbayes}[1]{\textcolor{aiMethBayes}{\textsf{#1}}}
\newcommand{\aimethgen}[1]{\textcolor{aiMethGen}{\textsf{#1}}}
\newcommand{\aimethdeep}[1]{\textcolor{aiMethDeep}{\textsf{#1}}}
\newcommand{\aimethphys}[1]{\textcolor{aiMethPhys}{\textsf{#1}}}
\newcommand{\aimethrepr}[1]{\textcolor{aiMethRepr}{\textsf{#1}}}
\newcommand{\aimethvis}[1]{\textcolor{aiMethVis}{\textsf{#1}}}

\definecolor{aiChCovariate}{HTML}{C2814C}  % Tussock (warm orange)
\definecolor{aiChUQ}{HTML}{92402D}         % Mule Fawn (deep red-brown)
\definecolor{aiChScale}{HTML}{7D7068}      % Dark Pumice/Taupe
\definecolor{aiChSparsity}{HTML}{5D4C76}   % Muted Purple
\definecolor{aiChMetrics}{HTML}{C06C84}    % Muted Pink

\newcommand{\aichcovariate}[1]{\textcolor{aiChCovariate}{\textsf{#1}}}
\newcommand{\aichuq}[1]{\textcolor{aiChUQ}{\textsf{#1}}}
\newcommand{\aichscale}[1]{\textcolor{aiChScale}{\textsf{#1}}}
\newcommand{\aichsparsity}[1]{\textcolor{aiChSparsity}{\textsf{#1}}}
\newcommand{\aichmetrics}[1]{\textcolor{aiChMetrics}{\textsf{#1}}}

\newglossaryentry{sbi}{%
  type=methods,
  name={\aimethbayes{Simulation-based inference (SBI)}},
  text={\aimethbayes{SBI}},
  description={Implicit-likelihood Bayesian inference using forward simulations and \acrshort{nde} techniques}
}

\newglossaryentry{neural-density-estimation}{%
  type=methods,
  name={\aimethbayes{Neural density estimation (NDE)}},
  text={\aimethbayes{NDE}},
  description={Neural networks trained to approximate probability densities, often used within \acrshort{acr:sbi}}
}

\newglossaryentry{npe}{%
  type=methods,
  name={\aimethbayes{Neural posterior estimation (NPE)}},
  text={\aimethbayes{NPE}},
  description={Direct neural approximation to the posterior distribution over parameters given simulated data}
}

\newglossaryentry{hierarchical-bayes}{%
  type=methods,
  name={\aimethbayes{Bayesian hierarchical modeling}},
  text={\aimethbayes{Hierarchical Bayes}},
  description={Hierarchical probabilistic models that share information across objects or populations}
}

\newglossaryentry{variational-inference}{%
  type=methods,
  name={\aimethbayes{Variational inference (VI)}},
  text={\aimethbayes{VI}},
  description={Optimization-based approach to approximating posterior distributions}
}

\newglossaryentry{ensembles}{%
  type=methods,
  name={\aimethbayes{Deep ensembles}},
  text={\aimethbayes{Ensembles}},
  description={Collections of models combined for improved performance and \acrshort{acr:uq}}
}

\newglossaryentry{gaussian-process}{%
  type=methods,
  name={\aimethbayes{Gaussian processes}},
  text={\aimethbayes{Gaussian processes}},
  description={Non-parametric Bayesian regression and emulation for scalar or functional outputs}
}

\newglossaryentry{diffusion-model}{%
  type=methods,
  name={\aimethgen{Diffusion / score-based models}},
  text={\aimethgen{Diffusion models}},
  description={Generative models that iteratively denoise data from a noise process, often used for images and fields}
}

\newglossaryentry{normalizing-flow}{%
  type=methods,
  name={\aimethgen{Normalizing flows}},
  text={\aimethgen{Normalizing flows}},
  description={Invertible neural networks used as flexible density models in likelihoods, posteriors, or simulators}
}

\newglossaryentry{vae}{%
  type=methods,
  name={\aimethgen{Variational autoencoders (VAEs)}},
  text={\aimethgen{VAEs}},
  description={Latent-variable generative models trained via \acrshort{vi}}
}

\newglossaryentry{emulator}{%
  type=methods,
  name={\aimethgen{Emulators}},
  text={\aimethgen{Emulators}},
  description={Surrogate models (often Gaussian process- or neural network-based) that approximate expensive simulations or likelihoods}
}

\newglossaryentry{neural-surrogate}{%
  type=methods,
  name={\aimethgen{Neural surrogates}},
  text={\aimethgen{Neural surrogates}},
  description={Neural network models trained to mimic the input--output behavior of complex physical simulations}
}

\newglossaryentry{cnn}{%
  type=methods,
  name={\aimethdeep{Convolutional neural networks (CNNs)}},
  text={\aimethdeep{CNNs}},
  description={Convolution-based neural networks for image-like data, widely used for classification and regression}
}

\newglossaryentry{transformer}{%
  type=methods,
  name={\aimethdeep{Transformers}},
  text={\aimethdeep{Transformers}},
  description={Attention-based neural architectures for sequences, time series, or generic sets of tokens (including vision transformers)}
}

\newglossaryentry{rnn}{%
  type=methods,
  name={\aimethdeep{Recurrent neural networks (RNNs)}},
  text={\aimethdeep{RNNs}},
  description={Sequence models (e.g.\ \acrshortpl{lstm}, \acrshortpl{gru}) for time series and light curves}
}

\newglossaryentry{gnn}{%
  type=methods,
  name={\aimethdeep{Graph neural networks (GNNs)}},
  text={\aimethdeep{GNNs}},
  description={Neural networks designed to operate on graph-structured data, such as cosmic webs or halo catalogs}
}

\newglossaryentry{deep-network}{%
  type=methods,
  name={\aimethdeep{Deep neural networks}},
  text={\aimethdeep{Deep networks}},
  description={Generic deep learning architectures not otherwise categorized}
}

\newglossaryentry{differentiable-programming}{%
  type=methods,
  name={\aimethphys{Differentiable programming}},
  text={\aimethphys{Differentiable programming}},
  description={Formulation of simulations and models as differentiable programs amenable to gradient-based inference}
}

\newglossaryentry{physics-informed}{%
  type=methods,
  name={\aimethphys{Physics-informed ML}},
  text={\aimethphys{Physics-informed ML}},
  description={Machine learning models that incorporate physical laws or constraints (e.g., \acrshortpl{pinn}, \acrshortpl{enn})}
}

\newglossaryentry{symbolic-regression}{%
  type=methods,
  name={\aimethphys{Symbolic regression}},
  text={\aimethphys{Symbolic regression}},
  description={Learning analytic mathematical expressions that describe data or physical relationships}
}

\newglossaryentry{neural-compression}{%
  type=methods,
  name={\aimethrepr{Neural compression}},
  text={\aimethrepr{Neural compression}},
  description={Learned low-dimensional representations (summary statistics or latents) of complex data}
}

\newglossaryentry{self-supervised}{%
  type=methods,
  name={\aimethrepr{Self-supervised learning}},
  text={\aimethrepr{Self-supervised learning}},
  description={Learning representations from unlabeled data by solving pretext tasks (e.g., masking, contrastive learning)}
}

\newglossaryentry{som}{%
  type=methods,
  name={\aimethrepr{Self-organizing maps (SOMs)}},
  text={\aimethrepr{SOMs}},
  description={Topology-preserving maps used for exploratory analysis and domain coverage characterization}
}

\newglossaryentry{anomaly-detection}{%
  type=methods,
  name={\aimethrepr{Anomaly detection}},
  text={\aimethrepr{Anomaly detection}},
  description={Identifying rare or out-of-distribution examples in data, crucial for new physics discovery}
}

\newglossaryentry{active-learning}{%
  type=methods,
  name={\aimethrepr{Active learning}},
  text={\aimethrepr{Active learning}},
  description={Strategies that adaptively select the most informative data points for labeling or follow-up}
}

\newglossaryentry{instance-segmentation}{%
  type=methods,
  name={\aimethvis{Instance segmentation}},
  text={\aimethvis{Instance segmentation}},
  description={Pixel-level segmentation of individual objects, relevant for deblending crowded scenes}
}

\newglossaryentry{object-detection}{%
  type=methods,
  name={\aimethvis{Object detection}},
  text={\aimethvis{Object detection}},
  description={Identifying and localizing objects in images (e.g., \acrshort{yolo})}
}

\newglossaryentry{deblending}{%
  type=methods,
  name={\aimethvis{Deblending algorithms}},
  text={\aimethvis{Deblending}},
  description={Separating overlapping light profiles from multiple sources}
}

\newglossaryentry{covariate-shift}{%
  type=challenges,
  name={\aichcovariate{Covariate Shifts}},
  text={\aichcovariate{Covariate shifts}},
  description={Distribution mismatches between training and target data (e.g.\ spectroscopic selection bias, sim-to-real gaps, model misspecification). Addressed in \autoref{sec4:model-misspec}, \autoref{sec4:hybrid-gen-phys}, and \autoref{sec:discovery}}
}

\newglossaryentry{uq}{%
  type=challenges,
  name={\aichuq{Uncertainty Quantification}},
  text={\aichuq{UQ}},
  description={Obtaining well-calibrated posteriors and propagating uncertainties to cosmological constraints. Addressed in \autoref{sec4:Bayes}, \autoref{sec4:sbi}, and \autoref{sec4:validation}}
}

\newglossaryentry{scalability}{%
  type=challenges,
  name={\aichscale{Scalability}},
  text={\aichscale{Scalability}},
  description={Handling LSST-scale data volumes, real-time alert processing, and high-dimensional inference. Addressed in \autoref{sec4:Bayes}, \autoref{sec4:sbi}, and \autoref{sec4:hybrid-gen-phys}}
}

\newglossaryentry{data-sparsity}{%
  type=challenges,
  name={\aichsparsity{Data Sparsity \& Rare Events}},
  text={\aichsparsity{Data sparsity}},
  description={Limited labeled samples, rare transients, class imbalance, and challenging edge cases like blending. Addressed in \autoref{sec:discovery}}
}

\newglossaryentry{metrics}{%
  type=challenges,
  name={\aichmetrics{Metrics \& Evaluation}},
  text={\aichmetrics{Metrics}},
  description={Task-relevant metrics, validation frameworks, benchmarking, and stress tests for DESC science. Addressed in \autoref{sec4:validation} and \autoref{sec4:physics-constraints}}
}

\newacronym{4most}{4MOST}{4-meter Multi-Object Spectroscopic Telescope}
\newacronym{act}{ACT}{Atacama Cosmology Telescope}
\newacronym{ads}{ADS}{\acrshort{sao} Astrophysics Data System}
\newacronym{adql}{ADQL}{Astronomical Data Query Language}
\newacronym{agn}{AGN}{active galactic nuclei}
\newacronym{agnsc}{AGNSC}{\href{https://agn.science.lsst.org/}{Active Galactic Nuclei Science Collaboration}}
\newacronym{ai}{AI}{artificial intelligence}
\newacronym{alcf}{ALCF}{Argonne Leadership Computing Facility}
\newacronym{alerce}{ALeRCE}{Automatic Learning for the Rapid Classification of Events}
\newacronym{alma}{ALMA}{Atacama Large Millimeter Array}
\newacronym{ampel}{AMPEL}{Alert Management, Photometry, and Evaluation of Light curves}
\newacronym{amsc}{AmSC}{American Science Cloud}
\newacronym{anacal}{AnaCal}{Analytic Calibration}
\newacronym{antares}{ANTARES}{Arizona--NOIRLab Temporal Analysis and Response to Events System}
\newacronym{api}{API}{application programming interface}
\newacronym{ascend}{ASCEND}{A Statement of Community Engagement and Needs for Digital Research Infrastructure}
\newacronym{ascr}{ASCR}{Advanced Scientific Computing Research}
\newacronym{atat}{ATAT}{Astronomical Transformer for time series And Tabular data}
\newacronym{bliss}{BLISS}{Bayesian Light Source Separator}
\newacronym{bnn}{BNN}{Bayesian neural network}
\newacronym{camb}{CAMB}{Code for Anisotropies in the Microwave Background}
\newacronym{candels}{CANDELS}{Cosmic Assembly Near-infrared Deep Extragalactic Legacy Survey}
\newacronym{ccd}{CCD}{charge-coupled device}
\newacronym{cc-in2p3}{CC-IN2P3}{\acrshort{in2p3} Computing Centre}
\newacronym{cd}{CD}{continuous delivery/deployment}
\newacronym{ci}{CI}{continuous integration}
\newacronym{cigale}{CIGALE}{Code Investigating GALaxy Evolution}
\newacronym{clmm}{CLMM}{Cluster Lens Mass Modeling tool}
\newacronym{cmb}{CMB}{cosmic microwave background}
\newacronym{cmnn}{CMNN}{color-matched nearest-neighbors}
\newacronym{acr:cnn}{CNN}{convolutional neural network}
\newacronym{cosmodc2}{CosmoDC2}{Cosmological Data Challenge 2}
\newacronym{cpu}{CPU}{central processing unit}
\newacronym{csd3}{CSD3}{Cambridge Service for Data Driven Discovery}
\newacronym{csp}{CSP}{Carnegie Supernova Project}
\newacronym{deepdisc}{DeepDISC}{Detection, Instance Segmentation, and Classification with Deep Learning}
\newacronym{decals}{DECaLS}{Dark Energy Camera Legacy Survey}
\newacronym{des}{DES}{Dark Energy Survey}
\newacronym{desc}{DESC}{\href{https://lsstdesc.org}{Dark Energy Science Collaboration}}
\newacronym{desi}{DESI}{Dark Energy Spectroscopic Instrument}
\newacronym{dl}{DL}{deep learning}
\newacronym{dnf}{DNF}{directional neighborhood fitting}
\newacronym{doe}{DOE}{Department of Energy}
\newacronym{dsps}{DSPS}{Differentiable Stellar Population Synthesis}
\newacronym{elasticc}{ELAsTiCC}{Extended LSST Astronomical Time Series Classification Challenge}
\newacronym{ellis}{ELLIS}{European Laboratory for Learning and Intelligent Systems}
\newacronym{enn}{ENN}{equivariant neural network}
\newacronym{esa}{ESA}{European Space Agency}
\newacronym{esnet}{ESNet}{Energy Sciences Network}
\newacronym{eucaif}{EuCAIF}{European Coalition for AI for Fundamental Physics}
\newacronym{eurohpc}{EuroHPC}{European High-Performance Computing Joint Undertaking}
\newacronym{fm}{FM}{foundation model}
\newacronym{fsps}{FSPS}{Flexible Stellar Population Synthesis}
\newacronym{galsc}{GSC}{\href{https://sites.google.com/view/lsstgsc/home}{Galaxies Science Collaboration}}
\newacronym{gan}{GAN}{generative adversarial network}
\newacronym{ggns}{GGNS}{gradient-guided nested sampling}
\newacronym{acr:gnn}{GNN}{graph neural network}
\newacronym{gpz}{GPz}{Gaussian Processes for Photo-$z$}
\newacronym{gpu}{GPU}{graphics processing unit}
\newacronym{gru}{GRU}{gated recurrent unit}
\newacronym{hats}{HATS}{Hierarchical Adaptive Tiling Scheme}
\newacronym{hmc}{HMC}{Hamiltonian Monte Carlo}
\newacronym{hod}{HOD}{halo occupancy distribution}
\newacronym{hpc}{HPC}{high-performance computing}
\newacronym{hpdf}{HPDF}{High Performance Data Facility}
\newacronym{hsc}{HSC}{Hyper Suprime-Cam}
\newacronym{hst}{\textit{HST}}{\textit{Hubble Space Telescope}}
\newacronym{ia}{IA}{Intrinsic Alignment}
\newacronym{iaifi}{IAIFI}{NSF Institute for Artificial Intelligence and Fundamental Interactions}
\newacronym{idac}{IDAC}{Independent Data Access Center}
\newacronym{iid}{IID}{independent and identically distributed}
\newacronym{imnn}{IMNN}{information-maximizing neural network}
\newacronym{in2p3}{IN2P3}{Institut National de Physique Nucl\'{e}aire et de Physique des Particules}
\newacronym{ir}{IR}{infrared}
\newacronym{iri}{IRI}{Integrated Research Infrastructure}
\newacronym{issc}{ISSC}{\href{https://issc.science.lsst.org/}{Informatics and Statistics Science Collaboration}}
\newacronym{jupiter}{JUPITER}{Joint Undertaking Pioneer for Innovative and Transformative Exascale Research}
\newacronym{jwst}{\textit{JWST}}{\textit{James Webb Space Telescope}}
\newacronym{kids}{KiDS}{Kilo Degree Survey}
\newacronym{kne}{KNe}{kilonovae}
\newacronym{knn}{kNN}{$k$-nearest neighbors}
\newacronym{lccf}{LCCF}{Leadership Class Computing Facility}
\newacronym{lcdm}{$\Lambda$CDM}{cold dark matter with cosmological constant}
\newacronym{lincc}{LINCC Frameworks}{LSST Interdisciplinary Network for Collaboration and Computing Frameworks}
\newacronym{llm}{LLM}{large language model}
\newacronym{lmc}{LMC}{Langevin Monte Carlo}
\newacronym{lsst}{LSST}{Legacy Survey of Space and Time}
\newacronym{lsst-da}{LSST-DA}{LSST Discovery Alliance}
\newacronym{lstm}{LSTM}{long short-term memory}
\newacronym{lumi}{LUMI}{Large Unified Modern Infrastructure}
\newacronym{madness}{MADNESS}{Maximum A posteriori with Deep NEural networks for Source Separation}
\newacronym{map}{MAP}{maximum a posteriori}
\newacronym{mas}{MAS}{multi-agent systems}
\newacronym{mchmc}{MCHMC}{micro-canonical Hamiltonian Monte Carlo}
\newacronym{mclmc}{MCLMC}{micro-canonical Langevin Monte Carlo}
\newacronym{mcmc}{MCMC}{Markov chain Monte Carlo}
\newacronym{ml}{ML}{machine learning}
\newacronym{moc}{MOC}{multi-order coverage}
\newacronym{moe}{MoE}{mixture-of-experts}
\newacronym{mvit}{MViT}{multiscale vision transformers}
\newacronym{nasa}{NASA}{National Aeronautics and Space Administration}
\newacronym{nde}{NDE}{neural density estimation}
\newacronym{nersc}{NERSC}{National Energy Research Scientific Computing Center}
\newacronym{nfw}{NFW}{Navarro--Frenk--White}
\newacronym{nle}{NLE}{neural likeliood estimation}
\newacronym{noirlab}{NOIRLab}{National Optical-Infrared Astronomy Research Laboratory}
\newacronym{acr:npe}{NPE}{neural posterior estimation}
\newacronym{nre}{NRE}{neural ratio estimation}
\newacronym{nsf}{NSF}{National Science Foundation}
\newacronym{nuts}{NUTS}{No U-Turn Sampler}
\newacronym{ode}{ODE}{ordinary differential equation}
\newacronym{olcf}{OLCF}{Oak Ridge Leadership Computing Facility}
\newacronym{onnx}{ONNX}{Open Neural Network Exchange}
\newacronym{opsim}{OpSim}{Operations Simulator}
\newacronym{pan-starrs}{Pan-STARRS}{Panoramic Survey Telescope and Rapid Response System}
\newacronym{pdf}{PDF}{probability density function}
\newacronym{photoz}{photo-$z$}{photometric redshift}
\newacronym{pinn}{PINN}{physics-informed neural network}
\newacronym{pisn}{PISN}{pair instability supernova}
\newacronym{pit}{PIT}{probability integral transform}
\newacronym{plasticc}{PLAsTiCC}{Photometric LSST Astronomical Time Series Classification Challenge}
\newacronym{psf}{PSF}{point-spread function}
\newacronym{qa}{QA}{question answering}
\newacronym{rag}{RAG}{retrieval-augmented generation}
\newacronym{rail}{RAIL}{Redshift Assessment Infrastructure Layers}
\newacronym{rd}{R\&D}{research \& development}
\newacronym{redmapper}{RedMaPPer}{Red-sequence Matched-filter Probabilistic Percolation}
\newacronym{resnet}{ResNet}{residual network}
\newacronym{resspect}{RESSPECT}{Recommendation System for Spectroscopic followup}
\newacronym{rhmc}{RHMC}{Riemann Hamiltonian Monte Carlo}
\newacronym{acr:rnn}{RNN}{recurrent neural network}
\newacronym{rsp}{RSP}{Rubin Science Platform}
\newacronym{sao}{SAO}{Smithsonian Astrophysical Observatory}
\newacronym{sassafras}{SASSAFRAS}{S/N Analysis of Simulated SpectrA for Rubin trAnsientS}
\newacronym{acr:sbi}{SBI}{simulation-based inference}
\newacronym{sbm}{SBM}{score-based model}
\newacronym{sde}{SDE}{stochastic differential equation}
\newacronym{sdss}{SDSS}{Sloan Digital Sky Survey}
\newacronym{sed}{SED}{spectral energy distribution}
\newacronym{sfh}{SFH}{star formation history}
\newacronym{sfr}{SFR}{star formation rate}
\newacronym{ssfr}{sSFR}{specific star formation rate}
\newacronym{skai}{SkAI}{NSF--Simons AI Institute for the Sky}
\newacronym{slac}{SLAC}{Stanford Linear Accelerator Center}
\newacronym{slacs}{SLACS}{Sloan Lens ACS (Advanced Camera for Surveys) Survey}
\newacronym{slsc}{SLSC}{\href{https://sites.google.com/view/lsst-stronglensing}{Strong Lensing Science Collaboration}}
\newacronym{smwlv}{SMWLVSC}{\href{https://rubin-smwlv.github.io/}{Stars, Milky Way, and Local Volume Science Collaboration}}
\newacronym{snana}{SNANA}{Supernova Analysis package}
\newacronym{sne}{SNe}{supernovae}
\newacronym{snia}{SN Ia}{Type Ia supernova}
\newacronym{snr}{SNR}{signal-to-noise ratio}
\newacronym{so}{SO}{Simons Observatory}
\newacronym{soar}{SOAR}{Southern Astrophysical Research Telescope}
\newacronym{acr:som}{SOM}{self-organizing map}
\newacronym{spherex}{\textit{SPHEREx}}{\textit{Spectro-Photometer for the History of the Universe, Epoch of Reionization, and Ices Explorer}}
\newacronym{sps}{SPS}{stellar population synthesis}
\newacronym{spt}{SPT}{South Pole Telescope}
\newacronym{ssl}{SSL}{self-supervised Learning}
\newacronym{sql}{SQL}{structured query language}
\newacronym{sssc}{SSSC}{\href{https://lsst-sssc.github.io}{Solar System Science Collaboration}}
\newacronym{stratlearn}{StratLearn}{stratified learning}
\newacronym{svm}{SVM}{support vector machine}
\newacronym{sz}{SZ}{Sunyaev--Zeldovich}
\newacronym{tides}{TiDES}{Time Domain Extragalactic Survey}
\newacronym{tpz}{TPZ}{Trees for Photo-$z$}
\newacronym{tvs}{TVSSC}{\href{https://lsst-tvssc.github.io/}{Transients and Variable Stars Science Collaboration}}
\newacronym{acr:uq}{UQ}{uncertainty quantification}
\newacronym{usdf}{USDF}{US Data Facility}
\newacronym{uv}{UV}{ultraviolet}
\newacronym{acr:vae}{VAE}{variational autoencoder}
\newacronym{vi}{VI}{variational inference}
\newacronym{wazp}{WaZP}{Wavelet Z-Photometric}
\newacronym{wcdm}{$w$CDM}{cold dark matter with cosmological equation of state}
\newacronym{wg}{WG}{working group}
\newacronym{wise}{\textit{WISE}}{\textit{Wide-field Infrared Survey Explorer}}
\newacronym{yolo}{YOLO}{You Only Look Once}
\newacronym{ztf}{ZTF}{Zwicky Transient Facility}

%% file: authors.tex
% \collaboration{The LSST Dark Energy Science Collaboration (LSST DESC)}
%\collaboration{\color{red}\textbf{Please explicitly validate your contribution information in the following document.} \\ \color{red}\textbf{It is our source of truth for authorship:}\\ \url{ https://tinyurl.com/ai-for-desc}.}

\collaboration{Version 1.0 -- January 2026}

\author{The LSST Dark Energy Science Collaboration (DESC)}
\noaffiliation

\author[0000-0002-5592-023X]{Eric Aubourg}
\affiliation{Université Paris Cité, CNRS, CEA, Astroparticule et Cosmologie,
F-75013 Paris, France}

\author[0000-0001-8868-0810]{Camille Avestruz}
\affiliation{Department of Physics, University of Michigan, Ann Arbor, MI 48109, USA}
\affiliation{Leinweber Institute of Theoretical Physics, University of Michigan, Ann Arbor, MI 48109, USA}

\author[0000-0001-7774-2246]{Matthew R. Becker}
\affiliation{Argonne National Laboratory, 9700 South Cass Avenue, Lemont, IL 60439, USA}

\author[0000-0002-7342-087X]{Biswajit Biswas}
\affiliation{Argonne National Laboratory, 9700 South Cass Avenue, Lemont, IL 60439, USA}

\author[0000-0002-5741-7195]{Rahul Biswas}
\affiliation{Independent}

\author[0000-0003-4922-7401]{Boris Bolliet}
\affiliation{Cavendish Astrophysics, University of Cambridge, Madingley Road, Cambridge CB3 0HA, UK}
\affiliation{Kavli Institute for Cosmology, University of Cambridge, Madingley Road, Cambridge CB3 0HA, UK}

\author[0000-0002-9836-603X]{Adam S. Bolton}
\affiliation{SLAC National Accelerator Laboratory, Menlo Park, CA 94025, USA}

\author[0000-0003-4383-2969]{Clecio R. Bom}
\affiliation{Centro Brasileiro de Pesquisas Físicas, Rio de Janeiro, Brazil}

\author[0009-0006-5127-7394]{Raphaël Bonnet-Guerrini}
\affiliation{Department of Computer Science, University of Milan, Milan, Italy}

\author[0000-0001-7387-2633]{Alexandre Boucaud}
\affiliation{Université Paris Cité, CNRS, Astroparticule et Cosmologie,
F-75013 Paris, France}

\author[0000-0002-1590-6927]{Jean-Eric Campagne}
\affiliation{Université Paris-Saclay, CNRS/IN2P3, IJCLab,  91405 Orsay, France}

\author[0000-0002-7887-0896]{Chihway Chang}
\affiliation{Department of Astronomy and Astrophysics, University of Chicago, Chicago, IL 60637, USA}
\affiliation{Kavli Institute for Cosmological Physics, University of Chicago, Chicago, IL 60637, USA}
\affiliation{NSF-Simons AI Institute for the Sky (SkAI), 172 E. Chestnut St., Chicago, IL 60611, USA}

\author[0000-0003-1281-7192]{Aleksandra \'Ciprijanovi\'c}
\affiliation{Fermi National Accelerator Laboratory, P.O. Box 500, Batavia, IL 60510, USA}
\affiliation{Department of Astronomy and Astrophysics, University of Chicago, Chicago, IL 60637, USA}
\affiliation{NSF-Simons AI Institute for the Sky (SkAI), 172 E. Chestnut St., Chicago, IL 60611, USA}

\author[0000-0001-9022-4232]{Johann Cohen-Tanugi}
\affiliation{Universit\'e Clermont-Auvergne, CNRS, LPCA, 63000 Clermont-Ferrand, France}

\author[0000-0002-8262-2924]{Michael W. Coughlin}
\affil{School of Physics and Astronomy, University of Minnesota, Minneapolis, MN 55455, USA}

\author[0000-0002-2495-3514]{John Franklin Crenshaw}
\affiliation{Kavli Institute for Particle Astrophysics and Cosmology, Stanford University, Stanford, CA  94305, USA}
\affiliation{Department of Physics, Stanford University, 382 Via Pueblo Mall, Stanford, CA 94305, USA}
\affiliation{SLAC National Accelerator Laboratory, Menlo Park, CA 94025, USA}

\author[0000-0002-7566-0412]{Juan C. Cuevas-Tello}
\affiliation{Engineering Faculty, Universidad Autonoma de San Luis Potosi, Zona Universitaria, San Luis Potosi, 78290, Mexico}

\author[0000-0001-8318-6813]{Juan de Vicente}
\affiliation{Centro de Investigaciones Energ\'eticas, Medioambientales y Tecnol\'ogicas (CIEMAT), Madrid, Spain}

\author[0000-0002-5296-4720]{Seth W. Digel}
\affiliation{SLAC National Accelerator Laboratory, Menlo Park, CA 94025, USA}
\affiliation{Kavli Institute for Particle Astrophysics and Cosmology, Stanford University, Stanford, CA 94305, USA}

\author[0000-0002-4773-1463]{Steven Dillmann}
\affiliation{Kavli Institute for Particle Astrophysics and Cosmology, Stanford University, Stanford, CA 94305, USA}
\affiliation{Stanford Artificial Intelligence Laboratory, Stanford University, Stanford, CA 94305, USA}
\affiliation{SLAC National Accelerator Laboratory, Menlo Park, CA 94025, USA}

\author[0000-0002-7982-3135]{Mariano Javier de Le\'on Dominguez Romero}
\affiliation{Instituto de Astronom\'ia Te\'orica y Experimental (IATE - UNC and CONICET CCT C\'ordoba), Observatorio Astron\'omico de C\'ordoba, Universidad Nacional de C\'ordoba, Laprida 854, X5000BGR, C\'ordoba, Argentina}

\author[0000-0001-8251-933X]{Alex Drlica-Wagner}
\affiliation{Fermi National Accelerator Laboratory, P.O. Box 500, Batavia, IL 60510, USA}
\affiliation{Department of Astronomy and Astrophysics, University of Chicago, Chicago, IL 60637, USA}
\affiliation{Kavli Institute of Cosmological Physics, University of Chicago, Chicago, IL 60637, USA}
\affiliation{NSF-Simons AI Institute for the Sky (SkAI), 172 E. Chestnut St., Chicago, IL 60611, USA}

\author[0000-0001-5717-2688]{Sydney Erickson}
\affiliation{Department of Physics, Stanford University, 382 Via Pueblo Mall, Stanford, CA 94305, USA}
\affiliation{SLAC National Accelerator Laboratory, Menlo Park, CA 94025, USA}

\author[0000-0003-4906-8447]{Alexander~T.~Gagliano}
\affiliation{The NSF AI Institute for Artificial Intelligence and Fundamental Interactions}
\affiliation{Center for Astrophysics \textbar{} Harvard \& Smithsonian, 60
Garden Street, Cambridge, MA 02138, USA}
\affiliation{Department of Physics and Kavli Institute for Astrophysics and Space Research, Massachusetts Institute of Technology, Cambridge, MA 02139, USA}

\author[0000-0002-7950-6076]{Christos Georgiou}
\affiliation{Institut de Física d'Altes Energies (IFAE), The Barcelona Institute of Science and Technology, Campus UAB, 08193 Bellaterra (Barcelona), Spain}

\author[0000-0002-2525-9647]{Aritra Ghosh}
\affil{Department of Astronomy \& DiRAC Institute, University of Washington, Seattle, WA 98195, USA}

\author[0000-0002-6741-983X]{Matthew Grayling}
\affiliation{Institute of Astronomy and Kavli Institute for Cosmology, University of Cambridge, Madingley Road, Cambridge, CB3 0HA, UK}

\author[0000-0003-3255-7340]{Kirill A. Grishin}
\affiliation{Université Paris Cité, CNRS, Astroparticule et Cosmologie, F-75013 Paris, France}

\author[0000-0003-1586-2773]{Alan Heavens}
\affiliation{Imperial Centre for Inference and Cosmology (ICIC), Imperial College London, Blackett Laboratory, Prince Consort Road, London SW7 2AZ, UK}

\author[0000-0002-1496-6514]{Lindsay R. House}
\affiliation{NSF-Simons AI Institute for the Sky (SkAI), 172 E. Chestnut St., Chicago, IL 60611, USA}
\affiliation{Data Science Institute, The University of Chicago, Chicago, IL 60615, USA}

\author[0000-0002-6024-466X]{Mustapha Ishak}
\affiliation{Department of Physics, The University of Texas at Dallas, Richardson, TX 75080, USA}

\author[0009-0001-6501-4564]{Wassim Kabalan}
\affiliation{Université Paris Cité, CNRS, Astroparticule et Cosmologie,
F-75013 Paris, France}

\author[0000-0001-8783-6529]{Arun Kannawadi}
\affiliation{Department of Physics, Duke University, Durham, NC 27708, USA}

\author[0000-0001-7956-0542]{François Lanusse}
\affiliation{Université Paris-Saclay, Université Paris Cité, CEA, CNRS, AIM, F-91191 Gif-sur-Yvette, France}

\author[0000-0002-7810-6134]{C. Danielle Leonard}
\affiliation{School of Mathematics, Statistics and Physics, Newcastle University, Newcastle upon Tyne, NE1 7RU, United Kingdom }

\author[0000-0002-8357-3984]{Pierre-Fran\c{c}ois L\'eget}
\affiliation{Department of Astrophysical Sciences, Princeton University, Princeton, NJ 08544, USA}

\author[0000-0003-2221-8281]{Michelle Lochner}
\affiliation{Department of Physics and Astronomy, University of the Western Cape, Bellville, Cape Town, 7535, South Africa}

\author[0000-0002-1200-0820]{Yao-Yuan Mao}
\affiliation{Department of Physics and Astronomy, University of Utah, Salt Lake City, UT 84112, USA}

\author[0000-0002-8873-5065]{Peter Melchior}
\affiliation{Department of Astrophysical Sciences, Princeton University, Peyton Hall, Princeton, NJ 08544, USA}

\author[0009-0005-7923-054X]{Grant Merz}
\affiliation{Department of Astronomy, University of Illinois Urbana Champaign, 1002 W. Green St., Urbana, IL, 61801, USA}

\author[0000-0001-7051-497X]{Martin Millon}
\affiliation{Institute for Particle Physics and Astrophysics, ETH Zürich, Wolfgang-Pauli-Strasse 27, CH-8093 Zurich, Switzerland}

\author[0000-0001-8211-8608]{Anais Möller}
\affiliation{Swinburne University of Technology, Hawthorn, Victoria 3122, Australia}

\author[0000-0001-6022-0484]{Gautham Narayan}
\affiliation{Department of Astronomy, University of Illinois Urbana Champaign, 1002 W. Green St., Urbana, IL, 61801, USA}
\affiliation{NSF-Simons AI Institute for the Sky (SkAI), 172 E. Chestnut St., Chicago, IL 60611, USA}

\author[0000-0002-0963-7310]{Yuuki Omori}
\affiliation{Department of Astronomy and Astrophysics, University of Chicago, Chicago, IL 60637, USA}
\affiliation{Kavli Institute for Cosmological Physics, University of Chicago, Chicago, IL 60637, USA}
\affiliation{NSF-Simons AI Institute for the Sky (SkAI), 172 E. Chestnut St., Chicago, IL 60611, USA}

\author[0000-0002-2519-584X]{Hiranya Peiris}
\affiliation{Institute of Astronomy and Kavli Institute for Cosmology, University of Cambridge, Madingley Road, Cambridge, CB3 0HA, UK}

\author[0000-0003-3544-3939]{Laurence Perreault-Levasseur}
\affiliation{D\'epartement de Physique, Universit\'e de Montr\'eal, 1375 Avenue Th\'er\`ese-Lavoie-Roux, Montr\'eal, QC, H2V 0B3, Canada}
\affiliation{Ciela - Montr\'eal Institute for Astrophysical Data Analysis and Machine Learning, Montréal, QC H2V 0B3, Canada}
\affiliation{Mila - Quebec Artificial Intelligence Institute, Montréal, QC H2S 3H1, Canada}

\author[0000-0002-2598-0514]{Andrés~A.~Plazas~Malagón}
\affiliation{Kavli Institute for Particle Astrophysics and Cosmology, Stanford University, Stanford, CA 94305, USA}
\affiliation{SLAC National Accelerator Laboratory, Menlo Park, CA 94025, USA}
\affiliation{Department of Astrophysical Sciences, Princeton University, Peyton Hall, Princeton, NJ 08544, USA}

\author[0000-0001-7772-0346]{Nesar Ramachandra}
\affiliation{Argonne National Laboratory, 9700 South Cass Avenue, Lemont, IL 60439, USA}

\author[0000-0002-0978-5612]{Benjamin Remy}
\affiliation{Department of Astronomy and Astrophysics, University of Chicago, Chicago, IL 60637, USA}
\affiliation{NSF-Simons AI Institute for the Sky (SkAI), 172 E. Chestnut St., Chicago, IL 60611, USA}

\author[0000-0002-9641-4552]{Cécile Roucelle}
\affiliation{Université Paris Cité, CNRS, Astroparticule et Cosmologie,
F-75013 Paris, France}

\author[0000-0002-7951-4391]{Jaime Ruiz-Zapatero}
\affiliation{Advanced Research Computing Centre, University College London, 90 High Holborn, London WC1V 6LJ, UK}

\author[0000-0003-2497-6334]{Stefan Schuldt}
\affiliation{Dipartimento di Fisica, Universit\`a  degli Studi di Milano, via Celoria 16, I-20133 Milano, Italy}
\affiliation{Finnish Centre for Astronomy with ESO (FINCA), University of Turku, FI-20014 Turku, Finland}
\affiliation{Department of Physics, P.O. Box 64, University of Helsinki, FI-00014
Helsinki, Finland}
\affiliation{INAF - IASF Milano, via A. Corti 12, I-20133 Milano, Italy}

\author[0000-0002-1831-1953]{Ignacio Sevilla-Noarbe}
\affiliation{Centro de Investigaciones Energ\'eticas, Medioambientales y Tecnol\'ogicas (CIEMAT), Madrid, Spain}

\author[0009-0009-1590-2318]{Ved G. Shah}
\affiliation{Department of Physics and Astronomy, Northwestern University, Evanston, IL, USA}
\affiliation{Center for Interdisciplinary Exploration and Research in Astrophysics, Northwestern University, Evanston, IL, USA}
\affiliation{NSF-Simons AI Institute for the Sky (SkAI), 172 E. Chestnut St., Chicago, IL 60611, USA}

\author[0000-0003-2539-8206]{Tjitske Starkenburg}
\affiliation{Department of Physics and Astronomy, Northwestern University, Evanston, IL, USA}
\affiliation{Center for Interdisciplinary Exploration and Research in Astrophysics, Northwestern University, Evanston, IL, USA}
\affiliation{NSF-Simons AI Institute for the Sky (SkAI), 172 E. Chestnut St., Chicago, IL 60611, USA}

\author[0009-0005-6323-0457]{Stephen Thorp}
\affiliation{Institute of Astronomy and Kavli Institute for Cosmology, University of Cambridge, Madingley Road, Cambridge, CB3 0HA, UK}

\author[0000-0002-8313-7875]{Laura Toribio San Cipriano}
\affiliation{Centro de Investigaciones Energ\'eticas, Medioambientales y Tecnol\'ogicas (CIEMAT), Madrid, Spain}

\author[0000-0003-3520-2406]{Tilman Tröster}
\affiliation{Institute for Particle Physics and Astrophysics, ETH Zürich, Wolfgang-Pauli-Strasse 27, CH-8093 Zurich, Switzerland}

\author[0000-0002-3415-0707]{Roberto Trotta}
\affiliation{Theoretical and Scientific Data Science, International School for Advanced Study, Via Bonomea 265, I-34136 Trieste, Italy}
\affiliation{Imperial Centre for Inference and Cosmology (ICIC), Imperial College London, Blackett Laboratory, Prince Consort Road, London SW7 2AZ, UK}

\author[0000-0001-8638-2780]{Padma Venkatraman}
\affiliation{Department of Astronomy, University of Illinois Urbana Champaign, 1002 W. Green St., Urbana, IL, 61801, USA}

\author[0000-0002-4186-6164]{Amanda Wasserman}
\affiliation{Department of Astronomy, University of Illinois Urbana Champaign, 1002 W. Green St., Urbana, IL, 61801, USA}
\affiliation{NSF-Simons AI Institute for the Sky (SkAI), 172 E. Chestnut St., Chicago, IL 60611, USA}

\author[0000-0001-5535-0452]{Tim White}
\affiliation{Department of Statistics, University of Michigan, Ann Arbor, MI 48109, USA}

\author[0000-0001-6002-5128]{Justine Zeghal}
\affiliation{D\'epartement de Physique, Universit\'e de Montr\'eal, 1375 Avenue Th\'er\`ese-Lavoie-Roux, Montr\'eal, QC, H2V 0B3, Canada}
\affiliation{Mila - Quebec Artificial Intelligence Institute, Montréal, QC H2S 3H1, Canada}

\author[0000-0002-5596-198X]{Tianqing Zhang}
\affiliation{Department of Physics and Astronomy and PITT PACC, University of Pittsburgh, Pittsburgh, PA 15260, USA}

\author[0000-0001-5969-4631]{Yuanyuan Zhang}
\affiliation{NSF NOIRLab, 950 N. Cherry Ave., Tucson, AZ 85719, USA}

%% file: contributors.tex
\section*{Contributors}
As a way to provide transparency in a large, multi-author writing effort, this white paper adopts the CRediT\footnote{CRediT -- Contributor Roles Taxonomy: \url{https://credit.niso.org/}} taxonomy to report individual contributions. In this white paper, authorship roles are defined as follows:
\begin{description}[font=\normalfont\itshape, nosep]
  \item[Conceptualization] Paper- or section-level intellectual framing: defining scope, structure, key messages, and narrative arc.
  \item[Project administration] Process coordination: soliciting and organizing inputs, managing timelines and revisions, integrating contributions, and ensuring internal consistency across contributors.
  \item[Writing -- Original Draft] Substantive preparation of original text, including drafting new material and/or synthesizing multiple contributions into a coherent section or subsection.
  \item[Writing -- Review \& Editing] Substantive review and revision of the manuscript text, including critical feedback, edits for clarity and correctness, and incorporation of comments during iterative drafting.
\end{description}
Where shown, bracketed scope tags (e.g., \texttt{[§4.1]}) indicate the section(s) associated with a listed role; \texttt{[all]} denotes contributions spanning the full manuscript. Bolded scopes identify primary contributions, and unbolded scopes identify secondary contributions.

\vspace{0.5em}
\begin{small}
\setlength{\LTleft}{0pt}
\setlength{\LTright}{0pt}
\renewcommand{\arraystretch}{1.25}
\newlength{\contribwidth}
\setlength{\contribwidth}{\dimexpr\textwidth-5.5cm-4\tabcolsep\relax}
\begin{longtable}{|l|l|}
\hline
Name & Contribution \\
\hline
\endfirsthead
\hline
Name & Contribution \\
\hline
\endhead
\hline
\endfoot
Eric Aubourg & \parbox[t]{\contribwidth}{Writing -- Original Draft [\S3.9, \S5.1]} \\
Camille Avestruz & \parbox[t]{\contribwidth}{Writing -- Original Draft [\textbf{\S3.4}, \S3.8]} \\
Matthew R. Becker & \parbox[t]{\contribwidth}{Conceptualization [all]; Writing -- Review \& Editing [all]} \\
Biswajit Biswas & \parbox[t]{\contribwidth}{Writing -- Original Draft [\S3.8]} \\
Rahul Biswas & \parbox[t]{\contribwidth}{Writing -- Original Draft [\S3.1]; Writing -- Review \& Editing [all]} \\
Boris Bolliet & \parbox[t]{\contribwidth}{Writing -- Original Draft [\S5.2]} \\
Adam S. Bolton & \parbox[t]{\contribwidth}{Conceptualization [\textbf{\S6}]; Project administration [\textbf{\S6}]; Writing -- Original Draft [\textbf{\S6}]; Writing -- Review \& Editing [\textbf{\S6}]} \\
Clecio R. Bom & \parbox[t]{\contribwidth}{Conceptualization [\textbf{\S5.2}]; Project administration [\textbf{\S5.2}]; Writing -- Original Draft [\textbf{\S5.2}]; Writing -- Review \& Editing [\textbf{\S5.2}]} \\
Rapha\"el Bonnet-Guerrini & \parbox[t]{\contribwidth}{Writing -- Original Draft [\S4.1]; Writing -- Review \& Editing [\S4.1]} \\
Alexandre Boucaud & \parbox[t]{\contribwidth}{Writing -- Original Draft [\S3.9, \S5.1]} \\
Jean-Eric Campagne & \parbox[t]{\contribwidth}{Writing -- Original Draft [\S3.1, \S3.2, \S4.1, \S4.2, \S5.1]} \\
Chihway Chang & \parbox[t]{\contribwidth}{Writing -- Review \& Editing [\S3.3]} \\
Aleksandra \'Ciprijanovi\'c & \parbox[t]{\contribwidth}{Writing -- Original Draft [\S3, \S4]; Writing -- Review \& Editing [\S3, \S4]} \\
Johann Cohen-Tanugi & \parbox[t]{\contribwidth}{Writing -- Original Draft [\S3.8]; Writing -- Review \& Editing [all]} \\
Michael W. Coughlin & \parbox[t]{\contribwidth}{Writing -- Original Draft [\S6]} \\
John Franklin Crenshaw & \parbox[t]{\contribwidth}{Writing -- Original Draft [\S3.1]} \\
Juan C. Cuevas-Tello & \parbox[t]{\contribwidth}{Writing -- Original Draft [\S3.2]} \\
Juan de Vicente & \parbox[t]{\contribwidth}{Writing -- Original Draft [\S3.1]} \\
Seth W. Digel & \parbox[t]{\contribwidth}{Writing -- Original Draft [\S1]; Writing -- Review \& Editing [\textbf{all}]} \\
Steven Dillmann & \parbox[t]{\contribwidth}{Writing -- Original Draft [\S5.1, \S5.2]; Writing -- Review \& Editing [\S5.1, \S5.2]} \\
Mariano Javier de Le\'on Dominguez Romero & \parbox[t]{\contribwidth}{Writing -- Review \& Editing [all]} \\
Alex Drlica-Wagner & \parbox[t]{\contribwidth}{Writing -- Original Draft [\S3.8]; Writing -- Review \& Editing [all]}\\
Sydney Erickson & \parbox[t]{\contribwidth}{Writing -- Original Draft [\S3.2]; Writing -- Review \& Editing [\S3.2]} \\
Alexander T. Gagliano & \parbox[t]{\contribwidth}{Conceptualization [\textbf{all}]; Project administration [\textbf{all}]; Writing -- Original Draft [\textbf{all}]; Writing -- Review \& Editing [\textbf{all}]} \\
Christos Georgiou & \parbox[t]{\contribwidth}{Writing -- Original Draft [\S2, \S3.6]} \\
Aritra Ghosh & \parbox[t]{\contribwidth}{Writing -- Original Draft [\S5, \S6]} \\
Matthew Grayling & \parbox[t]{\contribwidth}{Writing -- Original Draft [\S3.5]} \\
Kirill A. Grishin & \parbox[t]{\contribwidth}{Writing -- Original Draft [\S3]}\\
Alan Heavens & \parbox[t]{\contribwidth}{Writing -- Original Draft [\S3.3]} \\
Lindsay R. House & \parbox[t]{\contribwidth}{Writing -- Review \& Editing [\S1]} \\
Mustapha Ishak & \parbox[t]{\contribwidth}{Writing -- Original Draft [\S3.6]} \\
Wassim Kabalan & \parbox[t]{\contribwidth}{Writing -- Original Draft [\S3.7, \S4.1]} \\
Arun Kannawadi & \parbox[t]{\contribwidth}{Writing -- Original Draft [\S3.10]} \\
Fran\c{c}ois Lanusse & \parbox[t]{\contribwidth}{Conceptualization [\textbf{all}]; Project administration [\textbf{all}]; Writing -- Original Draft [\textbf{all}]; Writing -- Review \& Editing [\textbf{all}]} \\
C. Danielle Leonard & \parbox[t]{\contribwidth}{Writing -- Original Draft [\textbf{\S3.6}]} \\
Pierre-Fran\c{c}ois L\'eget & \parbox[t]{\contribwidth}{Writing -- Original Draft [\S3.5]} \\
Michelle Lochner & \parbox[t]{\contribwidth}{Conceptualization [\textbf{\S5.1}, \textbf{\S1}]; Project administration [\textbf{\S5.1}]; Writing -- Original Draft [\textbf{\S1}, \textbf{\S4.3}, \textbf{\S5.1}, \textbf{\S9}]; Writing -- Review \& Editing [all]} \\
Yao-Yuan Mao & \parbox[t]{\contribwidth}{Writing -- Original Draft [\S8]} \\
Peter Melchior & \parbox[t]{\contribwidth}{Conceptualization [\textbf{\S4}]; Project administration [\textbf{\S4}]; Writing -- Original Draft [\textbf{\S4}]; Writing -- Review \& Editing [\textbf{\S4}]} \\
Grant Merz & \parbox[t]{\contribwidth}{Writing -- Original Draft [\S3.8]} \\
Martin Millon & \parbox[t]{\contribwidth}{Writing -- Original Draft [\S3.2]; Writing -- Review \& Editing [\S3.2]} \\
Anais M\"oller & \parbox[t]{\contribwidth}{Conceptualization [\textbf{\S4}]; Project administration [\textbf{\S4}]; Writing -- Original Draft [\S3, \textbf{\S4}]; Writing -- Review \& Editing [\textbf{\S4}]} \\
Gautham Narayan & \parbox[t]{\contribwidth}{Writing -- Review \& Editing [all]} \\
Yuuki Omori & \parbox[t]{\contribwidth}{Writing -- Original Draft [\S3.3]}\\
Hiranya Peiris & \parbox[t]{\contribwidth}{Writing -- Original Draft [\S3.1, \S3.3, \S3.7, \S4.1, \S4.2]; Writing -- Review \& Editing [all]} \\
Laurence Perreault-Levasseur & \parbox[t]{\contribwidth}{Writing -- Original Draft [\S3.2]; Writing -- Review \& Editing [\S4]}\\
Andr\'es A. Plazas Malag\'on & \parbox[t]{\contribwidth}{Writing -- Original Draft [\S3.10]}\\
Nesar Ramachandra & \parbox[t]{\contribwidth}{Writing -- Original Draft [\S3.6]} \\
Benjamin Remy & \parbox[t]{\contribwidth}{Writing -- Original Draft [\S3.3, \S4.2]} \\
C\'ecile Roucelle & \parbox[t]{\contribwidth}{Writing -- Original Draft [\S3.9, \S5.1]} \\
Jaime Ruiz-Zapatero & \parbox[t]{\contribwidth}{Writing -- Original Draft [\S4.1]} \\
Stefan Schuldt & \parbox[t]{\contribwidth}{Writing -- Original Draft [\textbf{\S3.2}]; Writing -- Review \& Editing [all]} \\
Ignacio Sevilla-Noarbe & \parbox[t]{\contribwidth}{Writing -- Original Draft [\S3, \S7, \S8]} \\
Ved G. Shah & \parbox[t]{\contribwidth}{Writing -- Original Draft [\S3]; Writing -- Review \& Editing [all]} \\
Tjitske Starkenburg & \parbox[t]{\contribwidth}{Writing -- Original Draft [\S3, \S4]; Writing -- Review \& Editing [\S3, \S4]} \\
Stephen Thorp & \parbox[t]{\contribwidth}{Writing -- Original Draft [\S3.1, \S3.2, \S3.3, \S3.7, \S4.1, \S4.2, \S7]; Writing -- Review \& Editing [all]} \\
Laura Toribio San Cipriano & \parbox[t]{\contribwidth}{Writing -- Original Draft [\S3]} \\
Tilman Tr\"oster & \parbox[t]{\contribwidth}{Writing -- Original Draft [\S3.3, \S3.6]} \\
Roberto Trotta & \parbox[t]{\contribwidth}{Writing -- Original Draft [\S3, \textbf{\S4.1}, \S8]; Writing -- Review \& Editing [all]} \\
Padma Venkatraman & \parbox[t]{\contribwidth}{Writing -- Original Draft [\S3.2]; Writing -- Review \& Editing [\S3.2]} \\
Amanda Wasserman & \parbox[t]{\contribwidth}{Writing -- Original Draft [\S3.5]} \\
Tim White & \parbox[t]{\contribwidth}{Writing -- Original Draft [\S3.9]} \\
Justine Zeghal & \parbox[t]{\contribwidth}{Writing -- Original Draft [\textbf{\S3.3}, \S4.1]} \\
Tianqing Zhang & \parbox[t]{\contribwidth}{Project administration [\S3.8, \S3.9]; Writing -- Original Draft [\S3.1]; Writing -- Review \& Editing [\S3.8, \S3.9]} \\
Yuanyuan Zhang & \parbox[t]{\contribwidth}{Writing -- Original Draft [\textbf{\S3.4}]} \\
\end{longtable}
\end{small}

%% file: sections/executive_summary.tex
\newpage
\section{Executive Summary}
\label{sec:exec_summary}

The \acrlong{lsst} \acrlong{desc} (\acrshort{lsst} \acrshort{desc}) is an international collaboration whose mission is to measure the cosmic expansion history and the growth of structure using data from the Vera C. Rubin Observatory, thereby constraining the nature of dark energy and dark matter. Achieving these science goals requires jointly analyzing multiple cosmological probes---weak and strong gravitational lensing, galaxy clusters, Type Ia supernovae, and large-scale structure---each presenting distinct analysis challenges at LSST's unprecedented data volumes. Extracting robust cosmological constraints demands methods that deliver trustworthy uncertainty quantification, remain robust to systematic effects and model misspecification, and scale to the full petabyte-scale survey. These requirements motivate the integration of \acrfull{ai} and \acrfull{ml} into DESC pipelines. DESC's combination of community-accessible data, mature simulation infrastructure, and rigorous scientific standards makes the collaboration an excellent testbed for developing robust AI/ML practices for fundamental physics. 

Recognizing this situation, the DESC formed the \textit{AI for DESC Task Force} with the following charge:
\begin{itemize}
    \item Catalog the AI/ML needs, use cases, and projects in DESC.
    \item Identify current gaps in the adoption of AI/ML methodologies by leveraging expert domain knowledge.
    \item Identify the computational resources, storage, data access, and human research and managerial time needed to take full advantage of AI/ML-related opportunities.
    \item Identify either qualitatively or quantitatively the projected gains in DESC’s science that would result from pursuing AI/ML-related opportunities.
\end{itemize}
The response to the task force charge is presented in this white paper. It demonstrates the breadth and importance of AI and ML research within DESC, and highlights the challenges and promising pathways for future work.

In this Executive Summary, we synthesize key recommendations and opportunities into a coherent AI/ML strategy for DESC. Three core principles guide this strategy: 

\begin{itemize}
    \item \textbf{AI/ML tools should be carefully integrated into DESC pipelines} to facilitate scientific analyses while fulfilling the stringent requirements of precision cosmology and preserving scientific accountability and transparency.

    \item A \textbf{durable AI/ML ecosystem should be built} within DESC and maintained over the survey lifetime, for the collaborative development, validation, and deployment of production-grade AI/ML tools.

    \item AI/ML must be integrated into DESC in ways that \textbf{preserve and support the human-centric nature of research}, improve accessibility, strengthen collaboration quality, and amplify rather than supplant members' contributions.
\end{itemize}

We have defined a series of recommendations (R) and opportunities (O) in several key areas within DESC in support of these principles. \textit{Recommendations} are actions that the collaboration should undertake to meet its scientific requirements and ensure robust integration of AI and ML into DESC pipelines. \textit{Opportunities} indicate areas where DESC can extend beyond its requirements and assume a leadership role, influence broader community standards, or explore higher-risk, higher-reward efforts. We summarize these below, along with references throughout the paper where they are discussed.

\paragraph{Advancing Key Methodological Research Directions} Challenges such as uncertainty quantification, robustness to model misspecification, and novelty detection recur across DESC science cases. Progress on these foundational challenges will benefit all probes and merit dedicated effort.

\begin{itemize}
    \item \textbf{R1: Prioritize Fundamental Methodological Research.}  Foster collaboration-wide research in several critical areas: quantification of systematic and statistical uncertainties, simulation-based inference robustness, physics-informed modeling (hybrid generative-physical architectures), validation of neural posteriors, and novelty detection. Progress on these fundamental challenges will have an outsized impact across many DESC science cases. (\autoref{sec3:use_case_for_aiml}, \autoref{sec4:aiml_research})

    \item \textbf{O1: Methodological Leadership in Trustworthy AI.} The challenges DESC faces (robust inference under misspecification, calibrated uncertainty quantification at scale, physics-informed learning) are frontier problems in machine learning broadly, creating natural opportunities to attract specialist collaborators and position DESC as a leader in trustworthy AI for fundamental science. (\autoref{sec4:aiml_research}, \autoref{sec7:broader_coordination})
    
    \item \textbf{O2: DESC Simulation Assets as Community Benchmarks.} DESC's combination of petabyte-scale community data, stringent scientific requirements, and rich simulation assets---e.g.\ the \acrfull{plasticc}, \acrfull{elasticc}, and \acrfull{cosmodc2}---makes it an ideal testbed for pioneering robust AI/ML practices. Benchmarks and governance standards developed here can become reference implementations for fundamental physics, and attract colleagues in mathematics and computer science who see DESC's frontier challenges as compelling application areas for new methods. (\autoref{sec3:use_case_for_aiml}, \autoref{sec4:aiml_research}, \autoref{sec7:broader_coordination})
\end{itemize}

\paragraph{Foundation Models} Foundation models, which produce generalizable representations of large-scale, heterogeneous, and multi-modal datasets, are transforming AI capabilities. DESC must develop both the infrastructure to deploy them and the benchmarks to validate them for precision cosmology.
\begin{itemize}

\item \textbf{R2: Develop Shared Foundation Model Infrastructure.} Build a shared foundation model backbone for DESC, consistent across data modalities and of production-grade quality, and served behind stable APIs. (\autoref{sec5:emerging_tech}, \autoref{sec6:infra_requirements})

\item \textbf{R3: Establish DESC-specific Foundation Model Validation Standards.} Create benchmarks that go beyond industry practice: uncertainty calibration, robustness to systematics, sensitivity to training biases, stress tests under distribution shift (temporal, spatial, cross-survey). 
Develop astronomy-specific interpretability tools to verify the physically meaningful structure preserved within model representations. (\autoref{sec5:emerging_tech}, \autoref{sec:aiml_risks})

\item \textbf{O3: Leadership of Rubin-wide Development of Foundation Models.} DESC could play a central role in coordinating foundation model development across Rubin Science Collaborations and \acrfull{lincc}, leveraging distributed computing resources from universities to \acrfull{doe} or \acrfull{eurohpc} facilities. (\autoref{sec5:emerging_tech}, \autoref{sec7:broader_coordination})
\end{itemize}

\paragraph{Large Language Models \& Agentic AI} \Acrfullpl{llm} and agentic AI offer avenues to accelerate research and lower the barrier to entry for complex cosmological analyses in DESC. Harnessing this potential responsibly will require thoughtful governance and rigorous validation frameworks.

\begin{itemize}

  \item \textbf{R4: Establish Governance for LLMs and Agentic Systems.} Coordinate DESC-wide 
  activities involving LLMs and agents, establish best practices including evaluation, review, and tiger-team review of pilot studies. Include critical discussions of the technology's limitations and effects on human researchers, with input from experts across domains. Engage with Rubin Data Management to ensure that agentic AI can interface with data products effectively and reliably. (\autoref{sec5:emerging_tech}, \autoref{sec:aiml_risks})

  \item \textbf{R5: Build Natural Language Interfaces to DESC Resources.} Develop \acrfull{rag}-based interfaces to DESC documentation, simulations, and data products, lowering onboarding barriers and democratizing access to complex pipelines. (\autoref{sec5:emerging_tech}, 
  \autoref{sec6:infra_requirements})

  \item \textbf{O4: Pioneering Agentic AI for Scientific Rigor and Reproducibility.} An important application of this work could be 
  ``DESC research agents" that automate execution, documentation, and validation of analyses 
  against standardized benchmarks, coupling these systems to clear governance and tiger-team review procedures so that agentic workflows enhance transparency, provenance, and trust in DESC results. (\autoref{sec5:emerging_tech}, \autoref{sec:aiml_risks})
\end{itemize}

\paragraph{Infrastructure \& Software} DESC has a mature ecosystem of cosmological analysis pipelines. Building on this foundation, strategic development of AI software stacks, differentiable programming, and computing infrastructure can act as multipliers that benefit all science cases.

\begin{itemize}
  \item \textbf{R6: Establish a Durable AI Software Stack.} Adopt a coherent set of frameworks, tooling, and model export standards. The stack should be portable across DESC computational facilities, sustainable over the 10-year survey, and prioritize open governance to avoid proprietary lock-in. (\autoref{sec6:infra_requirements})

  \item \textbf{R7: Develop a Differentiable Programming Ecosystem.} Adoption of an interoperable differentiable programming ecosystem (e.g. based on JAX) will act as a multiplier, simultaneously enabling gradient-based sampling, GPU acceleration, hybrid physics-ML models, and end-to-end optimization across DESC pipelines. (\autoref{sec3:use_case_for_aiml}, \autoref{sec4:aiml_research}, \autoref{sec6:infra_requirements})

\item \textbf{R8: Secure Access to Emerging Computing Infrastructure.} Significant new AI-oriented 
computing is becoming available: DOE infrastructure such as the \acrfull{amsc}; the \acrfull{idac} 
network; and EuroHPC systems such as Leonardo in Italy, \acrfull{lumi} in Finland, and the \acrfull{jupiter} exascale system in Germany. 
DESC should engage early to shape these resources for cosmology and secure allocations for 
foundation model training at scales infeasible on current systems. 
(\autoref{sec6:infra_requirements}, \autoref{sec7:broader_coordination})
\end{itemize}

\paragraph{Organizational Structure \& Governance} The DESC is organized into computing, technical, and analysis \acrfullpl{wg}\footnote{\url{https://lsstdesc.org/pages/organization.html}}, with analysis working groups primarily aligned with key cosmological probes. Effective AI/ML integration across these groups requires consistent coordination mechanisms and clear standards for development, validation, and deployment.

\begin{itemize}
  \item \textbf{R9: Develop DESC-wide AI/ML Coordination Mechanisms.} Establish structures (e.g., standing working group, cross-WG task forces, regular interchange meetings) to share methodological innovations across probes, tackle common challenges collectively, and minimize duplication. Facilitate rapid dissemination through workshops, tutorials, and methodological discussions. (\autoref{sec3:use_case_for_aiml})

  \item \textbf{R10: Develop AI/ML Best Practice Guidelines.} Create guidelines to help DESC members develop robust AI/ML analyses, covering topics such as reproducibility, provenance tracking, validation checks, and comprehensive benchmarking—particularly for foundation models and other shared deliverables whose broad applicability demands thorough vetting before widespread adoption. (\autoref{sec3:use_case_for_aiml}, 
  \autoref{sec4:aiml_research}, \autoref{sec5:emerging_tech})
\end{itemize}

\paragraph{Human Capital \& Sustainability} The promise of AI/ML for accelerating cosmology with LSST will not be realized without training and support of DESC members. Sustainable adoption of AI/ML also requires attention to the growing computational demands and resulting footprint these methods entail. 
\begin{itemize}
  \item \textbf{R11: Focus on AI/ML for Augmenting Rather Than Replacing Understanding.} DESC must strengthen and maintain the technical literacy of the collaboration in AI/ML applications as tools for science rather than supplanting understanding. (\autoref{sec:aiml_risks})
  
  \item \textbf{R12: Track and Optimize Resource Footprint.} DESC should develop tools for monitoring and optimizing computational resource usage of AI/ML models, enabling the collaboration to maximize scientific productivity and make informed decisions about resource allocation and environmental impact. (\autoref{sec:aiml_risks})
\end{itemize}

\paragraph{External Coordination \& Partnerships} DESC operates within a rich ecosystem of other Rubin science collaborations, AI institutes, cosmology experiments, and alert brokers that filter streaming Rubin data. Deliberate coordination between these groups will amplify impact and avoid duplicated effort.
\begin{itemize}

  \item \textbf{R13: Coordinate Across Science Collaborations.} Partner with other LSST collaborations and other cosmology experiments. The former include 
  the \acrfull{issc}, \acrfull{tvs}, \acrfull{slsc}, \acrfull{agnsc}, and \acrfull{galsc}.  The latter include the \acrfull{desi}, the \acrfull{4most}, the \acrfull{esa} \textit{Euclid} Mission science teams, and the \textit{Nancy Grace Roman Space Telescope} science collaborations.  Areas of coordination should include methodological development, time-series and broker stress-testing, deblending/morphology benchmarks, and sharing tools and best practices. (\autoref{sec7:broader_coordination})

  \item \textbf{R14: Engage with AI Institutes and Networks.} \acrfull{nsf}--Simons AI Institutes 
  (with explicit LSST/cosmology themes), and European networks such as the \acrfull{eucaif}\footnote{\url{https://eucaif.org/}} and \acrfull{ellis}\footnote{\url{https://ellis.eu}}, are natural partners. 
  Build systematic engagement through co-funded postdocs, shared workshops, joint proposals, 
  and benchmark datasets. These efforts would connect DESC to the broader AI-for-science ecosystem. 
  (\autoref{sec7:broader_coordination})

    \item \textbf{R15: Develop the Human-Machine Interface.} Develop close connections between DESC, other LSST science collaborations, in-kind follow-up programs, alert broker teams, LSST data management, and citizen scientists, to facilitate active learning for classification, anomaly detection, and human-in-the-loop interpretability. (\autoref{sec4:aiml_research}, \autoref{sec7:broader_coordination})

  \item \textbf{O5: DESC Integration with the Broker Ecosystem.} DESC members are embedded in 
  all seven Rubin Community Broker teams—tight coordination gives direct leverage over SN~Ia sample 
  purity, selection effects, and host-galaxy priors, plus an on-ramp from research prototypes 
  to community-facing services. (\autoref{sec3:use_case_for_aiml}, \autoref{sec7:broader_coordination})
\end{itemize}

Implementing these recommendations and capitalizing on these opportunities would position DESC to fully exploit LSST's statistical power for cosmology while uncovering unexpected phenomena in the largest optical astronomical dataset ever collected. This will require sustained investment in researchers who bridge domain science and AI/ML methodology. Such investment would benefit not only DESC, but the broader effort to advance AI as a tool for fundamental scientific discovery.

%% file: sections/introduction.tex
\newpage
\section{Introduction}

The Vera C. Rubin Observatory's \acrshort{lsst} will produce unprecedented volumes of heterogeneous data (images, catalogs, alerts) whose full scientific exploitation demands continued methodological innovation. The mission of the \acrshort{desc} is to convert these data into robust constraints on the dark sector by jointly measuring the cosmic expansion history and the growth of structure, thereby shedding much-needed light on dark energy, dark matter, and possible deviations from general relativity. Delivering on these objectives requires methods that are statistically powerful, scalable, and operationally reliable. Recent advances in \acrshort{ai} and \acrshort{ml} show great promise for critical data analysis roles but still need to meet stringent requirements to be truly useful. \acrlong{ai} (\acrshort{ai}) refers broadly to systems capable of performing tasks that typically require human intelligence, including reasoning, perception, learning, and decision-making. \acrlong{ml} (\acrshort{ml}) is a subfield of AI in which algorithms learn patterns and relationships from data to make predictions or decisions, encompassing both classical methods (e.g., random forests, Gaussian processes) and deep learning (multi-layer neural networks). In the DESC context, ML methods learn mappings between variables (e.g., photometry to redshift, galaxy fields to cosmological parameters) and can be deployed at multiple stages of analysis. Their scientific utility, however, hinges on trustworthy uncertainty quantification and reproducible integration within DESC workflows; without these elements, ML methods cannot meet the stringent requirements of cosmological inference. In parallel, we use AI to denote systems capable of complex cognitive tasks---such as reasoning, knowledge synthesis, and natural language understanding---that can potentially orchestrate tools, generate code, and reshape scientific workflows. As with ML, turning recent AI advances into reliable accelerators of discovery remains an open problem that requires careful evaluation and governance.

The strategic question for DESC is therefore how to develop and integrate AI/ML \textit{the right way}, so that these approaches become dependable components of LSST-era analyses. Intrinsically, this question is relevant to a broad range of scientific endeavors and collaborations. Still, DESC is well positioned to pioneer robust AI/ML practices for fundamental physics as an international collaboration that works with community-accessible data, has a strong open-source culture, and pursues extremely demanding scientific objectives. In this white paper, we set out a strategic framework for how DESC should organize its AI/ML efforts, prioritize methodological investments, and respond effectively to new opportunities arising from rapid AI/ML progress. This paper is structured around four interconnected perspectives on AI/ML within DESC, each building on the previous to articulate a comprehensive strategy:

\paragraph{\autoref{sec3:use_case_for_aiml}: The Current Landscape: ML Across DESC Science} Machine learning is not a future aspiration for DESC: it is already deeply embedded in current science workflows. Here, we survey how ML methodologies intersect with DESC's primary cosmological probes: strong and weak gravitational lensing, galaxy clusters, \acrfull{snia} cosmology, large-scale structure, as well as cross-cutting analysis components including simulations, theory and modeling, deblending, \acrfull{photoz} estimation, and shape measurement. This inventory reveals a striking pattern: the same core methodologies (e.g. simulation based inference, differentiable programming, deep learning) appear repeatedly across disparate science cases, while the same fundamental challenges (e.g. uncertainty quantification, robustness to covariate shift and to model misspecification) represent concrete challenges across multiple working groups (see \autoref{fig:chord-diagram}).

\begin{figure}
    \centering
    \includegraphics[width=\linewidth]{figures/chord-diagram}
    \caption{Transversal connections between DESC science applications (left), AI/ML methodologies (top), and shared challenges (right), as surfaced by \autoref{sec3:use_case_for_aiml}. The recurring appearance of the same methods and challenges across disparate science cases motivates collaboration-wide coordination of AI/ML efforts rather than siloed development within individual working groups. An interactive version of this diagram is available at \url{https://lsstdesc.org/AI_For_DESC/figures/chord-diagram.html}}
    \label{fig:chord-diagram}
\end{figure}

\paragraph{\autoref{sec4:aiml_research}: Lifting the Limits of ML: Methodological Research Priorities} Building on the challenges surfaced in Section~3, we identify the key methodological research axes where targeted investment can lift current limitations and enable ML methods to meet the precision and reliability standards demanded by LSST-era cosmology. We organize these priorities around several interconnected themes:
\begin{itemize}
    \item \textbf{Bayesian inference and \acrfull{acr:uq}:} Developing fast and scalable inference techniques that may unlock promising high-dimensional hierarchical models. Improving methods for reliably estimating uncertainty arising from limited training data and model limitations.
    \item \textbf{\Acrfull{acr:sbi} and model misspecification:} Advancing \acrfull{nde} techniques, optimal summarization methods, and diagnostics for detecting and mitigating the biases introduced when training simulations imperfectly represent real observations or when training datasets fail to capture the full diversity of LSST data.
    \item \textbf{Physics-informed approaches:} Hybridizing explicit physical models with flexible generative components (flows, diffusion models) and advancing differentiable programming frameworks that embed cosmological theory directly into ML architectures, ensuring that learned components remain interpretable, physically consistent, and robust to extrapolation.
    \item \textbf{Novelty detection and discovery:} Developing representation learning and active human--AI collaboration frameworks capable of identifying rare, previously unmodeled phenomena in LSST's vast data volumes.
\end{itemize}
These research directions are not purely academic exercises; they directly address the technical barriers that limit the deployment of ML at scale for DESC's most ambitious analyses. We articulate not only what needs to be developed, but why these specific advances matter for cosmological inference and how DESC can contribute to the broader AI research ecosystem by presenting demanding, scientifically motivated benchmarks.

\paragraph{\autoref{sec5:emerging_tech}: Looking Forward: Foundation models and Agentic AI}
While Sections~3 and~4 focus on current ML applications and their refinement, Section~5 adopts a forward-looking perspective, examining how two emerging AI paradigms (\textit{data foundation models} and \textit{LLM-based agentic systems}) have the potential to reshape large sections of DESC workflows in ways that go qualitatively beyond incremental improvements to existing methods.
\begin{itemize}
    \item \textbf{\Acrfullpl{fm}}, trained at scale on heterogeneous data modalities (images, spectra, time series, catalogs), offer the promise of \textit{reusable representations} that can be rapidly fine-tuned or directly deployed across a wide range of downstream tasks (classification, regression, anomaly detection, simulation-based inference) without retraining from scratch for each application. For DESC, this paradigm shift could enable unified, survey-scale feature extractors that serve as common backbones for weak lensing, photometric redshifts, transient classification, and more, dramatically reducing duplication of effort while ensuring cross-probe consistency. However, realizing this vision requires careful attention to uncertainty propagation, robustness to distribution shifts, architectural choices suited to astronomical data, and rigorous, community-governed benchmarking to ensure that foundation models meet DESC's validation standards.

    \item \textbf{LLM-driven agentic systems} are rapidly evolving from research prototypes into tools capable of orchestrating complex scientific workflows: querying databases, generating and executing code, synthesizing literature, and autonomously iterating on analyses. These systems offer tantalizing possibilities for accelerating exploratory research, onboarding new collaboration members, and scaling human oversight across large analysis campaigns. Yet they also introduce new risks: biased recommendations, irreproducible results, and erosion of scientific understanding if deployed without governance. Section~5 outlines both the transformative potential and the implementation requirements (provenance tracking, human-in-the-loop validation, benchmark design, and clear policies on data rights and model transparency) necessary to integrate agentic AI into DESC in ways that guarantee scientific rigor.
\end{itemize}
The forward-looking stance of Section~5 is deliberate: DESC must not only respond to today's ML methods but actively shape the trajectory of emerging AI technologies by setting clear scientific requirements, contributing demanding use cases to the broader AI research community, and pioneering governance frameworks that other collaborations can learn from.

\paragraph{\autoref{sec6:infra_requirements}, \ref{sec7:broader_coordination}, \ref{sec:aiml_risks}: Operationalizing AI/ML: Infrastructure, Coordination, and Risk Management}
Even the most sophisticated AI/ML methods will have limited impact if they cannot be reliably deployed, maintained, and integrated into DESC's production pipelines. The final sections of this white paper address the operational foundations required to translate research prototypes into dependable cosmological infrastructure. Section~6 details the \textit{infrastructure requirements} across software, computing, and data:
\begin{itemize}
    \item \textbf{Software:} Establishing a robust, collaboration-endorsed AI software stack (frameworks, experiment tracking, model registries, continuous integration/continuous deployment for models) that ensures reproducibility, portability across DESC computing facilities, and long-term sustainability over the LSST decade. This includes strategies for integrating AI components into DESC analysis pipelines and for managing the rapidly evolving ecosystem of LLMs and agentic frameworks.
    \item \textbf{Computing:} Securing the \acrfull{gpu} allocations, distributed training capabilities, and co-located data access necessary for foundation model development, large-scale simulation-based inference campaigns, and real-time alert stream processing. This involves strategic coordination with national labs such as the \acrfull{alcf} and \acrfull{olcf}, emerging initiatives (\acrshort{amsc}, \acrshortpl{hpdf}), and international partners (\acrshort{eurohpc}, \acrshortpl{idac}).
    \item \textbf{Data:} Ensuring that LSST data products, multi-survey training datasets, and simulation outputs are accessible, well-documented, and equipped with the interfaces---e.g., \acrfullpl{api}, streaming services, tokenization strategies---required for efficient AI/ML workflows. This includes establishing shared repositories, benchmark datasets, and provenance standards.
\end{itemize}

Section~7 broadens the scope to examine \textit{opportunities for coordination beyond DESC}: with the other Rubin LSST Science Collaborations, Stage-IV experiments (in particular \acrshort{desi}, \acrshort{4most}, Roman, Euclid), AI institutes (e.g., NSF--Simons institutes, CosmicAI), European networks (e.g., \acrshort{eucaif}, \acrshort{ellis}), and the Rubin alert broker ecosystem. These partnerships offer opportunities for shared training data, cross-survey foundation models, joint benchmark development, and access to specialized compute resources and expertise. DESC is uniquely positioned to act as both a consumer and a driver of AI methodologies within this broader ecosystem, articulating the demanding requirements of precision cosmology while contributing validated methods and datasets that benefit the wider community.

Finally, Section~8 confronts the \textit{risks and challenges} inherent in DESC's increasing reliance on AI/ML: model miscalibration, opaque failure modes, reproducibility challenges, data governance complexities, and the potential erosion of human scientific understanding. We outline concrete mitigation strategies (validation protocols, redundancy in critical analyses, provenance tracking, training programs, and governance structures) that apply the same rigor to AI components as to any other element of the cosmological inference pipeline.

\paragraph{\autoref{sec:conclusion}: Conclusion} Viewed as a whole, this paper shows that AI/ML is already central to DESC science (\autoref{sec3:use_case_for_aiml}), but unlocking its full potential requires targeted methodological research (\autoref{sec4:aiml_research}), proactive engagement with emerging technologies (\autoref{sec5:emerging_tech}), and robust operational foundations (\autoref{sec6:infra_requirements}--\ref{sec:aiml_risks}). The transversality of methods and challenges across DESC working groups demands deliberate coordination to prevent fragmented effort and ensure that best practices, validated tools, and lessons learned propagate rapidly throughout the collaboration. By articulating this vision (grounded in current capabilities, guided by research priorities, forward-looking in its engagement with foundation models and agentic AI, and operationally realistic about infrastructure and risk) DESC can position itself not only to meet its own science goals but to pioneer robust AI/ML practices for fundamental physics that serve as a model for the broader community. 

%% file: sections/sec3_ai_in_desc.tex
\newpage
\section{The Current Landscape of ML Across DESC Science}
\label{sec3:use_case_for_aiml}

The science goals of \acrshort{desc} place unusually stringent demands on statistical methodology. Extracting percent-level constraints on dark energy, dark matter, and tests of gravity from Rubin \acrshort{lsst} data requires not only exquisite control of observational systematics, but also analysis pipelines that can efficiently exploit information distributed across billions of galaxies, multiple probes, and heterogeneous data modalities (images, catalogs, time series, simulations). While \acrshort{ai} (in the sense of \acrshortpl{llm} and agents) has not yet significantly started to impact DESC, \acrshort{ml} is already embedded in many of these workflows and its importance will only grow as analyses become more ambitious and data volumes increase.

In this section, we survey existing intersections of \acrshort{ml} methods with DESC science, organized by application area, including photometric redshifts, strong and weak lensing, and galaxy clusters; supernovae and transients; cosmological theory and simulations; deblending; and shape measurement. Each subsection highlights both current capabilities and open challenges, to clarify where targeted methodological investment will yield the most significant scientific returns for DESC.

\textit{A note on reading this section:} In the subsections that follow, each DESC science application is accompanied by a summary box highlighting the \acrshort{ml} methodologies it employs and the challenges it faces. As you read through these applications, pay attention to how often the same methods (\acrshort{acr:sbi}, differentiable programming, \acrshort{nde}) and the same challenges (covariate shifts, uncertainty quantification, scalability, data sparsity, metrics) recur across seemingly disparate topics. This reveals a fundamental pattern: a small set of transversal methodologies and challenges cuts across the entirety of DESC's \acrshort{ml} applications, as illustrated in Figure~\ref{fig:chord-diagram}. This pervasive transversality has direct implications for how DESC should organize its AI/ML efforts, which we synthesize at the end of this section and develop further in Section~\ref{sec4:aiml_research}.

\subsection{Photometric redshifts} \label{sec3:photo-z}
\begin{ThemeBoxA}[]
\themebullet \themekey{Methodology.} \meth{gaussian-process}, \meth{neural-density-estimation}, \meth{som}, \meth{transformer}, \meth{hierarchical-bayes}, \meth{emulator}, \meth{neural-surrogate}, \meth{diffusion-model} \\
\themebullet \themekey{Challenges.} \challenge{covariate-shift}, \challenge{uq}, \challenge{scalability}, \challenge{metrics} \\
\themebullet \themekey{Opportunities.} Multi-survey training, simulation infrastructure, hierarchical inference
\end{ThemeBoxA}

The inference of \acrshortpl{photoz} represents a foundational challenge for \acrshort{lsst}, where the vast majority of tens of billions of detected galaxies will lack spectroscopic redshift measurements due to both observational time constraints and the intrinsic faintness of the sample. Photometric redshifts are derived by establishing empirical or physically motivated mappings between broadband photometry (including colors, magnitudes), morphology, and redshift. This process is fundamentally limited by our incomplete knowledge of galaxy \acrfullpl{sed}, the spatial and temporal variations of stellar populations in galaxies, and the nature and distribution of attenuating dust. However, the accuracy and reliability of photo-$z$ estimation is critical across virtually all extragalactic LSST science cases, including weak gravitational lensing, large-scale structure, galaxy cluster cosmology, and supernova surveys. To achieve the \acrshort{desc} goals for constraining the dark energy equation of state, calibration of photo-$z$ estimates must reach the $0.002 \times (1+z)$ level for the first year of LSST data \citep{v1DESC-SRD}. Achieving these benchmarks necessitates not only accurate point predictions but also \textit{well-calibrated uncertainty quantification}, motivating an emphasis on probabilistic methods. In addition, model benchmarking requires a unified framework for per-galaxy photo-$z$ algorithms, ensemble calibration algorithms, mock data generation, and performance evaluation. These tasks are fulfilled by the \acrlong{rail} \citep[\acrshort{rail};][]{RAIL_2025}, a photo-$z$ library developed by \acrshort{lincc} and DESC.

\paragraph{Supervised Photo-$z$ Estimation} The empirical approach to  photo-$z$ inference is to learn a mapping from observed broadband photometry or imaging to redshift, leveraging spectroscopic training samples. From catalog-level photometry, empirical regressors such as random forests in \acrlong{tpz}~\citep[\acrshort{tpz};][]{carrasco2013tpz}, gradient boosting machines \citep[\href{https://github.com/rizbicki/FlexCoDE}{\tt FlexZBoost};][]{izbicki2017converting,2020Dalmasso_FlexZBoost}, and Gaussian processes~\citep[\acrshort{gpz};][]{almosallam2016gpz} have demonstrated competitive performance by constructing mappings from color-magnitude space to redshift. Neural networks, including both fully connected and specialized architectures, have proven particularly adept at capturing complex relationships in high-dimensional photometric data \citep{Collister2004}. On the other hand, nearest-neighbor approaches such as \acrlong{knn} and \acrlong{cmnn}~\citep[\acrshort{knn} and \acrshort{cmnn};][]{graham2017photometric} take the training as a reference sample and compute the average redshift of neighbors of the target galaxy. In a similar way, \acrlong{dnf}~\citep[\acrshort{dnf};][]{2016DeVicente_DNF} performs a regression on the neighborhood, which allows the construction of a local linear model for each galaxy. \\ 
Contemporary approaches have evolved from point estimators to full probabilistic models capable of capturing the full conditional distribution $p(z \mid \text{photometry})$, with \acrshort{nde} techniques \citep[e.g., \href{https://github.com/jfcrenshaw/pzflow}{\tt PZFlow};][]{2024Crenshaw_PZFlow} enabling flexible, well-calibrated redshift \acrfullpl{pdf} via maximum likelihood training. Complementing catalog-based methods, image-based inference circumvents the information bottleneck imposed by aperture photometry by operating directly on multi-band pixel data, delegating feature extraction to deep neural networks that leverage morphology and spatial structure inaccessible to catalogs. The \acrfull{deepdisc} framework \citep{Merz23DeepDISC,Merz25DeepDISCpz} exemplifies this approach, integrating object detection, segmentation, and redshift estimation into a unified pipeline using \acrlong{mvit} \citep[\acrshort{mvit};][]{mvitv2} as the backbone feature extractor coupled with mixture density networks for probabilistic PDF estimation. Beyond redshift estimation, tomographic bin assignment for 3$\times$2pt analyses is itself amenable to supervised learning; the DESC Tomography Optimization Challenge \citep{Zuntz:2021} benchmarked multiple algorithms for this task, and subsequent work has shown that neural network classifiers can identify galaxies likely to be correctly binned, improving cosmological constraints \citep{Moskowitz:2023}. \\
Despite their sophistication, \textit{all supervised \acrshort{ml} methods remain fundamentally limited by the quality and representativeness of their spectroscopic training samples}: spectroscopic incompleteness, magnitude-limited surveys, and selection biases induce systematic offsets and distortions in the learned photo-$z$ mapping, particularly at faint magnitudes and high redshifts where spectroscopic follow-up is most incomplete \citep{newman2022}. This represents the main challenge for photo-$z$ today and has motivated dedicated calibration strategies.
A further problem is the ``implicit prior'' imposed by each photo-$z$ method \citep{schmidt2020}.
These priors, which have a large impact on photo-$z$ estimates, are opaque and difficult to quantify, making it difficult to compare and combine photo-$z$ posteriors provided by different methods.

\paragraph{Calibration Strategies to Account for Covariate Shifts}  \Acrfullpl{acr:som} have emerged as the preeminent unsupervised learning technique for diagnosing and mitigating the biases caused by covariate shifts by performing non-linear dimensionality reduction of photometric feature vectors onto a discrete two-dimensional grid \citep{Masters2015}. SOM-based calibration approaches, such as those deployed in \acrfull{des} Year 3 \citep{Myles2021} and \acrfull{kids} analyses \citep{wright2020a, wright2020b, wright2025, vanDenBusch2022}, directly assign photometric galaxies the empirical redshift distribution of spectroscopic galaxies in their SOM cell while down-weighting or even rejecting regions of color space poorly represented in the spectroscopic catalog. More sophisticated SOM-guided data augmentation strategies selectively populate under-represented SOM cells with simulated galaxies from mock catalogs improving ML model performance where spectroscopic coverage is deficient \citep{Moskowitz2024, Zhang2025}. An alternative approach to SOM for covariate shift mitigation (and without using data augmentation) is represented by stratification by the propensity score (defined as the probability of a covariate vector to be admitted as part of the training set) of both training and target data. Within each propensity score group, supervised photo-$z$ can proceed with any method of choice. This \acrfull{stratlearn} approach is theoretically guaranteed (under some conditions) to cancel covariate shift~\citep{Autenrieth_2023}. It has demonstrated state-of-the-art performance in the \acrshort{plasticc} \citep{PLAsTiCC1810.00001} \acrshort{snia} classification challenge, a factor of $\sim 2$ improvement in photo-$z$ calibration from the cosmic shear KiDS+VIKING-450 dataset~\citep{Autenrieth_2024} and a reduced fraction of catastrophic errors and one order of magnitude improvement of the bias for simulated photo-$z$ reconstruction~\citep{Moretti_2025}.

\paragraph{Hybrid Template-Based Estimators}
In contrast to empirical photo-$z$ estimators, a broad class of ``template-based'' photo-$z$ estimators \citep[e.g.,][]{eazy,lephare} attempt to circumvent the problem of covariate shift using physical models of galaxy SEDs.
These estimators trade the problem of covariate shift for the problem of model misspecification.
Hybrid methods, however, attempt to combine the strengths of empirical and template-based estimators by deriving SED templates in a physics-informed, data-driven manner \citep{budavari2000,Csabai2000}.
These models have been shown to deliver higher-quality photo-$z$ estimates than traditional template-based estimators while suffering less from covariate shift than pure empirical methods \citep{crenshaw2020,li2025}.
They do not perform as well in-distribution as pure empirical methods, however, and still rely on spectroscopic calibration sets.
It may be possible to remedy these defects by implementing hybrid, physics-informed models in deep learning frameworks to enable self-supervised learning without reliance on spectroscopic data sets \citep{2021Boone_ParSNIP}.

\paragraph{Population-Level Hierarchical Forward Modeling} Traditional photo-$z$ workflows estimate individual galaxy redshifts and aggregate these posteriors to derive population-level quantities such as ensemble redshift distributions $n(z)$ -- a computationally expensive bottom-up approach prone to biases when combining noisy individual posteriors \citep{Leistedt2016, Malz2021, Malz2022, Alsing2023}. Population-level inference inverts this paradigm by directly targeting the population distribution $P(\boldsymbol{\theta})$ over redshift and physical galaxy parameters (stellar mass, star formation rate, metallicity) as the primary inference objective, leveraging the collective constraining power of the entire photometric dataset while naturally incorporating physical priors on galaxy evolution. These methods rely on forward modeling: generating synthetic photometry from physical parameters via \acrlong{sps} \citep[\acrshort{sps}; for reviews, see, e.g.,][]{Conroy:2013, Iyer:2026} models and comparing the distribution of model photometry to observed data. Classical SPS calculations---as implemented by, e.g., \acrlong{fsps} (\acrshort{fsps}; \citealp{Conroy:2009, Conroy:2010, ConroyGunn:2010}) and the \href{https://github.com/bd-j/prospector/}{\tt prospector} model family \citep{Leja:2017, Johnson:2021, Wang:2023}---are computationally prohibitive for large samples, motivating neural network emulators like \href{https://github.com/justinalsing/speculator/tree/master}{\tt speculator} \citep{SPECULATOR} that achieve $\sim10^{3}$--$10^{4}\times$ speedups with negligible accuracy loss. The \href{https://github.com/Cosmo-Pop/pop-cosmos}{\tt pop-cosmos} framework \citep{Alsing:2024, Thorp:2024, Thorp:2025, Deger:2025} exemplifies this approach: it defines a probability distribution over a 16-dimensional SPS parameterization using a score-based diffusion model calibrated on $\sim$420,000 galaxies from COSMOS2020 \citep{Weaver:2022} with 26-band photometry spanning deep \acrfull{uv} to mid-\acrfull{ir}. 
This model enables direct estimation of tomographic redshift distributions, and, when used as a data-driven prior in SED fitting, highly accurate individual galaxy redshift inference.

\paragraph{Benchmarking and Evaluation Frameworks}
As \acrshort{ai} methods become increasingly central to cosmological analyses, it is critical to develop robust frameworks for testing and validation that ensure reproducibility and enable systematic comparison of different approaches.
For this purpose, the DESC has developed RAIL, an open-source, Python-based framework to support large-scale photometric-redshift workflows for LSST.
RAIL is a library that hosts many per-galaxy algorithms (e.g., \href{https://github.com/rizbicki/FlexCoDE}{\tt FlexZBoost}, \href{https://github.com/ltoribiosc/DNF_photoz}{\tt DNF}, \href{https://github.com/jfcrenshaw/pzflow}{\tt PZFlow}, \href{https://github.com/grantmerz/deepdisc}{\tt DeepDISC}), ensemble calibration algorithms, (e.g., self-organizing maps). RAIL also provides comprehensive infrastructure that
(i) supplies a unified \acrshort{api} and modular pipeline stages to train, apply, and compare a broad range of redshift estimators (catalog-based, image-based, probabilistic),
(ii) embeds evaluation modules and metrics for both individual-galaxy redshift and ensemble PDFs ~\citep{RAIL_2025}, 
(iii) generates realistic mock data for supervised learning algorithms by applying photometric noise, reference redshift selection, and error to an input catalog, and (iv) enables data challenges to test the robustness of photo-z estimators to a wide array of systematic errors. RAIL's standardized framework facilitates reproducible results and fair benchmarking across different methods, essential for validating AI techniques in preparation for LSST data.

\subsection{Strong Lensing}
\label{sec:strong_lensing}
\begin{ThemeBoxA}[]
\themebullet \themekey{Methodology.} \meth{gaussian-process}, \meth{cnn}, \meth{rnn}, \meth{transformer}, \meth{sbi}, \meth{diffusion-model}, \meth{variational-inference}  \\
\themebullet \themekey{Challenges.} \challenge{data-sparsity}, \challenge{covariate-shift}, \challenge{uq} \\
\themebullet \themekey{Opportunities.} Multi-survey cross-matching (Roman+LSST+Euclid), population-level inference, automated discovery, subhalo constraints from anomalous flux ratios
\end{ThemeBoxA}

Strong gravitational lensing is a rare astrophysical phenomenon where the light of a distant object, the source, is deflected by the gravity of an intervening structure, the lens, forming multiple images of the background source. In galaxy-galaxy strong lensing, both the source and the lens are individual galaxies, while on larger scales, the lens could range from a group to an entire galaxy cluster. Despite their rarity (a result of the stringent alignment required between source, lens, and observer), lensed systems are powerful cosmological probes that can constrain dark energy and probe dark matter on sub-galactic scales. Since elliptical galaxies dominate the deflector population, strong lenses also enable studies of their mass profiles, stellar content, and dark matter halos. Furthermore, lensing magnification enables the study of high-redshift sources, offering insights into early galaxy evolution \citep{1992grle.book.....S}.

\acrshort{lsst} will be transformative for strong lensing science. Forecasts predict the discovery of ${\sim}120{,}000$ systems \citep{collett15}, two orders of magnitude more than currently known. This large sample will provide the statistical power required for precision cosmology: time-delay lenses and large samples of static lenses have been shown to enable competitive dark energy measurements \citep{shajib_SL_2025}. The sample will also include significant numbers of currently rare systems, such as double-source-plane lenses, lensed supernovae, and cluster-scale lenses with multiple background sources. \acrshort{lsst}'s six-filter imaging (see \autoref{sec3:photo-z}) will enable photo-$z$ estimation for both lens and source populations, as well as classification of time-delay lenses.

One particularly powerful application is time-delay cosmography, which uses strongly lensed transients to measure the Hubble constant ($H_0$)~\citep[e.g.,][]{tdcosmo2025}. This approach yields a geometrical measurement independent of both the local distance ladder \citep[e.g., SH0ES;][]{riess22} and early-universe measurements from the \acrlong{cmb} \citep[\acrshort{cmb}; see, e.g.][]{planck20}. \acrshort{lsst} will provide time-domain coverage for ${\sim}100\times$ more systems than current surveys \citep{wojtak2019, goldstein2019, arendse2024, erickson_de_2025, abe_2025}, dramatically increasing the sample of lensed \acrfull{sne} and \acrfull{agn} available for cosmography. \acrshort{ai}/\acrshort{ml} models have been proposed for time-delay estimation from light curves, particularly kernel-based methods \citep[e.g.][]{cuevas06,cuevas2010uncovering,otaibi16}.

\paragraph{Supervised Detection in the Low Data Regime} Given the rarity of strong lensing events, identifying them among billions of cutouts is inherently challenging for visual inspection in wide-field surveys. While early automated detection approaches relied on curvature-based features \citep{estrada2007systematic} and arc-characterizing descriptors \citep{de2012metodo, bom2017neural}, the advent of \acrfullpl{acr:cnn} led to state-of-the-art performance in lens finding \citep{2017MNRAS.472.1129P, 2018A&A...611A...2S, 2019MNRAS.482..807P, 2018MNRAS.473.3895L}. Building on this progress, \citet{metcalf19} launched a lens-finding challenge using Bologna Lens Factory simulations based on the Millennium data \citep{Lemson2006}, showing that LSST-like ground-based multi-band images are well suited for this task. A subsequent Euclid-like challenge produced a winning algorithm combining multi-resolution CNNs \citep{bom2022developing}, later validated on real data by \citet{melo25} using Legacy Survey and \acrfull{hst} images to mimic LSST--\textit{Euclid} synergy. Leveraging multiple networks classifications as ensemble lens classifiers showed improved results over a single network classifier \citep[e.g.,][]{andika23, schuldt23b, gonzalez25}, while \cite{holloway_2024} incorporated citizen science annotations to enhance the performance. Because too few real lenses exist for supervised training, realistic mock datasets are essential. Works such as \citet{2017MNRAS.472.1129P} or \citet{schuldt21} proposed simulating only the lensing effect on real galaxy images, a practice now standard in the ongoing LSST \acrshort{desc} and \acrshort{slsc} challenge (Bom et al., in prep.). In preparation for LSST, \acrfull{hsc} data (\citealp{aihara18}; with similar filters and pixel scale) have been used to develop and compare models \citep[see e.g.,][]{shu22, andika23, canameras24, jaelani24, more24}. While early efforts focused on simple CNN or \acrfull{resnet} architectures, more advanced architectures such as vision transformers have also been applied \citep{2025Gonzalez_ViTLenses}. Foundation models such as Zoobot \citep{ZoobotRelease2023} have also achieved strong results on \textit{Euclid} imaging \citep{walmsley25, lines25}, and will soon be adopted for LSST (see Sect.~\ref{sec:foundation_models}). Finally, ML methods are now expanding beyond galaxy-scale lenses to systems involving entire clusters \citep{schuldt25, 2025arXiv251103064E} and galaxy–galaxy lenses within clusters \citep{angora23}.

\paragraph{Simulation-Based Inference (SBI)} Beyond lens finding, \citet{hezaveh17} pioneered the use machine learning models to predict characteristics of strong lensing systems. Specifically, \citet{hezaveh17} showed that simple CNNs can be used to predict parameters of the lens (Einstein radius, complex ellipticity, and the coordinates of the center of the lens) from images from \textit{HST} with a precision comparable to that of traditional methods. \citet{Perreault:2017} proposed using approximate \acrfullpl{bnn} to obtain calibrated estimations of the marginal posterior of these lens parameters, an approach applied to \acrfull{alma} observations in~\citet{Morningstar2018}. Such ML estimates can also be used as starting points for classical model fitting methods \citep[e.g.][]{2025arXiv250315329E}, providing significant speedups. \citet{Legin2021, Legin2023} compared this approach to \acrfull{nle}, demonstrating potential for better calibration in 2-stage \acrshort{acr:sbi} methods. In~\cite{Poh2025}, it was demonstrated that BNNs and \acrfull{acr:npe} can be used to infer parameters that describe the lens system even in ground-based \acrshort{des}-like imagining. In subsequent years, significant progress was made in using HSC images to prepare for LSST \citep[e.g.,][]{pearson19, schuldt21, gentile23, schuldt23a, schuldt23b, gawade25}. Following earlier work by \citet{Park2021,Wagner-Carena2021}, \citet{erickson25} applied (sequential) NPE within a hierarchical framework to model strongly lensed quasars, testing on real systems discovered by DES and followed-up with \textit{HST} high-resolution imaging, and \citet{venkatraman_2025} applied hierarchical NPE modeling to simulated LSST $i$-band coadds.  Ongoing DESC work examines how modeling an uncertain sample of static galaxy-galaxy lenses with ML enables new cosmological constraints (Holloway et al. in prep.), leveraging the method demonstrated by \citet{li_gg_SL}. And in~\cite{Jarugula2024}, authors presented a scalable approach for inferring the dark energy equation-of-state parameter from a population of strong gravitational lens images using \acrfull{nre}. \citet{filipp25} investigated the robustness of neural ratio and neural posterior estimators to distributional shifts for dark matter substructure inference from strong lensing, finding that these methods can exhibit significant biases when the test data deviates from the training distributions. Initial tests of using domain adaptation~\citep{Farahani2020ABR} for improving robustness of CNNs and neural posterior estimators when predicting characteristics of strong lens systems in the presence of distributional shift were performed in~\cite{Swierc2023,Swierz2024} and~\cite{Agarwal2024}.

\paragraph{High Dimensional Inverse Problem for Lens Modeling}  The task of lens modelling, that is, predicting surface brightness of background sources and density maps of foreground lenses is, in its simplest form, a non-linear inverse problem involving a handful of parameters ($\sim 10-20$). However, as the quality and resolution of data increases, such parametric description of lensed object becomes too simplistic, and more complex parametrization become necessary to avoid biases. An example of such a parametrization that is particularly well-adapted to \acrshort{ml} applications are pixelated images of sources surface brightness and projected densities of lenses. Traditionally, it has been difficult to characterize appropriate priors analytically on such high-dimensional spaces~\citep[see, e.g.,][]{Suyu2006, WarrenDye2005, Birrer2015, Delaunay2009, Nightingale2018}, however, recent advances in high-dimensional inference with deep learning has made progress on this front possible. \\
An initial attempt at solving the source reconstruction problem was presented in~\cite{Morningstar2019}, and extended in ~\cite{AlexAdam2023} to enable joint modeling of generalized pixellated lens densities and sources surface brightness. However, while these models provided high-fidelity \acrfull{map} estimates, they lacked the crucial ability to quantify uncertainties. Approaches based on \acrfull{vi} have also been proposed in~\citet{Chianese2020, Karchev2022GP, Mishra-Sharma2022}.\\
Advances of, e.g., \cite{Song_2019, Ho_2020, Song:2020, Song_2021, 2022Yang_Diffusion}, have shown that generative models used as expressive, data-driven priors are a promising alternative to address this problem. \cite{adam2022posterior, Karchev2022Diffusion} used \acrfullpl{sbm} as flexible priors in an explicit inference framework to produce posterior samples of background galaxy sources. In \cite{Barco2025blindinversion}, this method was extended to allow joint sampling the source and lens parameters for smooth, parametric lenses. Such methods have been shown to alleviate known biases in lens parameters induced by misspecified traditional priors, and methods have been proposed to empirically adapt initially biased SBM priors to correct for, e.g., population-level evolution of galaxy morphologies~\citep{Barco_2025}, and to empirically extend misspecified physical models~\citep{Payot2025}. More recently, \cite{RonanSLACLenses} has shown that SBM priors can be leveraged in a Gibbs sampling scheme to reanalyze \textit{HST} data from lenses observed by \acrfull{slacs}. Ongoing challenges include increasing the sampling efficiency of these methods to allow modelling a large fraction of the strong lenses expected with LSST.

\paragraph{Leveraging LSST time-series data} LSST will generate an overwhelming number of transient alerts, making the discovery and characterization of strongly lensed short-lived transients (e.g., supernovae) both difficult and time-critical. Kernel-based methods and probabilistic machine learning models such as Gaussian processes will likely play a major role in time-delay inference for lensed quasars \citep[e.g.,][]{cuevas06, cuevas2010uncovering, hojjati14, otaibi16, tak17} and supernovae \citep[e.g.,][]{hayes24, hayes25} discovered by LSST. Temporal deep learning models will also be essential in this area. For instance, \citet{morgan2021deepzipper} developed DeepZipper, which integrates \acrfull{lstm} networks with \acrshortpl{acr:cnn} to jointly process temporal and spatial information for identifying strongly lensed supernovae in time-domain surveys. \citet{bag24} developed a model using unresolved light-curve data from difference imaging, while \citet{bag25} extended this to full multi-band time series with a 2D convolutional \acrshort{lstm} network. \citet{huber24} instead used an LSTM network to predict time delays between such transients directly from their light curves, whilst \citet{goncalves25} used an ensemble of CNNs to directly estimate $H_0$ from time series of lensed supernova images and \citet{Campeau2023} demonstrated the potential of NREs to infer $H_0$ from time delays and lens models. Beyond short-lived transients, \citet{jimenez25} modeled microlensing in lensed quasar light curves, and \citet{fagin_light_curves} introduced a latent \acrfull{sde} framework to jointly model AGN variability and transfer functions, potentially extendable to lensed AGNs for joint inference of time delays and disk parameters.

\subsection{Weak Lensing}
\label{sec3:wlss}
\begin{ThemeBoxA}[]
\themebullet \themekey{Methodology.} \meth{sbi}, \meth{neural-compression}, \meth{differentiable-programming}, \meth{hierarchical-bayes}, \meth{diffusion-model} \\
\themebullet \themekey{Challenges.} \challenge{covariate-shift}, \challenge{uq}, \challenge{scalability} \\
\themebullet \themekey{Opportunities.} Multi-resolution joint processing (Roman+LSST), probabilistic deblending, DeepDISC instance segmentation, physics-informed priors for galaxy morphology
\end{ThemeBoxA}

As light from background galaxies travels through the Universe, its path is deflected by the gravitational potential of foreground matter, inducing subtle shape distortions of observed galaxies that can be statistically measured. This effect, referred to as weak gravitational lensing, provides a direct probe of the total matter distribution in the Universe, making it a powerful tool for constraining cosmological parameters such as the matter density $\Omega_m$, the amplitude of matter fluctuations $\sigma_8$, and the dark energy equation of state $w$ in a \acrfull{wcdm} model.
With its unprecedented depth, image quality, and sky coverage, \acrshort{lsst} will provide the most precise mapping of the large-scale structure of the Universe to date. This level of precision has two key implications:
(1) It  presents a major opportunity to refine cosmological constraints, motivating the development of advanced inference methods that can fully exploit this high-quality data; (2) It demands rigorous control of systematic uncertainties to ensure unbiased cosmological constraints.

\paragraph{Systematics modeling / mitigation}
Systematic errors, such as imperfect shear calibration, \acrshort{photoz} uncertainties, spatially varying selection effects, and \acrfull{psf} residuals, must be well characterized. To date, the precision of current cosmological surveys has permitted the use of state-of-the-art prescriptions that capture the dominant effects of these systematics (e.g., \citealp{Mandelbaum:2018}). However, as forthcoming large-scale structure and weak-lensing data from LSST achieve substantially higher statistical precision, a more accurate and detailed characterization of these systematics will be required, at a level of complexity that renders purely analytical treatments intractable.  
\acrshort{ml} methods offer a complementary pathway by learning complex nonlinear mappings from observational features---e.g., properties of individual galaxies, local image quality metrics, PSF residuals, depth maps, shape measurement parameters--- to the resulting systematic bias or residual error \citep{Tewes:2018she,Rezaie2020,Pujol:2020wrk}. Neural network or other ML algorithms can be trained on simulated or calibration data for which the true systematic shifts are known, and can then be tuned to predict, flag or correct for the systematic effect when applied to real survey data \citep{Fluri:2022rvb}. In doing so, these methods enable rapid and flexible removal of systematic contamination from the cosmological signal, thereby yielding a cleaner, more robustly inferred signal.

\paragraph{Clean catalog construction} Systematics such as \acrlong{ia} \citep[\acrshort{ia}; e.g.,][]{Mandelbaum:2006, Mandelbaum:2011, Troxel:2015, Joachimi:2015} can be mitigated by constructing clean source and/or lens catalogs from galaxy populations where these effects are known to be negligible. Scalable inference of galaxy properties that correlate with active galaxy populations (such as \acrlong{ssfr}, \acrshort{ssfr}) can enable the construction of IA-mitigated galaxy samples. For instance, machine-learned generative priors can be leveraged (e.g., \href{https://github.com/Cosmo-Pop/pop-cosmos}{\tt pop-cosmos}; \citealp{Alsing:2024, Thorp:2024, Thorp:2025, Deger:2025}) to estimate per-galaxy sSFR and construct clean catalogs of star-forming galaxies with conservative cuts based on this parameter. This approach, especially when combined with amortized \acrshort{acr:npe}, is scalable to LSST-sized datasets and is expected to outperform color-based selections, which are affected by contamination. Moreover, generative models of the galaxy population can be applied in a weak lensing context to directly infer the redshift distributions of source catalogs subject to tomographic binning and sample selection criteria, provided that the color-redshift relation is realistic and robust. This provides an alternative to approaches such as \acrshort{acr:som} calibration.

\paragraph{Neural Compression and Simulation-Based Inference}
Traditional weak-lensing analyses follow a two-step pipeline: compress high-dimensional shear or convergence fields into summary statistics, then perform Bayesian inference on these summaries to obtain posteriors over cosmological parameters. The matter power spectrum and shear–shear correlation functions remain workhorse statistics (\acrshort{kids}-1000 cosmic shear: \citealt{Asgari2021}; \acrshort{des} Y3 cosmic shear: \citealt{Amon_2022,Secco_2022}), but in the LSST era significant non-Gaussian information will become available. This has motivated the use of higher-order moments such as the bispectrum and trispectrum \citep[e.g.,][]{desy3_moments}, as well as peak counts \citep[e.g.,][]{Marques_2024}, persistent homology \citep{prat2025}, and Minkowski functionals \citep[e.g.,][]{PhysRevD.85.103513}. While powerful, these handcrafted summaries are not guaranteed to capture all cosmological information. An alternative enabled by ML is to train neural networks that compress the maps directly into low-dimensional summaries. Several strategies have been explored in the weak lensing literature for training such networks \citep[see][for a comparison]{neural_summary_lanzieri_2025}; among these, information-theoretic criteria (such as those used in Information-Maximizing Neural Networks \citep[IMNNs;][]{2018PhRvD..97h3004C,Makinen:2021} and Variational Mutual Information Maximization \citep[VMIM;][]{Jeffrey:2021}) can yield summaries that closely approximate sufficient statistics, achieving near-optimal compression. Hybrid methods that combine physics-based summaries (e.g., power spectra) with learned summaries can leverage the strengths of both \citep{Makinen2024}. Because the likelihood of these learned summaries is unknown, \acrshort{nde} techniques such as normalizing flows are then used to approximate the posterior within a \acrshort{acr:sbi} framework \citep{Alsing:2019}. This strategy was first demonstrated on survey data in \citet{Jeffrey:2021} and subsequently applied to recent surveys \citep[e.g.,][]{jeffrey2024darkenergysurveyyear, kramsta2025}, with \citet{jeffrey2024darkenergysurveyyear} reporting more than a factor of two improvement in dark energy parameter precision compared to power spectrum inference. However, the practical limit to \acrshort{acr:sbi} is not the ability to extract information from the data through learned summaries, but rather the difficulty of producing simulations realistic enough to be compared to observations without incurring biases from model misspecification. In addition, neural summaries are notoriously difficult to interrogate: monitoring them for covariate shifts, unmodeled systematics, and failures in specific regions of the data space is challenging, thereby complicating the construction of robust null tests and diagnostic pipelines. That said, posterior predictive checks against conventional summary statistics (e.g., the power spectrum or higher-order moments) can help detect some forms of model misspecification, though they are not guaranteed to catch all issues. 
 
\paragraph{Hierarchical Bayesian Field-Level Inference} With the advent of \acrshort{gpu}-accelerated probabilistic programming, it has become feasible to model the full weak-lensing field in a hierarchical framework that links Gaussian initial conditions of the matter density to observed shear maps through an explicit forward simulation model. Proof-of-concept studies have demonstrated this approach in simplified weak-lensing settings \citep[e.g.,][]{porqueres2023fieldlevelinferencecosmicshear}, showing substantial gains in constraining power relative to power-spectrum analyses. DESC members have contributed key building blocks for such end-to-end pipelines, including differentiable lensing lightcone constructions \citep{Lanzieri_2023} and accurate, differentiable ray-tracing schemes \citep{Zhou2024}. However, scaling these methods to a full LSST analysis remains extremely challenging: the survey volume and the resolution required for the forward model place stringent demands on memory, compute, and algorithmic efficiency. Ongoing work aims at lifting this bottleneck through distributed simulations across multiple GPUs \citep{Kabalan_jaxDecomp_2025}. In parallel with full forward modeling of the large-scale structure, \acrshort{desc} members have also explored map-based hierarchical inference using lognormal fields \citep{boruah2022mapbasedcosmologyinferencelognormal, Zhou2024Prd}, which is far less computationally demanding but whose ultimate accuracy is constrained by the limitations of the lognormal approximation. As an alternative approach, DESC members have proposed using diffusion models to learn the forward model of the density field implicitly from simulations and combine this learned prior with an explicit likelihood to constrain observed shear data \citep{remy2023}, enabling fast reconstruction of high-fidelity mass maps. 

\subsection{Galaxy Clusters}
\begin{ThemeBoxA}[]
\themebullet \themekey{Methodology.} \meth{sbi}, \meth{object-detection}, \meth{cnn}, \meth{hierarchical-bayes}\\
\themebullet \themekey{Challenges.} \challenge{covariate-shift}, \challenge{uq}, \challenge{scalability} \\
\themebullet \themekey{Opportunities.} Combination of imaging and catalog data, Hierarchical Modeling
\end{ThemeBoxA}

Galaxy clusters trace the most massive peaks of the matter density field and form relatively late in cosmic history, making their abundance and internal properties highly sensitive to the growth of structure and to dark energy. Cosmological constraints from clusters have traditionally relied on measurements of the cluster mass function and its redshift evolution, anchored by calibrated relations between mass and observable proxies. \acrshort{ml} methods are now entering this pipeline at multiple stages (cluster finding, mass–observable calibration, and population-level inference) offering new ways to combine imaging, catalog, and multi-wavelength data while retaining control over systematics and uncertainties.

\paragraph{Cluster Finding from Images and Catalogs}
The first step in cluster cosmology is robust identification of cluster candidates. Non-ML algorithms, such as \acrlong{redmapper} \citep[\acrshort{redmapper};][]{rykoff2014, rykoff2016}, the \acrfull{wazp} cluster finder \citep{aguena2021}, and the Euclid cluster finders AMICO and PZWAV \citep{2019A&A...627A..23E}, have been widely used to identify optically selected clusters in galaxy catalogs. In parallel, deep-learning–based cluster finders have emerged in the \acrlong{sz} \citep[\acrshort{sz}; e.g.,][]{bonjean2020, lin2021, hurier2021, Meshcheryakov2022} and optical domains \citep{chan2019, grishin2023, grishin2025, tian2025}. A key advantage of these approaches is that they can operate directly on images, rather than on pre-processed catalogs, and thus potentially exploit features (e.g., diffuse emission, subtle color–magnitude structure, environment) that are not captured in standard catalog-level summaries. For example, a \acrlong{yolo} (\acrshort{yolo}; \citealp{redmon2015, redmon2016, redmon2018}) architecture trained on images centered on SDSS RedMaPPer clusters was shown to recover not only the training sample but also previously missed systems that were later confirmed in external X-ray catalogs \citep{grishin2023}. Recent \acrshort{desc} work applied YOLO-CL to DC2 simulations \citep{grishin2025}, training on both observed SDSS RedMaPPer clusters and simulated massive halos. While the model performs well on images centered on known clusters, blind application to DC2 as a survey shows degraded completeness and purity, highlighting the need for further development and architectural updates, including modern YOLO variants and transformers (Tran et al., in prep.), before deployment on \acrshort{lsst}.

\paragraph{Mass-Observable Relations and Weak-Lensing Mass Calibration}
Cosmological analyses require accurate and precise relations between cluster mass and observables (richness, SZ signal, X-ray luminosity/temperature, velocity dispersion). Deep neural networks are being explored as flexible mass estimators across multiple wavebands, including X-ray signatures \citep{ntampaka2019, krippendorf2024, iqbal2025}, the dynamics of member galaxies \citep{ho2019, ho2021, ho2022, wangthiele2025}, and SZ measurements \citep{deandres2022}. For LSST, photometric galaxy data contribute primarily through weak-lensing mass estimates that anchor mass–observable relations. The DESC \acrlong{clmm} (\acrshort{clmm}; \citealp{aguena2021clmm}) currently infers weak-lensing masses from radial shear profiles using traditional likelihoods and \acrfull{mcmc}, on the assumption of parametric mass models such as \acrlong{nfw} (\acrshort{nfw}; \citealp{nfw97}). Ongoing work within DESC explores alternative \acrshort{acr:sbi} approaches at this level, which can in principle incorporate more realistic shear profiles, complex noise, and selection effects without requiring an explicit closed-form likelihood.

\paragraph{Simulation-Based Inference for Cluster Cosmology}
SBI provides a computationally efficient method for deriving posteriors for cluster- and population-level parameters directly from simulated data vectors. This is particularly attractive for analyses that must jointly model individual clusters and the cluster population via hierarchical frameworks, where traditional likelihood-based methods become increasingly costly and brittle as the parameter space and model complexity grow. Ongoing work within DESC indicates that SBI can recover cluster weak-lensing mass posteriors consistent with those from MCMC, provided that model misspecification is not worse than in the explicit-likelihood case (Gill et al., in prep.). In addition, SBI has been shown to derive relevant constraints directly from cluster counts \citep{reza2022, reza2024, zubeldia25}, and this approach is now being developed on DESC simulations as an alternative, simulation-native pathway to cluster cosmology that can be integrated into the DESC Cluster Cosmology Pipeline. Compared with Stage-III experiments, DESC cluster analyses will require richer astrophysical modeling and the exploration of larger parameter spaces; SBI offers the algorithmic flexibility and, in many regimes, the computational efficiency required to meet these demands.

% \newpage
\subsection{Supernova Cosmology and Transients}
\label{sec3:td}
\begin{ThemeBoxA}[]
\themebullet \themekey{Methodology.} \meth{hierarchical-bayes}, \meth{rnn}, \meth{transformer}, \meth{active-learning}, \meth{ensembles}, \meth{sbi}, \meth{anomaly-detection}, \meth{vae}, \meth{gaussian-process}\\
\themebullet \themekey{Challenges.} \challenge{covariate-shift}, \challenge{data-sparsity}, \challenge{scalability}, \challenge{uq} \\
\themebullet \themekey{Opportunities.} PLAsTiCC/ELAsTiCC simulation infrastructure, DESC leadership in alert broker integration (ALERCE, Fink, ANTARES)
\end{ThemeBoxA}

\acrshort{lsst} is expected to detect $\sim$10 million transient and variable objects each night, a thousand-fold increase over current surveys. The sheer volume and cadence of detections renders traditional spectroscopic classification infeasible for most events. This presents a major bottleneck to the identification of pure \acrshort{snia} samples for cosmological distance measurements, and to constraining the explosion physics of populations of rare and novel phenomena for the first time. To achieve reliable cosmological constraints and meet \acrshort{desc} goals of reducing the systematic uncertainties from light curve modeling below 3\% of those obtained from SALT2 \citep{2007Guy_SALT2}, analysis techniques demand \textit{well-calibrated uncertainty quantification}, \textit{adaptive and scalable performance}, and \textit{robustness to covariate shifts and data corruption}.

\paragraph{Spectrophotometric Modeling} SNe~Ia are broadly homogeneous and viable standard candles, but diversity in their spectro-temporal properties and persistent host-dependent effects \citep{2006ApJ...648..868S,2010MNRAS.406..782S,2010ApJ...722..566L,2010ApJ...715..743K,2013ApJ...764..191H} still limit standardization precision. Modern \acrshort{ml} approaches to standardization now focus on data-driven, differentiable, and multi-modal models rather than hand-engineered linear corrections. Already progress in these directions can be seen in modeling using probabilistic auto-encoders~\citep{2022ApJ...935....5S} and \acrlongpl{acr:vae} (\acrshortpl{acr:vae}; e.g., \href{https://github.com/LSSTDESC/parsnip}{\texttt{ParSNIP}}~\citealp{2021Boone_ParSNIP}), that predict time-evolving \acrshortpl{sed} from light curves, and use a differentiable forward model to compare in observation space. High quality, well-calibrated data has been instrumental in these endeavors, which can be augmented by the LSST/Rubin samples; however, strategies that account for the shifts induced by calibration errors must still be investigated. 

More recent efforts involve \acrshort{ml} models as emulators \citep[e.g.][]{chen2020, kerzendorf2021, magee2024} for radiative transfer codes such as TARDIS~\citep{kerzendorf2014}, that can be used to infer \emph{physical parameters} such as the total luminosity, nickel mass, the composition and the asymmetry of the ejecta given multi-modal inputs such as light curve, time-series spectroscopy, and Rubin and high-resolution space-based imaging (e.g.\ from the \acrshort{esa} \textit{Euclid} mission and the \textit{Nancy Grace Roman Space Telescope}). These physically motivated emulators are also capable of several tasks that would have previously required individual specialized models. These include predicting the future evolution of \acrshort{sne} of all types, classifying spectra while being agnostic to the imbalance in extant training samples, and can help distinguish SN Ia from impostors that might otherwise contaminate the cosmological sample, as well as actively schedule follow-up spectroscopy.

\paragraph{Photometric Classification} The methodological evolution of photometric classifiers from feature-based approaches~\citep[e.g.][]{2018Narayan_ANTARES} to end-to-end learning has been driven by the challenge of processing irregular, heteroskedastic observations: \href{https://github.com/LSSTDESC/snmachine}{\tt SNmachine} \citep{SNmachine2016} leveraged a range of feature sets, from physics-based through to non-parametric approaches, coupled with a variety of traditional \acrshort{ml} techniques to achieve high classification accuracy; \href{https://github.com/helenqu/scone}{\tt SCONE}'s Gaussian process interpolation \citep{SCONE:2021} creates regular 2D representations from sparse observations (see also \href{https://github.com/kboone/avocado}{\tt Avocado}; \citealp{2019Boone_avocado}); while transformer architectures \citep{2023Pimentel_Attention,2024Allam_Attention,ATAT2024} leverage self-attention mechanisms to handle missing data naturally. Hybrid, physics-informed approaches have also been explored to extract latent features from light curves using generative modeling, which are then used for classification \citep{2021Boone_ParSNIP}. Classification tools have already proven their utility in LSST survey optimization for supernova metrics, with realistic LSST survey cadences (e.g., \href{https://github.com/LSSTDESC/snmachine}{\tt SNMachine}, \citealt{Alves2022, Alves2023}).   

\acrshort{des} pioneered the use of neural networks for photometric classification of SNe~Ia for cosmological analysis \citep{Moller:2020,SCONE:2021,Moller:2022,Vincenzi:2023,DESSN:2024}. Propagating the prediction uncertainties from these models through to cosmological constraints remains an open problem. \href{https://github.com/supernnova/SuperNNova}{\tt SuperNNova} \citep{Moller:2020} addresses the former with a \acrfull{acr:rnn} providing calibrated probabilities essential for contamination modeling in dark energy constraints \citep{Vincenzi:2023}. In DES, \acrshortpl{bnn} and ensemble methods were also tested \citep{Moller:2024}; for DESC, we can advance fully Bayesian approaches to photometric classification for LSST.

\paragraph{Forward Modeling of the Time-Domain Landscape}
Observational modeling led by DESC has catalyzed the development of neural approaches to photometric classification. \acrshort{plasticc} \citep{PLAsTiCC1810.00001, 2023Hlozek_PLasTicc} provided 3.49M test light curves across 18 transient classes using simulations from the \acrlong{snana} (\acrshort{snana}; \citealp{kessler2009}) with cadences and realistic observing conditions from the LSST \acrlong{opsim} \citep[\acrshort{opsim};][]{2014Delgado_OpSim}. The challenge established weighted logarithmic loss metrics prioritizing SNe Ia and \acrfull{kne} for DESC science goals \citep{2019Malz_Metric}, with winning solutions employing gradient boosting and ensemble neural networks requiring engineered features \citep{2019Boone_avocado,2023Hlozek_PLasTicc}. The dataset remains a foundational benchmark for time-series representation learning in astrophysics years after its development \citep{Fraga:2024,Masson:2024,CadizLeyton:2025, CadizLeyton:2025:MoE,ROMAE2025}. Building on PLAsTiCC, \acrshort{elasticc} \citep{2023AAS_ELAsTiCC, knop2023} stress-tested end-to-end broker infrastructure with $\sim$50M alerts streamed in real-time to seven community brokers from September 2022--January 2023. DESC remains the only Rubin Science Collaboration to have tested the alert infrastructure from end-to-end on this scale, as it is critical for deploying \acrshort{ai} models live, to support online learning and other tasks to optimize the scientific return from Rubin.

Similar to ELAsTiCC and PLAsTiCC, \acrfull{sassafras} is a novel dataset of simulated LSST spectroscopic follow-up with 2.1M spectra across 14 transient types and three telescopes -- Gemini, the \acrfull{soar}, and \acrshort{4most} -- using SNANA simulations with realistic noise from each of the three telescopes. These simulations contain an even distribution of spectra per class and have a wide range of redshift distribution between 0.023--1, minimizing the bias inherent from uneven distribution of spectra seen in real data. One group at \acrfull{skai} is currently utilizing SASSAFRAS to train a spectroscopic classifier and then transfer learn with real data to create a state of the art classifier.  Spectroscopic classifiers trained on SASSAFRAS will be essential for confirming transient labels for active learning algorithms as outlined below.

\paragraph{Online Learning for Spectroscopic Optimization}
While archival photometric classification will suffice for the bulk of LSST cosmology analyses, inference over partially obtained data remains crucial for prioritizing spectroscopic targeting before an event has ended. \Acrfull{gru}-based recurrent and convolutional neural networks have successfully classified partial-phase synthetic light curves \citep{2019Muthukrishna_RAPID,SCONE:2021,2023Gagliano_FirstImpressions, shah2025b_oracle}, but performance on observed data remains modest. Transformer-based methods are being increasingly used \citep{ATAT2024}, with synthetic pre-training playing a growing role in bridging the simulation gap \citep{2025Gupta_Sims}. Within DESC, host-galaxy correlations have been shown to improve early classification \citep{2021Gagliano_GHOST}; this has driven data-driven modeling of host-galaxy correlations for the ELAsTiCC challenge \citep{2023Lokken_SCOTCH}, although spurious host-galaxy associations and the small postage stamps of the field contained within the LSST alert packets may limit utility of real-time inference using these data.

Beyond real-time classification, active learning faces unique astronomical challenges: objects must be selected for spectroscopic follow-up before informative light curve data are obtained, the untargeted population is substantially dimmer than the spectroscopically confirmed sample used in training, and labeling costs vary dramatically with object brightness and sky position. The \acrfull{resspect}\footnote{\url{https://resspect.readthedocs.io/en/latest/}}, an initial approach to active learning for transient science, implements uncertainty sampling with random forest classifiers on Bazin \citep{2011Bazin_LCModel} parametric features, but requires a minimum of five observations per filter, limiting early-time selection \citep{2020Kennamer_RESSPECT}. More recent implementations have refined active learning for early-time SN Ia identification, demonstrating effective follow-up optimization with simulations \citep{Ishida:2019}, real-data a posteriori \citep{Leoni:2022} and real-time observational campaigns \citep{Moller:2025}, the latter revealing the need for training sets containing events beyond supernovae. \href{https://github.com/MichelleLochner/astronomaly}{\tt Astronomaly} \citep{2021Lochner_Astronomaly} introduces personalized anomaly detection by combining isolation forests with human relevance scoring, addressing the fundamental subjectivity of an anomaly label. The approach has been shown to double the rate of anomaly discovery in radio transients \citep{2025Andersson_astronomaly}. However, active learning remains fundamentally limited by the lack of an informative initial training set, such that early random sampling can produce biased or unrepresentative data that propagates through subsequent iterations of learning. This is a fundamental challenge for novelty detection in LSST data.

\paragraph{Prompt Processing with the Transient Alert Brokers}
The seven Rubin Community Brokers implement diverse classification pipelines. \acrfull{alerce} employs a \acrshort{acr:cnn} for top-level classification from alert postage stamps \citep{2021Carrasco_Stamp}, and a hierarchical random forest applied to photometric features for classification along a 15-class taxonomy \citep{2021Sanchez_AlerceLC}. \acrfull{ampel} uses a four-tier system with predominantly gradient-boosted random forests \citep{2025Nordin_AMPEL}, and Fink deploys multiple classifiers for early and late-time classification \citep{Fraga:2024,Leoni:2022, Moller:2020, Moller:2025, fink}. \acrfull{antares} employs multi-stage filtering with community-contributed Python classes for tagging sources \citep{2018Narayan_ANTARES}, while Lasair integrates a boosted decision tree classifier from host galaxy properties \citep{2020Smith_ATLAS} with \acrfull{moc}-based watchmaps for coordination with 4MOST's \acrfull{tides}, which will be providing 35,000 transient spectra for SN~Ia cosmology \citep{2024Williams_Lasair}. These brokers have demonstrated sub-second latency in processing millions of alerts during the ELAsTiCC campaign, with classification probabilities reported via standardized Avro schemas that enable systematic evaluation of heterogeneous ML architectures. DESC members are involved in all seven Rubin Community Brokers, offering substantial potential for shared software infrastructure for processing the LSST alert stream.

\paragraph{Cosmological Inference using Type Ia Supernovae}

Hierarchical Bayesian models have been applied to SN~Ia for various goals, including cosmological inference \citep[e.g.,][]{2011March_BAHAMAS, 2016Shariff_BAHAMAS, 2015Rubin_Unity, 2025Rubin_Unity, 2018Feeney_H0}, constructing empirical SED models \citep[\href{https://github.com/bayesn/bayesn}{\tt BayeSN};][]{2009Mandel_BayeSN, 2011Mandel_BayeSN, 2022Mandel_BayeSN, 2021Thorp_BayeSN, 2023Ward_BayeSN, 2024Grayling_BayeSN, Uzsoy_2024}, modeling intrinsic colors and dust extinction \citep{2017Mandel_SimpleBayeSN, 2022Thorp_BayeSN, 2024Thorp_BayeSN}, handling uncertain photometric classifications \citep{2007Kunz_BEAMS, 2012Hlozek_BEAMS} and redshifts \citep{2017Roberts_zBEAMS}, and modeling the spectrophotometric standards used in photometric calibration \citep{2025Boyd_DA, 2025Popovic_Dovekie}. However, in the LSST era such models will need to incorporate complex effects that cannot easily be treated analytically; e.g., selection effects, photometric classification and photometric redshifts. Work is ongoing to enhance our statistical models using \acrshort{acr:sbi}, to leverage the flexibility of neural networks to capture these complex effects \citep[e.g.,][]{2024_Boyd_Flows, 2024Karchev_SIDEreal, 2025Karchev_CIGARS}. SBI will enable scalable and principled statistical inference of cosmological parameters with LSST. 

\subsection{Theory and Modeling}
\begin{ThemeBoxA}[]
\themebullet \themekey{Methodology.} \meth{emulator}, \meth{gaussian-process}, \meth{neural-surrogate}, \meth{differentiable-programming}, \meth{symbolic-regression}, \meth{sbi} \\
\themebullet \themekey{Challenges.} \challenge{covariate-shift}, \challenge{scalability} \\
\themebullet \themekey{Opportunities.} Flexible emulation frameworks, model selection and hypothesis testing, efficient construction of realistic mock datasets, gradient-based sampling.
\end{ThemeBoxA}

The role of theory and modeling within \acrshort{desc} is to provide the essential bridge between cosmological parameters and the statistical observables derived from \acrshort{lsst} data. Accurate theoretical models are required to translate measured galaxy shapes, positions, and fluxes into constraints on dark energy, dark matter, and gravity. This entails constructing predictive models of large-scale structure, galaxy bias, baryonic physics, and lensing observables that can be robustly compared with data while marginalizing over astrophysical and observational systematics. As the scale and precision of LSST data demand modeling at unprecedented accuracy and speed, \acrshort{ml}-based emulators, differentiable theory libraries, and \acrshort{acr:sbi} approaches are increasingly central to this effort, enabling fast and robust connections between data and theory.

\paragraph{Fast Surrogates for Cosmological Likelihoods} 

Emulation and related methods for creating fast-surrogate models using ML and \acrshort{ai} will be of crucial importance in accelerating inference pipelines for cosmological analyses in DESC.  Emulation is an indispensable tool for integrating aspects of modeling which by nature require simulation too slow to ever consider incorporating directly in sampling (e.g., those requiring $N$-body or hydrodynamical simulations). At the same time, even for aspects of modeling which are more moderate in evaluation cost (seconds rather than many hours), emulation allows individual likelihood evaluations to be dramatically accelerated. This is one of the key ways we can make computationally feasible the sampling in high-dimensional parameter spaces which will be required for DESC analyses.

Work on these emulation techniques has included directly building emulation tools, particularly outside of \acrshort{wcdm} models \citep{ramachandra2021matter}, emulating intrinsic alignment correlations \citep{Pandya2025IAEmu}, as well as using emulators to efficiently evaluate modeling choices for LSST data \citep{boruah2024machine}. The DESC theoretical modeling package \href{https://github.com/LSSTDESC/CCL}{\tt pyCCL} \citep{chisari2019core} natively supports key matter-power-spectrum emulation tools \href{https://bitbucket.org/rangulo/baccoemu/src/master/}{\tt baccoemu} \citep{arico2021bacco} and \href{https://github.com/lanl/CosmicEmu}{\tt CosmicEmu} \citep{lawrence2017mira}. However, developing frameworks for DESC to analyze models outside $w$CDM in the nonlinear regime of LSST data remains a challenge that needs to be addressed \citep{Ishak:2019BwCDM}, for which AI can play a major role.  

Looking to the future, DESC would benefit from developing mechanisms to enable emulation that has more flexibility with respect to modeling components. Our current methods of building a single emulator from scratch per modeling set-up is high cost (computationally and in terms of person power). This does not scale well for enabling adaptation to new systematics modeling or inference in models outside of $w$CDM. Considering approaches which use meta-learning (e.g., \citealt{macmahon2025meta}) or which are philosophically aligned with foundation models would be of value. 

Another area of growth where fast emulators can make major impact is in model selection and hypothesis testing of beyond-$w$CDM models. Instead of being limited by the cost of theoretical model evaluations, future analyses will be constrained by how efficiently inference pipelines can navigate and compare competing cosmological models. Integrating these emulators within agentic AI systems (\autoref{sec:llm_agentic})---which can autonomously refine training data, adapt inference strategies, leverage tools for evidence computation and simulation-based inference, and even propose new model extensions---will further accelerate discovery. For DESC, this synergy will transform the capacity to test gravity, dark energy, and dark-sector interactions, turning high-quality data into a powerful engine for identifying new physics. 

\paragraph{Differentiable Programming for Accelerated Sampling}

Traditional cosmological inference pipelines often rely on computationally expensive numerical methods such as the \acrlong{camb} \citep[\acrshort{camb};][]{Lewis_2000} or HaloFit  \citep{Smith_2003, Takahashi_2012}, limiting the use of gradient-based sampling methods. Differentiable cosmological codes address this by enabling efficient computation of gradients with respect to model parameters, unlocking samplers such as \acrfull{hmc} and the \acrlong{nuts} \citep[\acrshort{nuts};][]{HoffmanGelman2014}. 

The \href{https://github.com/DifferentiableUniverseInitiative/jax_cosmo}{\texttt{jax-cosmo}} library \citep{JAX-COSMO} provides a differentiable and hardware-accelerated framework for cosmological computations. With a NumPy-compatible \acrshort{api} and close integration with tools such as NumPyro \citep{phan2019composable} and JAXopt \citep{jaxopt_implicit_diff}, \texttt{jax-cosmo} offers a practical foundation for building scalable, fully differentiable cosmological models. \cite{CosmoPower} combined \texttt{jax-cosmo} with neural-network emulators in \href{https://github.com/dpiras/cosmopower-jax}{\tt CosmoPower-JAX}, enabling high-dimensional Bayesian inference through automatic differentiation and GPU acceleration. The \href{https://github.com/fkeruzore/halox}{\texttt{halox}} package \citep{halox} uses \texttt{jax-cosmo} for cosmological calculations such as power spectra and distance measures in its modeling of dark-matter halo statistics (such as halo mass function and halo bias). \cite{2025arXiv250707833S} used \texttt{jax-cosmo} to test a differentiable Fisher-information approach based on score matching. Recently too, \cite{bartlett2025symbolic} have developed a symbolic emulator that leverages genetic programming-based symbolic regression to derive compact, analytic expressions for cosmological observables, including the radial comoving distance, linear growth factor, and nonlinear matter power spectrum. 

\subsection{Cosmological and Survey Simulations}
\label{sec3:sims}
\begin{ThemeBoxA}[]
\themebullet \themekey{Methodology.} \meth{emulator}, \meth{diffusion-model}, \meth{differentiable-programming}, \meth{sbi}, \meth{gnn}\\
\themebullet \themekey{Challenges.} \challenge{covariate-shift}, \challenge{scalability}, \challenge{metrics} \\
\themebullet \themekey{Opportunities.} Joint modeling of galaxies and environments, inference at catalog and field level, modular components for DESC simulators, survey-scale generative models, survey design, systematics mitigation, pipeline stress tests.\end{ThemeBoxA}

Cosmological simulations are a fundamental tool not only for validating analysis pipelines but also, increasingly, for providing “theory” samples for \acrshort{acr:sbi} frameworks. Producing mock survey data at the scale and accuracy required for \acrshort{lsst} science remains a major challenge, which \acrshort{ml} can help address by emulating expensive numerical predictions and by providing data-driven models of otherwise poorly constrained aspects of the galaxy population. Neural emulators have long been used as fast surrogates for non-differentiable components of cosmological forward models, e.g., summary statistics of $N$-body simulations as in \href{https://github.com/lanl/CosmicEmu}{\tt CosmicEmu} \citep{CosmicEmu, moran2023mira}, \href{https://github.com/AemulusProject}{\tt Aemulus} \citep{2019ApJ...875...69D}, \href{https://github.com/dpiras/cosmopower-jax}{\tt CosmoPower-JAX} \citep{CosmoPower}, or \href{https://github.com/21cmfast/21cmEMU}{\tt 21cmEMU} \citep{21CMEMU}, and more recently for accelerating \acrshort{sps} calculations via models such as \href{https://github.com/justinalsing/speculator}{\tt speculator} and {\tt ProMage }\citep{SPECULATOR,ProMage}. Extending these approaches to additional components of the simulation pipeline holds the promise of greatly increasing the dynamical range, realism, and flexibility of LSST mock catalogs at manageable computational cost.

\paragraph{Population-Level Generative Models for Realistic Galaxy Catalogs}
Diffusion-based generative models operating in the space of physical galaxy parameters provide a powerful route to building realistic mock catalogs that remain anchored in deep-field observations. The \href{https://github.com/Cosmo-Pop/pop-cosmos}{\texttt{pop-cosmos}} framework \citep{Alsing:2024,Thorp:2025,Deger:2025} defines a score-based diffusion model over a high-dimensional SPS parameterization---\acrfull{sfh}, metallicity, dust, nebular emission, etc.---calibrated on $\sim$420,000 galaxies from COSMOS2020 \citep{Weaver:2022} spanning 26 bands from \acrshort{uv} to mid-\acrshort{ir}. Rather than directly emulating observed fluxes, \texttt{pop-cosmos} learns a data-driven prior $p(\boldsymbol{\theta}_{\mathrm{SPS}},z)$ over physical parameters and redshift that reproduces the joint distribution of observed photometry. This model encodes realistic priors on star-formation histories over cosmic time, and learns the evolution of the star-forming sequence. When coupled to survey-specific selection functions and noise models, \texttt{pop-cosmos} can therefore generate realistic mock galaxy catalogs that inherit both empirical constraints from deep multi-wavelength data and the flexibility of generative modeling.

\paragraph{Differentiable Empirical Galaxy–Halo Forward Modeling}
Complementary to purely catalog-level generative approaches, differentiable galaxy–halo forward models seek to describe galaxy populations as conditional generative processes on top of dark-matter structure. The
\href{https://github.com/ArgonneCPAC/diffsky/}{\tt Diffsky} framework \citep{OpenUniverse2024} rebuilds the traditional “halo $\rightarrow$ SFH $\rightarrow$ SED’’ chain using differentiable, physically interpretable blocks. \href{https://github.com/ArgonneCPAC/diffmah}{\tt Diffmah} \citep{Diffmah} provides a JAX-based, few-parameter model of halo mass assembly $M_{\rm halo}(t)$, replacing noisy merger trees with smooth, analytic, differentiable growth histories. On top of this, \href{https://github.com/ArgonneCPAC/diffstar/}{\tt Diffstar} \citep{Diffstar} models in situ star-formation histories with a small set of parameters (e.g., star-formation efficiency, gas-consumption timescale, quenching time), while \href{https://github.com/ArgonneCPAC/diffstarpop}{\tt DiffstarPop} \citep{Diffstarpop} lifts this to the population level by learning the statistical link between SFH parameters and halo assembly across suites of reference simulations. Finally, \acrlong{dsps} \citep[\href{https://github.com/ArgonneCPAC/dsps/}{\tt DSPS};][]{DSPS} maps these SFHs and associated metallicity/dust parameters to SEDs and photometry entirely within JAX. Together, {\tt Diffmah} + {\tt Diffstar}/{\tt DiffstarPop} + \texttt{DSPS} replace merger trees, non-differentiable semi-analytic recipes, and black-box SPS calls with a modular, probabilistic, fully differentiable stack whose low-dimensional, physically meaningful parameters can be calibrated and explored with gradient-based methods and SBI, while still generating large, realistic synthetic catalogs.
\acrshort{ai}-based models are in development to generate multiband galaxy images from these synthetic catalogs, conditioned on the parameters of the \texttt{Diffsky} suite.

\paragraph{Differentiable Cosmological N-body Solvers} Recent years have seen the emergence of particle–mesh $N$-body solvers implemented in modern, \acrshort{gpu}-accelerated, deep-learning frameworks that support automatic differentiation \citep[e.g.,][]{FlowPM, pmwd, DISCO-DJ, JAXPM}. Automatic differentiation enables hierarchical Bayesian inference directly over forward simulations of large-scale structure, opening a path toward full-field inference and near-optimal extraction of cosmological information. The main obstacles to deploying these methods at LSST scale are the computational and engineering demands of simulating survey-sized volumes in a differentiable way. Computing derivatives through the simulation implies non-trivial memory costs that are difficult to satisfy under the constraints of GPU accelerators. Several complementary strategies are being developed to address this challenge, including multi-node domain decomposition for distributed simulations \citep{Kabalan_jaxDecomp_2025}, techniques to reduce the memory cost of gradient evaluation \citep{Li__2024}, and improved time integrators that achieve a given accuracy with fewer time steps \citep{Rampf2025}. Beyond enabling full-field inference, differentiable simulations also enable the combination of physics-based solvers with learned components, yielding hybrid schemes that can improve the speed and accuracy of particle–mesh simulations \citep{Lanzieri_2023, Payot2023}.

\paragraph{Hydrodynamical-Simulation-Based Mappings of Galaxy Properties}
A complementary strategy is to treat state-of-the-art hydrodynamical simulations as high-fidelity “teachers’’ and use ML to distill their complex, small-scale physics into fast, effective models defined directly on dark-matter fields. Rather than specifying parametric galaxy–halo or SPS models, these approaches learn mappings from halo or large-scale-structure descriptors to galaxy properties as realized in the simulations. Recent work on \acrshort{ia} illustrates this paradigm. A traditional approach, followed by \citet{VanAlfen2024} is to develop an empirical IA model constrained by hydrodynamical simulations within a flexible \acrfull{hod}-like framework. A more ML-oriented approach is to learn an end-to-end emulator of galaxy properties. \citet{Jagvaral2025} introduce a geometric deep-learning approach in which galaxy shapes and orientations from IllustrisTNG \citep{IllustrisTNG} are modeled using E(3)-equivariant graph neural networks defined on the cosmic web, capturing the conditional distribution of shapes and orientations given halo mass, environment, and tidal field. This yields a fast, simulation-calibrated surrogate that reproduces intrinsic-alignment statistics at the percent level, enabling embedding of hydro-level realism into DESC mock catalogs without rerunning computationally expensive hydrodynamical simulations.

\subsection{Object Classification}
\begin{ThemeBoxA}[]
\themebullet \themekey{Methodology.}  \meth{ensembles}, \meth{gnn}, \meth{active-learning}, \meth{transformer}, \meth{self-supervised} \\
\themebullet \themekey{Challenges.} \challenge{covariate-shift}, \challenge{data-sparsity}, \challenge{scalability}, \challenge{uq} \\
\themebullet \themekey{Opportunities.} Multi-survey training, Data-driven Priors, Generative Models 
\end{ThemeBoxA}
The classification of astronomical objects is an area that has seen a surge in \acrshort{ml} and \acrshort{ai} applications over the past decade.
In regards to \acrshort{desc} science, we call out three potentially relevant areas: the detection and removal of bogus/non-astrophysical sources, the classification of galaxy types, and star/galaxy separation.
While the first area is critical for minimizing the pollution of source catalogs, it falls mostly under the responsibility of the Rubin Project's data management team and is addressed at the instrument signature removal stage of the Rubin pipeline \citep{Bosch18hsc}.
Galaxy type classification has been a strong application of citizen science and \acrshort{ai}/\acrshort{ml} applications, with an increasing demand and opportunity expected from the depth that will be achieved by Rubin \citep[e.g.,][]{2024A&A...683A..42C}.
However, galaxy type classification is not currently identified as a main concern for DESC beyond what is necessary to ensure a clean sample of galaxies for cosmological analyses (e.g., identifying merging galaxies that could confuse some of cosmological measurements).

On the other hand, star/galaxy separation is expected to be crucial to DESC science, as it may influence catalog completeness, weak-lensing shear estimation, galaxy clustering estimates, calibration of photometric redshifts, and object selection for spectroscopy. At the imaging depth of \acrshort{lsst}, galaxies vastly outnumber stars, and a nontrivial fraction of those galaxies are compact and blue, making them morphologically and photometrically similar to point sources, especially under variable seeing and in crowded fields \citep{fadely12}. 

The nominal approach to star/galaxy separation implemented by the Rubin Science Pipeline uses a cut-based approach based on object “extendedness”, which is constructed from a comparison of the \acrshort{psf}- and model-based measurements \citep[see][and references therein]{slater_morphological_2020}. 
This classifier performs well at bright magnitudes but struggles at the faint end, leading to either very large contamination in the faint star sample (i.e., orders of magnitude more galaxies than stars) or significant (i.e., nearly total) incompleteness for stars. 
Prior work from DESC members, working on precursor surveys like \acrshort{des}, improved on simple cut-based analyses by using feature-based machine learning—decision trees, random forests, boosted ensembles, and shallow neural networks trained on catalog-level features such as colors, shape moments, and PSF--model magnitude differences \citep[e.g.,][]{2015MNRAS.450..666S, sevilla18, baqui21, bechtol2025}.
However, these methods are ultimately limited by deblending errors, incomplete PSF modeling, and the loss of informative spatial structure in catalog summaries.

Parallel investigations have examined models that ingest multi-band image cutouts together with catalog features, allowing networks to learn morphology directly while leveraging color-based information (e.g., proximity to the stellar locus, color-color degeneracies, and uncertainty-aware colors; \citealp{2017MNRAS.464.4463K}). 
Efforts increasingly incorporate PSF awareness \citep[e.g.][]{Patel2025NPE}, multi-epoch data (i.e., variability), and deblending context to improve robustness across seeing conditions and sky regions, and explore semi/self-supervised representation learning to exploit LSST’s vast unlabeled datasets. 
Looking ahead, promising methodologies include end-to-end probabilistic deep learning that propagates PSF and noise models into calibrated class probabilities \citep{2019MNRAS.490.3952B}, which can include prior probabilities based on spatial location and spectrum \citep{lopez19}; vision transformers and \acrfullpl{acr:gnn} that integrate multi-visit information; multi-modal architectures combining images, colors, variability, and proper motion; domain adaptation and label-shift correction to handle spatial and temporal heterogeneity; active learning using sparse spectroscopic labels; and simulation-based training on realistic survey mocks. Emphasis on \acrshort{acr:uq}, continual and federated learning across data releases, and physics-informed constraints should yield classifiers that are both scalable and scientifically reliable at the LSST depth.

\subsection{Deblending}
\begin{ThemeBoxA}[]
\themebullet \themekey{Methodology.} \meth{vae}, \meth{npe}, \meth{instance-segmentation}, \meth{diffusion-model}, \meth{normalizing-flow}, \meth{cnn}, \meth{object-detection}, \meth{som} \\
\themebullet \themekey{Challenges.} \challenge{data-sparsity}, \challenge{metrics}, \challenge{uq} \\
\themebullet \themekey{Opportunities.} Multi-survey training, Data-driven Priors, Generative Models 
\end{ThemeBoxA}

Turning pixels into objects is a fundamental problem in astronomical survey pipelines. Object detection and deblending of \acrshort{lsst} data is a crucial step in producing catalogs useful for \acrshort{desc} science.  Given its unprecedented depth for a ground-based survey, LSST will face new challenges in its detection pipelines compared to previous legacy surveys \citep{Melchior21blending}.  Blending, or the overlapping of source light profiles, is an imaging systematic that affects all downstream analysis, as it becomes difficult (in reality intractable) to disentangle photons from a given source in a blend.  This problem is exacerbated with increased observing depth, as more light is collected from sources that are overlapping due to line-of-sight projections or physical interactions. Traditional object detection pipelines for wide-field surveys use a maximum likelihood estimator method \citep{Bosch18hsc} to identify peaks in intensity corresponding to sources in an image.  This method, while statistically justified, is still subject to failure modes, wherein \acrshort{ai} can provide alternative and complimentary methods.  Similarly, traditional deblending algorithms typically rely on models and assumptions about source light profiles that may not provide sufficient flexibility for the billions of sources LSST will observe \citep{Melchior18scarlet}.  DESC has been exploring and developing AI methods to aid in these challenging problems, which are crucial to understand and mitigate.

\paragraph{Catalog-level Blend Identification}  While a majority of sources observed by LSST are expected to have some level of blending, a particularly pernicious case is that of unrecognized blends.  These are sources in a blended scene that are indeed distinct (determined from high-resolution space-based observations), but are only recognized as a single source by cataloging and deblending pipelines.  Unrecognized blends impact measured properties such as galaxy shapes \citep{Dawson16ublends}, photometric redshifts \citep{Liang25catblend}, and more.  Estimates of the level of unrecognized blends in LSST  range from $\sim15$--$30\%$, with analysis of early LSST data compared to catalogs from the \acrshort{hst} \acrfull{candels} yielding an unrecognized blend rate of 18\% \citep{sitcom128}.  Blends that remain at the catalog level are definitionally unrecognized blends but may still be detectable as outliers via their multi-band photometry or their shapes. Random forests and \acrshortpl{acr:som} along with various anomaly detection algorithms were tested in \cite{Liang25catblend} who showed that unrecognized blends can be detected at a cost to the sample size. These algorithms were used to assign an unrecognized blend probability, however improvements can be made for specific science cases. For example, using the blend entropy (Ramel et al., in prep) can improve cluster cosmology by removing the most problematic unrecognized blends for cluster analysis. Designing better blending metrics like blend entropy and incorporating them into \acrshort{ml} algorithms like \acrshortpl{acr:gnn} is the main goal of the software package, \href{https://github.com/LSSTDESC/friendly}{friendly} . 

\paragraph{Image-level Deblending Using Deep Learning} Deep learning algorithms designed for object detection and deblending provide an alternative method to traditional pipelines that may help improve catalog completeness and source property measurements. For instance, the \acrfull{bliss} framework uses \acrshort{acr:npe} to infer probabilistic catalogs by training a \acrshort{acr:cnn} directly on multi-band images \citep{hansen2022}. BLISS produces well-calibrated posterior approximations for various source properties, and point estimates based on these posterior approximations outperform the standard LSST pipeline in source detection, flux measurement, star/galaxy classification, and galaxy shape estimation \citep{Duan25NPE}. The method is robust to spatially varying backgrounds and point-spread functions, provided these features are present in the simulated training images \citep{Patel2025NPE}. The \acrshort{deepdisc} instance segmentation \citep{Merz23DeepDISC} framework produces object catalogs and segmentation masks from image data, and is being tested with joint \textit{Roman}--Rubin data to incorporate multimodal information for downstream detection and deblending improvement. Both \href{https://github.com/astrodeepnet/debvader}{\tt DebVader} \citep{Arcelin21debvader} and \acrlong{madness} \citep[\acrshort{madness};][]{Biswas25mad} use \acrshortpl{acr:vae} to handle blending. They use self-supervised training to learn the structure of isolated galaxies. Through additional training of a deblending encoder they learn to isolate a galaxy from a blend. A specialized decoder can directly measure the characteristics of the galaxy (shape, \acrshort{photoz}) without reconstructing explicitly the image of the isolated galaxy. MADNESS adds a normalizing flow to the architecture to improve performance by modelling the latent-space distribution of galaxies, thereby providing an explicit likelihood for posterior optimization. Ongoing work uses a multimodal VAE to learn both from imaging and spectroscopy, adding more information in the latent space to improve the galaxy characteristics measurement, especially photo-$z$. The very principles of deblending VAEs alleviate the impact of unrecognized blends, and ongoing work on the use of probabilistic catalogs where the number of detected galaxies is itself non-deterministic will reduce it even more.  Even if sources are in such close proximity that LSST imaging will not be able to recognize the overlap and detect the blended group as one source, it is feasible to model the detected sources first, compute the residuals from the fit, and run detection again on the residuals. Because the residuals can have complicated structure, it is beneficial to perform the detection on multi-band residuals, where unrecognized sources appear as colored, localized over- or underdensities. Recognizing them, as well as their likely centers is possible and fairly effective with computer vision architectures like \acrshort{yolo} \citep{sowmya_kamath_2020_3721438}.

\paragraph{Data Driven Priors for Deblending with Explicit Likelihoods}

Generative models such as normalizing flows and diffusion models can be trained on unblended galaxies (potentially limited amounts of space-based data) and then serve as data-driven priors for galaxy morphologies \citep{Lanusse19gen}. Posterior optimization and sampling becomes possible for inverse problems with explicit likelihood functions (such as inpainting, deconvolution, and deblending). This is particularly effective in low \acrfull{snr} cases where the deblender \href{https://github.com/pmelchior/scarlet}{\tt scarlet}, \citep{Melchior18scarlet} which is the default deblending method in the Rubin Science Pipelines, is outperformed by a new, prior-augmented version from \href{https://github.com/pmelchior/scarlet2/}{\tt scarlet2} \citep{Sampson24scarlet2}. The same approach can also perform transient photometry in the presence of a host galaxy without the need for difference imaging \citep{Ward25scarlet2}. Additionally, posterior optimization in latent space by the MADNESS deblender \citep{Biswas25mad}, using data-driven priors, also outperformed scarlet.

\subsection{Shape Measurement}
\begin{ThemeBoxA}[]
\themebullet \themekey{Methodology.} \meth{differentiable-programming}, \meth{deep-network}, \meth{sbi} \\
\themebullet \themekey{Challenges.} \challenge{covariate-shift}, \challenge{uq}, \challenge{data-sparsity}, \challenge{scalability} \\
\themebullet \themekey{Opportunities.} Joint optimization of detection, deblending, and shear, hybrid analytic–neural estimators,  realistic multi-instrument and multi-epoch scene modeling, active learning, unified shear–photo-z modeling
\end{ThemeBoxA}

Weak-lensing shape measurement is one of the most critical and challenging components of the \acrshort{desc} analysis pipeline: small percent-level biases in ensemble shear propagate directly into the cosmological parameters targeted by \acrshort{lsst}. Meeting DESC requirements therefore demands methods that are simultaneously accurate enough to control multiplicative and additive shear biases, computationally efficient enough to process billions of galaxies, and amenable to calibration. A unifying theme in recent work is the exploitation of differentiability and \acrshort{gpu} acceleration, both in explicit shear calibration schemes and in forward models.

\paragraph{Analytic Calibration and Differentiable Shear Estimators}
The \acrfull{anacal} framework  \citep{Li2023,Li2025_bias} demonstrates how differentiability can be used to obtain high-precision shear responses and noise-bias corrections without relying on large external simulation campaigns. By representing galaxy properties and pixelized images using differentiable basis functions, AnaCal yields analytic shear responses for detection, selection, and shape measurement, achieving LSST-grade accuracy with sub-millisecond inference per galaxy. More broadly, other calibration schemes such as metacalibration \citep{Huff2017,Sheldon2017} stand to benefit from differentiable image models and measurement operators: with gradients available throughout the pipeline, shear response and noise-bias corrections can be computed more quickly and robustly.

\paragraph{Deep Learning–Based Shape Estimators}
Modern deep-learning architectures provide a natural path to an end-to-end differentiable shear estimator that can simultaneously integrate detection, deblending, shear estimation, and robustness to image defects. Neural networks are inherently GPU-accelerated, highly parallel, and differentiable, making them well suited to high-throughput shear inference at LSST scale. Such an approach was originally demonstrated in \citep{Ribli2019} and is being explored in a DESC context using the \acrshort{deepdisc} architecture \citep{Merz23DeepDISC}, which was originally designed as a general purpose architecture for detection and segmentation and which can estimate gravitational shears within a single GPU-resident and differentiable model. Being automatically differentiable, this estimator can be calibrated using the schemes mentioned above.

\paragraph{Hierarchical Forward Modeling with Differentiable Image Simulators}
A complementary strategy frames shape measurement as a hierarchical forward-modeling problem, in which cosmological parameters, population-level distributions of galaxy properties, and individual galaxy shapes are inferred jointly from the pixel data \citep{Schneider2015}. In this view, a forward model generates simulated images given a set of hierarchical parameters, and inference proceeds by comparing these simulations to the observed images. Such approaches are made practical thanks to the \href{https://github.com/GalSim-developers/JAX-GalSim}{\tt JAX-GalSim} effort which re-implements key \href{https://github.com/GalSim-developers/GalSim}{\tt GalSim} \citep{ROWE2015121} functionalities in JAX, making this forward model fully differentiable and GPU-accelerated while supporting vectorized batch simulations of thousands of galaxies at once. In ongoing DESC efforts, \texttt{JAX-GalSim} is used to implement the hierarchical shear-inference framework of \citet{Schneider2015}. Instead of traditional \acrshort{mcmc}, gradient-based samplers (\acrshort{nuts}), GPU acceleration, and batching can be used to yield roughly an order-of-magnitude speedup while keeping multiplicative shear biases within LSST requirements. While the aforementioned approach relies on analytic surface brightness profiles to model galaxies, more realism can be achieved through projects like \href{https://github.com/pmelchior/scarlet2/}{\tt scarlet2} \citep{Sampson24scarlet2} which extends the modeling to non-parametric morphologies and blended scenes observed with multiple instruments, providing a JAX-based, differentiable scene-modeling framework in which gradients of the likelihood with respect to source parameters and hyperparameters are readily available.

\subsection{Synthesis and Recommendations}
\label{sec3:synthesis}

\acrshort{ml} has become a foundational component of the collaboration's scientific infrastructure. It appears at every stage of the analysis pipeline: from pixel-level data processing (deblending, shape measurement) through derived observables (\acrshort{photoz}, cluster masses) to cosmological inference itself (weak lensing, \acrshort{sne}, clusters). Adoption is not driven by bespoke, application-specific methods. Instead, a small set of core methodologies and a consistent set of fundamental challenges recur across the entire application landscape, as visualized in Figure~\ref{fig:chord-diagram}. These cross-cutting patterns have direct implications for how \acrshort{desc} should organize its \acrshort{ai}/\acrshort{ml} efforts, prioritize methodological investments, and structure collaboration-wide infrastructure.

Among shared approaches, a few key methodologies appear repeatedly: \textit{\acrshort{acr:sbi}}, enabling parameter estimation from lensing to supernovae, though constrained by the fidelity of forward-models; \textit{differentiable programming frameworks} such as \href{https://github.com/DifferentiableUniverseInitiative/jax_cosmo}{\texttt{jax-cosmo}}, \href{https://github.com/GalSim-developers/JAX-GalSim}{\tt JAX-GalSim}, \href{https://github.com/pmelchior/scarlet2/}{\tt scarlet2}, for unlocking gradient-based inference at scale;  \textit{\acrshort{nde} and generative models} such as \href{https://github.com/Cosmo-Pop/pop-cosmos}{\texttt{pop-cosmos}}, \href{https://github.com/astrodeepnet/debvader}{\tt DebVader}, and \href{https://github.com/LSSTDESC/madness}{\tt MADNESS}, which provide flexible probabilistic representations with demonstrated cross-application reusability; \textit{emulators} such as \href{https://github.com/dpiras/cosmopower-jax}{\tt CosmoPower-JAX} and \href{https://github.com/justinalsing/speculator}{\texttt{speculator}}, which trade training cost for orders-of-magnitude speedups; and \textit{active learning}, for maximizing scientific return from limited expert annotations. These techniques recur because the scientific challenges in DESC (extracting maximal information under stringent systematic control, scaling to billions of objects, marginalizing over complex nuisance parameters) demand the same classes of solutions regardless of the specific probe. The fundamental challenges in the use of these techniques are equally shared between domains: \textit{covariate shifts}, spectroscopic selection bias in photo-$z$, sim-to-real gaps in \acrshort{sne}, model misspecification limiting \acrshort{acr:sbi}, and domain adaptation across surveys; \textit{\acrshort{acr:uq}}, obtaining well-calibrated posteriors and propagating uncertainties to cosmological constraints; \textit{scalability}, from billions of galaxies to $10^7$ nightly alerts, demanding algorithmic and infrastructure innovations; \textit{data sparsity and rare events}, including limited labeled samples, rare transients, class imbalance, and challenging edge cases like blending; and \textit{metrics and evaluation}, defining task-relevant metrics, validation frameworks, and stress tests aligned with DESC science requirements. Addressing these methodologies and challenges in a general, reusable way (rather than independently within each working group) has multiplicative impact: improving SBI's robustness to covariate shifts benefits not only clusters but also photo-$z$, lensing, and SNe; developing robust UQ enhances reliability for deblending, shape measurement, and generative simulations simultaneously.\\

The shared methodological approaches and barriers documented above demand a coordinated response. DESC should \textit{establish collaboration-wide AI/ML coordination mechanisms} (e.g., standing working group, cross-\acrshort{wg} task forces, interchange meetings) to ensure methodological innovations are rapidly evaluated for applicability across probes, common challenges are tackled collectively, and duplication is minimized. DESC should also \textit{invest in shared infrastructure and benchmarks} (reusable libraries, e.g., \href{https://github.com/DifferentiableUniverseInitiative/jax_cosmo}{\texttt{jax-cosmo}} and \href{https://github.com/GalSim-developers/JAX-GalSim}{\tt JAX-GalSim}, standardized interfaces, e.g., \acrshort{rail} for photo-$z$, and validation frameworks that stress-test methods under covariate shift and model misspecification) recognizing that these investments have multiplicative returns. DESC should \textit{develop collaboration-wide best practices and validation standards} for ML methods intended for cosmological inference, establishing requirements for UQ calibration such as coverage tests and \acrfull{pit} histograms, distribution-shift diagnostics, stress tests under deliberate misspecification. Finally, mechanisms to \textit{facilitate rapid dissemination of knowledge} (e.g., AI/ML workshops, shared tutorials, method spotlights in collaboration meetings) would accelerate the transfer of innovations across working groups.

%% file: sections/sec4_methodology.tex
\newpage
\section{Methodological Research Priorities to Advance ML for Precision Cosmology}
\label{sec4:aiml_research}

Extracting robust cosmological constraints from \acrshort{lsst} requires not only advanced algorithms but also a coherent methodological foundation that bridges simulation, data processing, and inference. Each of these pillars must meet unprecedented demands in scale, accuracy, and interpretability, demands that challenge the limits of both physical modeling and \acrshort{ml}. Beyond simply applying existing \acrshort{ai} techniques, \acrshort{desc} must develop methods tailored to the structure of astronomical data, the physics of observables, and the statistical rigor required for precision cosmology.

The scientific ambitions of LSST thus motivate AI/ML research in several key areas.
First, Bayesian inference and \acrshort{acr:uq} must evolve to handle the high-dimensional, hierarchical models that describe cosmic fields and galaxy populations, while maintaining interpretability and calibration across vast data volumes.
Second, \acrshort{acr:sbi} and related implicit-likelihood methods must confront the challenge of model misspecification and covariate shifts, ensuring that learned posteriors remain valid when simulations imperfectly represent real observations.
Third, physics-informed modeling, through differentiable programming and hybrid generative–physical architectures, offers a path toward interpretable and physically consistent deep learning, capable of representing both known and unknown components of the Universe.
Fourth, discovery and anomaly detection are essential to LSST’s potential for unexpected science, requiring representation learning and active human-AI collaboration to identify rare and previously unmodeled phenomena.

This section examines these research directions in detail, outlining both recent progress and outstanding challenges. We emphasize not only algorithmic innovation but also the validation, calibration, and interpretability principles required to integrate AI into cosmological analysis pipelines.

The fundamental question is: \textit{What would convince us of a cosmological result obtained with AI?}
Answering this question defines the research agenda for AI/ML in DESC and ensures that ML becomes not merely a computational shortcut, but a scientifically trustworthy component of cosmological inference.

\subsection{Bayesian Inference and Uncertainty Quantification}

\acrshort{acr:uq} represents perhaps the most critical challenge for deploying deep learning in precision cosmology. It is an area where \acrshort{ai} in the sciences demands solutions that differ from those in many commercial settings. Robust UQ must distinguish between aleatoric uncertainties (irreducible measurement noise) and epistemic uncertainties (model limitations and incomplete knowledge), as these have fundamentally different implications for cosmological inference and systematic error budgets.
\acrshort{ml} is poised to be revolutionary for inference, but only if current challenges are satisfactorily addressed. 

\subsubsection{Explicit Likelihood-Based Bayesian Inference}
\label{sec4:Bayes}

\begin{ThemeBoxA}
\themebullet \themekey{Related Methodologies.} \meth{hierarchical-bayes}, \meth{variational-inference}, \meth{gaussian-process} \\
\themebullet \themekey{Addresses.} \challenge{uq}, \challenge{scalability}
\end{ThemeBoxA}
    The inferential paradigm in astrophysical and cosmological data analysis has been for the past two decades primarily Bayesian, as this offers conceptual, methodological, and computational benefits~\citep{Trotta_2008}. The massive increase in data size and complexity afforded by \acrshort{lsst} will require a new step forward in inferential methodology, as LSST data will challenge the computational feasibility of current inferential engines. We cover in this section the case of \textit{explicit inference}, where we have the ability to directly evaluate the log-likelihood of our probabilistic models, and potentially its gradients. 

\paragraph{Accelerating Posterior Inference} \acrshort{mcmc} methods have been the workhorses of likelihood-based Bayesian parameter inference to date. Notable examples are Gibbs sampling \citep{Casella1992} and parallelized versions of Metropolis--Hastings (e.g., the affine-invariant sampler by ~\citealp{goodman2010ensemble, Foreman_Mackey_2013}). For cases where multimodality and/or strong parameter degeneracies are important, nested sampling~\citep{RN599,Ashton_2022} in its many variants (e.g., \texttt{MultiNest}~\citealp{Feroz:2007kg, Feroz:2009}; \texttt{PolyChord}~\citealp{Handley:2015,Handley:2015fda}; \texttt{dynesty}~\citealp{RN1654}; \texttt{DNEst4}~\citealp{Brewer:2016scw}), recently improved with gradients~\citep{lemosGGNS}, accelerated with neural emulators~\citep{Lovick:2025wdj}, or normalizing flows~\citep{RN1653} have been key to ensuring reliable inference.  

However, as analyses become increasingly complex, involving large numbers of nuisance parameters and expensive likelihood evaluations, the cost of running cosmological inference with conventional techniques becomes prohibitive. 

One avenue to speed up inference is to leverage access to the gradients of the log-posterior. When such gradients are available, a number of inference methods can benefit from them, including the well-established \acrshort{hmc} \citep{neal2011hmc}, as well as more modern generalizations such as \acrshort{nuts} \citep{HoffmanGelman2014}. Remarkably, physical ideas continue to lead the development of gradient-based inference techniques. \acrlong{rhmc} \citep[\acrshort{rhmc}][]{RHMC} simulates trajectories in arbitrary geometries by following the geodesics of the likelihood manifold. This makes RHMC schemes extremely robust sampling algorithms for high-dimensional inference in the face of severely non-gaussian posteriors. However, RHMC has not seen a wide spread adoption due to instability of second-order auto-differentiation needed to compute the curvature of the likelihood. In a similar vein, relativity-inspired HMC schemes \citep{RelativtyHMC} have recently been proposed that particularly target Minkowski geometries. This effectively introduces a maximum speed for the simulated particle, slowing it down in the areas of most challenging geometry, achieving most of the goals of RMHC without many of its hurdles. The recent Ray-tracing sampler \citep{2025arXiv251025824B} take a related approach, using light refraction as the guiding analogy to steer samples toward high-likelihood regions while providing a unifying framework in which HMC, LMC, and related methods emerge as special cases. As an alternative to Hamiltonian dynamics, \acrlong{lmc} (\acrshort{lmc}; \citealp{LangevineMC}) is based on a Langevin diffusion (an \acrshort{sde} whose invariant distribution is the target posterior) and in practice simulates a discretization of this process to generate approximate posterior samples. In this framework, the Metropolis adjustment that ensures the target distribution is sampled can be replaced with a bias requirement on the solution of the SDE, leading to significant speed ups \citep{GrumittLangevin}. A final avenue of improvement is to revise the assumed partition function of the particles simulated by the inference algorithm. Traditional HMC schemes assume that the distribution of particles being simulated follows a canonical partition function. However, more efficient sampling schemes can be constructed by exploring other partition functions. Microcanonical or energy-conserving HMC \citep[\acrshort{mchmc},][]{Robnik2022MicrocanonicalHM} explores the posterior distribution using a single energy shell. In a way similar to relativistic schemes, this is achieved by modifying the momentum of the particle to slow down at the regions of high-posterior density \citep{SteegHMC}, leading to a far more efficient sampling. \Acrlong{mclmc} \citep[\acrshort{mclmc};][]{MCLMC} is a sampling algorithm that combines all the ideas described above to great success. MCLMC already has been deployed to perform inference on physics problems such as lattice field theory simulations \citep{MCLMC} and even for cosmology where it has been shown to speed up field-level inference by an order of magnitude \citep{2023arXiv230709504B,simononfroy2025benchmark}. This makes MCLMC the cutting edge of gradient-based inference schemes and a promising tool to speed up analyses within \acrshort{desc}.   

Alternatively, one can replace a complicated posterior distribution with a more tractable one. \acrshort{vi} aims to find an approximation to the posterior distribution $p$ by a ``surrogate" parametrized distribution $q_\phi$, whose parameters $\phi$ are trained to minimize the Kullback-Leibler divergence between $q$ and $p$ (see, e.g., \citealp{Uzsoy_2024, JAX-COSMO}, using the JAX-powered NumPyro framework, \citealp{phan2019composable}). Here gradients are only needed for $q$, not for $p$.

Another area of research focuses on \textit{neural sampling methods}, which leverage in various ways neural networks to accelerate sampling while attempting to preserve asymptotic correctness guarantees. For example, normalizing flows have been used to re-parameterize the sampling space and cure complex geometries \citep{2022PNAS..11909420G}. Another recent line of research also leverages ideas from diffusion models and uses a neural score model to accelerate sampling \citep{2025arXiv250411713H}. 

As shown above, such inference strategies depend on differentiable components and will benefit greatly when likelihood codes are rewritten in frameworks that support automatic differentiation. An additional advantage of making probabilistic models compatible with such frameworks is that they usually support \acrshort{gpu} acceleration and vectorization, which opens up yet another avenue for acceleration---e.g.,the Numpyro \citep{phan2019composable, bingham2019pyro} and BlackJax \citep{cabezas2024blackjax} libraries written in JAX. Affine-invariant samplers \citep[see][]{Foreman_Mackey_2013} are particularly suited to vectorization on GPU hardware, as has been demonstrated in astronomy contexts (e.g.,\citealp{Thorp:2024, Thorp:2025}, using the \href{https://github.com/justinalsing/affine}{\tt affine} sampling code). 

\paragraph{Bayesian model comparison} Estimation of the Bayesian evidence, the central quantity for model comparison, remains challenging when the models being compared are very high dimensional. Nested sampling has been established as one of the main methods for Bayesian evidence computation, but in its original formulation it suffers from the curse of dimensionality: the efficiency of the constrained sampling step decreases rapidly as the dimensionality of parameter space increases. This has been somewhat mitigated by recent developments such as \texttt{PolyChord}~\citep{Handley:2015}, which can be used in a few hundreds of dimensions; \texttt{dynesty}~\citep{RN1654}, which uses dynamical allocation of live points \citep[see also][]{higson19}; and \acrlong{ggns}~\citep[\acrshort{ggns};][]{lemosGGNS}, which exploits gradients, generative flows and differentiable programming to achieve better efficiency and accuracy in up to $\sim 200$ dimensions.
A suite of other methods for the evaluation of the high-dimensional average of the likelihood over the prior are also being explored, sometimes combining density estimation with neural techniques  \citep[e.g.,][]{RN590,mcewen2023,Srinivasan_2024}. However, they remain confined to moderately low-dimensional parameter spaces, of order a few tens of dimensions. 

The frontier represented by evidence estimation in very large dimensional (of order $10^3$ or more) parameter spaces from real data remains largely untouched outside of synthetic demonstration examples where the ground truth is known. \acrshort{nre} shows promise in this respect, in that evidence estimation can be obtained from an NRE architecture by adding a suitable inferential head that is trained only on model labels, thus implicitly marginalizing over all parameters in the model. Such an approach naturally also generalizes to performing Bayesian model averaging. An example of this method is~\cite{SimSIMS}, where six models for empirical corrections for \acrshort{snia} data are compared from \acrfull{csp} observations \citep{krisciunas17} within a Bayesian hierarchical model setting with $\sim 4,000$ latent variables. 

\paragraph{Hierarchical Bayesian Models in Extremely High Dimensions} The manyfold increase in data size requires in many cases a more sophisticated model to capture previously unimportant effects; this in turn increases the dimensionality of the parameter space (especially in hierarchical models, where the latent space dimensionality scales with the number of objects within the model); the likelihood might become intractable, or previously used approximations, such as approximate Gaussianity or linear propagation of errors~\citep{Karchev2022}, neglecting of Eddington bias~\citep{Karchev_STARNRE}, might break down. 

 \autoref{sec3:wlss} introduced so-called full-field inference for cosmological surveys, in which not only cosmological parameters are inferred but also the initial conditions that seed the evolution of the large-scale structure of the Universe~\citep{porqueres2023fieldlevelinferencecosmicshear}. The fidelity of the forward simulation directly controls the accuracy of posterior constraints on cosmological parameters; consequently, this approach requires exploring extremely high-dimensional parameter spaces (millions to billions of parameters). Sampling such spaces is intractable for traditional MCMC and instead calls for gradient-based methods, as noted above. This, in turn, demands a forward simulation that is both fast and differentiable (see \autoref{sec3:sims}) to make full-field inference at LSST scale attainable.

It is worth noting that such hierarchical full-field inference models are substantially more computationally expensive than alternative \acrshort{acr:sbi} methods (see next section~\ref{sec4:sbi}), but offer several advantages. First, analyzing statistical errors directly in data space is more interpretable than working with the compressed summary statistics typical of SBI workflows; even with optimal compression, signals can mix and model misspecification becomes difficult to detect. Here, systematic contamination can be treated as additional parameters to be sampled \citep{2019porqueres}, becoming a machine-aided report of contaminations that have a characteristic pattern on the sky. Second, hierarchical Bayesian inference is designed for extensible, modular models in which new physics can be added—e.g., augmenting the simulation with a baryonification model—whereas SBI would require retraining neural density estimators from scratch. Taken together, these properties make hierarchical Bayesian inference well suited to joint inference of cosmology, systematics, and redshift-distribution uncertainties—capabilities that are considerably more difficult with implicit approaches. Additionally, hierarchical inference provides a digital twin of the Universe, which has multiple scientific applications but also provides a unique way of testing the results by cross-validating with independent data \citep{stopyra24}.

\subsubsection{Implicit Likelihood Bayesian Posterior Inference}
\label{sec4:sbi}

\begin{ThemeBoxA}
\themebullet \themekey{Related Methodologies.} \meth{sbi}, \meth{npe}, \meth{normalizing-flow} \\
\themebullet \themekey{Addresses.} \challenge{uq}, \challenge{scalability}
\end{ThemeBoxA}

The other paradigm is implicit inference, in which we do not assume direct access to the likelihood function, but only have access to samples from the joint distribution $p(x, \theta)$ of data samples $x$ and parameters of interest $\theta$. It should be noted that this situation covers both the case of \acrshort{acr:sbi}, and the case where $x$ and $\theta$ are available from a training sample of real observations (the canonical example being \acrshort{photoz} estimation from a set of spectroscopic observations). 

In particular, SBI is rapidly emerging as a powerful alternative to traditional fitting techniques for Bayesian models. The key idea is to replace an explicit likelihood function by forward simulating (under the model) parameter-data pairs, which are then used to train a neural network to perform inference \citep[e.g.,][]{Alsing2018, AlsingWandelt2018, AlsingWandelt2019,savchenko2024,lyu2025}. The advantages are that the (potentially intractable) likelihood can, in principle, incorporate physical effects of arbitrary complexity, which would otherwise be difficult to model (including, e.g., selection effects and complex parameter dependencies). In some variants~\citep{Miller2021} the 1- or 2-dimensional {\em marginal} distribution for the parameters of interest is targeted directly, thus circumventing the need to evaluate the high-dimensional joint posterior over all parameters in the model; such approaches are naturally suited to Bayesian evidence estimation~\citep{SimSIMS}. 

Additionally, inference can be {\em amortized} within a certain prior range, meaning that once trained the network can deliver almost instantaneous posteriors for a wide range of parameter values, a critical benefit when dealing with billions of galaxies \citep{2022ApJ...938...11H}.  This speed-up also permits posterior calibration methods (e.g., guaranteed coverage), which are computationally unfeasible with traditional posterior evaluation methods.

\paragraph{Neural Density Estimation (NDE) methods} The fundamental building blocks of these methods is \acrshort{nde}, where a neural network is used to estimate a distribution, or a ratio of distributions. Various kinds of methods exist: \acrshort{nle} (e.g., \citealp{Papamakarios2019, Lueckmann2019, Alsing:2019}), \acrshort{acr:npe} (e.g., \citealp{PapamakariosMurray2016, Lueckmann2017}) and \acrshort{nre} are among the most popular (for an overview see~\citealp{Alsing:2019, Cranmer2020, Lueckmann2021}). Implementations of NLE and NPE both learn a density based on simulated parameter-data pairs (see, e.g., \citealp{Alsing:2019}), with a variety of different approaches used for learning the multivariate joint or conditional density. Approaches to this include Gaussian mixtures \citep[e.g.,][]{Alsing2018}, mixture density networks \citep[e.g.,][]{PapamakariosMurray2016}, and normalizing flows (e.g., \citealp{Papamakarios2017, Papamakarios2019, Alsing:2019, Jeffrey:2021, 2022ApJ...938...11H}; for a review see \citealp{Kobyzev2021, Papamakarios2021}). More sophisticated density estimators used in generative modeling -- such as continuous normalizing flows \citep{Grathwohl2018, Chen2018}, score-based diffusion models \citep{Song:2020}, flow-matching models \citep{Lipman:2022}, and transformers \citep{Transformers2017} -- are also well suited to NLE and NPE tasks \citep[e.g.,][]{DiazRiveroDvorkin2020, Geffner:2022, Wildberger:2023, Gloeckler:2024} alongside the generative modelling tasks they are commonly used for \citep[e.g.,][]{Alsing:2024, Cuesta:2024, Thorp:2025}. 

\paragraph{Optimal Neural Summarization} 
To ensure the robustness of implicit inference, the process is usually divided into two main steps enabling each neural network to focus on a specific task: (1) compression of high-dimensional data into informative summary statistics, and (2) performing Bayesian inference using neural density estimation methods on this low dimensional but highly informative statistic. To maximize information extraction and improve constraints on cosmological parameters, the community has increasingly adopted neural network–based summarization techniques. While any neural network can be trained on the regression task of inferring parameters given data \citep[e.g.,][]{Gupta2018MAE, kacprzak2022deeplss, lu2023cosmological}, it is unclear how much of the information contained in the data is extracted. In particular, \cite{neural_summary_lanzieri_2025} demonstrate that standard regression loss functions do not guarantee the systematic construction of sufficient statistics. \Acrlongpl{imnn} (\acrshortpl{imnn}; \citealp{2018PhRvD..97h3004C}) directly address this problem by learning summary functions that maximize the Fisher information. They can produce nearly exact posteriors and are thus approximately sufficient statistics of the data. Another approach is to derive a loss function directly from the definition of sufficiency, i.e., by maximizing the mutual information between the summary statistics and parameters of interest \citep{Jeffrey:2021, chen2021neural}.

\paragraph{Controlling Epistemic Errors in Inference Results} One fundamental limitation of NDE methods is that their reliance on a neural network to model at some level the likelihood of the data is inherently imperfect. In the asymptotic regime of infinite data and flexible neural network, the approximation to the target posterior will converge, but in practice we are never guaranteed to find ourselves in this regime, and must therefore take into account and mitigate \textit{epistemic errors}. Several strategies have been developed over the years to quantify and mitigate this epistemic uncertainty on inference results.  \acrshort{mcmc} sampling over network parameters provides gold-standard uncertainty estimates but at usually prohibitive computational cost \citep[e.g.][]{2025arXiv251025824B}. In addition, detecting convergence of the chain remains difficult, usually necessitating drawing more samples than ultimately needed. Because such approach is extremely expensive, other approaches have been developed. \textit{\acrshortpl{bnn}} approximate the posterior distribution over network parameters through \acrshort{vi}, providing principled uncertainty estimates at reduced computational cost. However, the tradeoff between approximation quality and speed remains concerning, especially in the highly multi-modal loss landscapes of deep neural networks. In such settings, scalable variational methods often collapse to a single mode of the posterior rather than exploring the full diversity of solutions \citep{fort2020losslandscape}, which limits the quality of their uncertainty estimates. One of the most used VI methods in astronomy is Monte Carlo dropout which utilizes the dropout layer commonly introduced in deep neural networks to prevent correlated activation as one of the computationally cheapest approximations to Bayesian inference; however, its theoretical justification and empirical accuracy are questionable \citep{LeFolg:2021}. Nonetheless, comparisons with other methods have shown promise for astronomical applications: strong lensing \citep{Perreault:2017}, supernova time-series classification \citep{Moller:2020,Moller:2022ICML}, and star time-series classification \citep{Astromer2023, CadizLeyton:2025, CadizLeyton:2025:MoE}. Other VI methods such as Bayes by Backprop and SWAG have been sparsely used for time-series classification and regression with mixed results \citep{Moller:2020,Cranmer:2021}. Another strategy is Deep
Ensembles in which multiple networks are trained from different random initializations to provide uncertainty estimates via the variance of their predictions \citep{Makinen:2021,Moller:2022,Moller:2024}. Unlike variational methods, ensembles capture uncertainty by effectively sampling from different modes of the loss landscape, resulting in more robust and better-calibrated uncertainty estimates. However, they are more computationally demanding than MC dropout, and do not mitigate errors arising from model misspecification. Comparative studies exploring the trade-offs between these \acrshort{acr:uq} methodologies for achieving superior uncertainty evaluation are a growing focus in the field \citep{CadizLeyton:2025}.

\subsubsection{Model Mispecification and Covariate Shifts}
\label{sec4:model-misspec}

\begin{ThemeBoxA}
\themebullet \themekey{Related Methodologies.} \meth{sbi} \\
\themebullet \themekey{Addresses.} \challenge{covariate-shift}
\end{ThemeBoxA}

From a technical standpoint, \acrshort{acr:sbi} has achieved impressive results: \acrshort{acr:npe} with normalizing flows performs well in low-dimensional regimes~\citep[e.g.,][]{srinivasan2025}, \acrshort{nle} scales satisfactorily to higher dimensions, and marginal \acrshort{nre} has shown success across diverse applications~\citep[e.g.,][]{alvey2024,list2023,Franco_Abell_n_2024,Saxena_2024}. Yet, these demonstrations rely primarily on simulated data and therefore represent best-case scenarios; direct validations of SBI on real data remain scarce~\citep{2024Karchev_SIDEreal,Lueber2025}.

The robustness of SBI depends critically on the realism and completeness of the simulations that underpin it. Simulations must reproduce all relevant aspects of the observations, including astrophysical, instrumental, and observational effects. Any unmodeled process leads to domain shift or model misspecification, which can severely bias inference. This is particularly problematic for NRE, which relies on accurate joint modeling of data and parameters~\citep{filipp25}, while NLE is comparatively more interpretable since it operates directly in data space. Even when the theoretical model is sound, the observational and noise models must be equally faithful—a condition often unmet given the traditional divide between theoretical and observational cosmology. Bridging this gap is essential for SBI to succeed. Efforts are underway to diagnose and quantify model misspecification through simulation-based calibration and related approaches~\citep{2018arXiv180406788T,2023PMLR..20219256L,CoLT,montel2025,kelly2025simulation}. While model misspecification is a key vulnerability of SBI, it is not fundamentally different (if more difficult to diagnose and cure) than the similar risk incurred when using explicit, likelihood-based models: missing components of the model w.r.t. the true data-generating process will lead to potentially severe bias in the resulting inference. SBI is a relatively new technique, and therefore appropriate diagnostic tools are still being developed to ensure its robustness and reliability. 

\paragraph{Training Set Representativity}
A fundamental challenge across deep learning applications in astronomy is the representativity of training data. Models trained on simulations may fail to generalize to real observations, while those trained on current surveys may struggle with the deeper, higher-resolution \acrshort{lsst} data. Techniques such as fake source injection—embedding simulated objects into real images—can mitigate these gaps~\citep{2016MNRAS.457..786S,2018PASJ...70S...6H}, though their success depends on how realistic the injected sources are. The problem is particularly acute for rare phenomena, where training examples are intrinsically limited. To improve generalization, domain adaptation, transfer learning, and hierarchical Bayesian methods are being explored. E.g., \cite{Swierz2024} use domain adaptation to obtain more robust data summaries that can generalize well between simulated data and mock observations, enabling more accurate neural density estimation. Principled approaches such as stratified learning (discussed in Section~\ref{sec3:photo-z}) can mitigate covariate shift with little modification in the learning procedure, but other methods often require substantial experimentation and modifications of training procedures~\citep{2025arXiv251019168K}. Such corrections must themselves be treated as part of the inference pipeline and undergo rigorous calibration and uncertainty quantification.

\paragraph{Physics Hardening}
When available datasets are incomplete or non-representative, physics-informed augmentation can enhance robustness. For example, the \acrshort{desc} \acrshort{elasticc} challenge~\citep{knop2023} injected transients simulated using semi-analytic models (e.g., \acrshort{sne}, \acrshort{kne}) to make classifiers more resilient to underrepresented classes, and \cite{Moskowitz2024} augmented spectroscopically-incomplete training samples with simulated photometry to improve photometric redshift estimation. Latent representations derived from \acrshort{sps} models can also be used to generate synthetic photometry for missing or incomplete observations~\citep[e.g., \href{https://github.com/Cosmo-Pop/pop-cosmos}{\tt pop-cosmos}][]{Alsing:2024,Thorp:2025,Deger:2025}, and facilitate comparisons with hydrodynamical simulations without observational systematics. However, these methods inherit the assumptions and uncertainties of the underlying theoretical models—such as uncertain nebular emission strengths in SPS~\citep{Byler2017_Nebular,Li2025_cue,Morisset2025_Nebular,2025arXiv250103133N}—which can themselves introduce model misspecification~\citep{Leistedt2023,Jespersen2025_opticalIR}. Addressing these limitations requires deeper astrophysical modeling of galaxy formation and evolution, as well as diagnostic tools for identifying misspecification in high-dimensional generative models~\citep[e.g.,][]{Thorp:2025:QQ}. Because physics-informed generative models (e.g., those that capture information within SPS parameterizations) can be used to synthesize observables that the model has not been trained on, such models can be validated not only against unseen data from a test set but also on new types of observations and other surveys \citep[e.g.,][]{Alsing2023, Alsing:2024, Thorp:2024, Thorp:2025, Deger:2025}.

\subsubsection{Validating Inference Results}
\label{sec4:validation}

\begin{ThemeBoxA}
\themebullet \themekey{Addresses.} \challenge{uq}, \challenge{metrics}
\end{ThemeBoxA}

While the quality of neural posteriors can be tested \citep{2018arXiv180406788T, 2023PMLR..20219256L}, and while statistical tests can be performed to determine the probability that the distribution learned by a generative model is identical to that of the training data \citep{PQMass}, an open issue is the determination and procurement of a sufficient volume of training data for those tests to be sufficiently sensitive and for the learned distribution to be accurate.

Models trained on the same data but with different algorithms exhibit distinct probability calibration characteristics that must be evaluated and corrected. Similarly, identical algorithms trained on different training sets require independent calibration assessment. Common diagnostics include reliability diagrams used in time-series classification \citep{Moller:2020} or non-conformity scores from conformal inference techniques \citep{Xie:2025}, both of which compare predicted probabilities against observed frequencies. Detection of anomalies, i.e., the classification whether a signal is anomalous enough to be reported, is particularly vulnerable because it probes the tails of a learned distribution.  For regression tasks, calibration ensures that predicted uncertainties accurately capture the true error distribution. Poorly calibrated uncertainties can introduce systematic biases in downstream cosmological analyses, leading to incorrect parameter constraints. Recalibration methods for lens modeling are presented by, e.g.,  \citet{Perreault:2017}, \citet{Karchev2022GP}, and \citet{gentile23}. Regarding generative models, \cite{Campagne_2025} propose a “two-models” framework to evaluate their statistical consistencies trained on independent subsets of galaxy images. The results emphasize the need for large-enough datasets to enable calibration and validation strategies specific to each generative architecture (e.g., generative adversarial networks, normalizing flows and score-based diffusion), since apparent visual quality and morphological variable distributions alone do not guarantee statistical reliability.

\subsection{Physics-Informed Approaches}
\label{sec4:physics-informed}

From a high-level point of view, neural networks are never perfectly trustworthy and are often hardly interpretable. This motivates a general desire to build \textit{physics-informed} models, which can leverage as many explicit physical constraints as possible, thus limiting the potential failure modes of \acrshort{ai} components. 

\subsubsection{Hybridization of Generative Modeling and Physical Models}
\label{sec4:hybrid-gen-phys}

\begin{ThemeBoxA}
\themebullet \themekey{Related Methodologies.} \meth{normalizing-flow}, \meth{diffusion-model}, \meth{neural-surrogate}, \meth{differentiable-programming} \\
\themebullet \themekey{Addresses.} \challenge{covariate-shift}, \challenge{scalability}
\end{ThemeBoxA}

A promising direction for scientific inference is the hybridization of explicit, physics-based models with generative components that flexibly represent unknown or intractable distributions. In this framework, flow-based or diffusion models serve as probabilistic priors over complex latent variables, such as galaxy morphology or small-scale baryonic processes, while the rest of the model remains physically interpretable and simulation-driven. These generative priors have already proven powerful in astronomical inference, e.g., in the estimation of galaxy properties and photometric redshifts at scale (e.g.,  \href{https://github.com/Cosmo-Pop/pop-cosmos}{\tt pop-cosmos}; \citealp{Alsing:2024, Thorp:2024, Thorp:2025, Deger:2025}), and in the generation of high-fidelity, field-level H\,{\sc i} maps from dark matter simulations~\citep{Mishra:2025}. More broadly, generative models naturally support amortized inference frameworks, where neural posterior estimators are trained on samples from the generative process, enabling accurate Bayesian inference without \acrshort{mcmc} sampling but at the cost of greater upfront training effort.

\paragraph{Differentiable Programming} To fully integrate such probabilistic components with physical simulations, modern astrophysical codes are increasingly being reimplemented in automatic-differentiation libraries. Differentiable simulators eliminate the approximation errors of emulators and provide exact gradient information for optimization and uncertainty quantification. Examples include differentiable particle-mesh cosmological solvers~\citep{FlowPM,DISCO-DJ}, theoretical cosmology computations in \href{https://github.com/DifferentiableUniverseInitiative/jax_cosmo}{\texttt{jax-cosmo}}~\citep{JAX-COSMO}, galaxy–halo connection models in \href{https://github.com/ArgonneCPAC/diffsky}{\texttt{Diffsky}}/\href{https://github.com/ArgonneCPAC/diffstar/}{\texttt{Diffstar}}~\citep{Diffstar}, stellar population synthesis in \href{https://github.com/ArgonneCPAC/dsps/}{\texttt{DSPS}}~\citep{DSPS}, halo-model calculations in \href{https://github.com/fkeruzore/halox}{\texttt{halox}}~\citep{halox}, and differentiable image simulations in \href{https://github.com/GalSim-developers/JAX-GalSim}{\texttt{JAX-GalSim}}~\citep{jaxgalsim}. 

While the widely used \href{https://github.com/GalSim-developers/GalSim}{\texttt{GalSim}} library~\citep{ROWE2015121} produces realistic galaxy images, its non-differentiable design limits efficient gradient-based inference. In contrast, the emerging \href{https://github.com/GalSim-developers/JAX-GalSim}{\texttt{JAX-GalSim}} library reimplements core \texttt{GalSim} functionalities in JAX, yielding \acrshort{gpu}-accelerated, fully differentiable forward models that enable direct gradient computation for both population-level and individual-level parameters. Similarly, \href{https://github.com/pmelchior/scarlet2}{\tt scarlet2} offers a JAX-based, differentiable framework for non-parametric source morphologies and blended scenes observed by multiple instruments. Both libraries support vectorized batch simulations, crucial for large-scale hierarchical inference, and allow gradients of differentiable likelihoods to be computed automatically for maximum-likelihood estimation, variational inference, or gradient-based MCMC.

\paragraph{High-Dimensional Inverse Problems with Explicit Likelihood and Data-Driven Priors} 
High-dimensional inverse problems are central to cosmology, from galaxy deblending and strong-lensing source reconstruction to recovering the dark matter field from noisy data. In these settings, the forward process, such as instrumental response, noise model, or lensing distortion, is well understood and can be encoded in an explicit likelihood. The underlying components, however, such as galaxy morphologies or the non-Gaussian dark matter structure, lack closed-form descriptions and require expressive statistical models. Generative models, such as diffusion models, can learn realistic priors from high-dimensional observations or simulations. Combining such data-driven priors with explicit likelihoods yields a principled framework: the prior enforces realistic structure, while the likelihood anchors inference to the data, even under low \acrshort{snr} conditions where fully amortized approaches may drift. Recent works have demonstrated this hybrid approach for galaxy source reconstruction, strong lens modeling, and superresolution \citep{adam2022posterior,Barco2025blindinversion,2025Adam_SBProfiles} and dark matter field inference \citep{remy2023}. Moreover, the presence of an explicit likelihood enables learning data-driven priors directly from observations, through iterative refinement using posterior samples \citep{rozet2024learning, Barco_2025}, and can even allow for the correction of model misspecification \citep{Payot2025}. A remaining challenge is efficient posterior sampling, as inference with diffusion priors entails solving \acrfullpl{ode}, which is computationally demanding, although it can be performed practically at scale \citep[see, e.g.,][]{Thorp:2024, Thorp:2025}.

\subsubsection{Imposing Consistency with Physical Equations and Symmetries}
\label{sec4:physics-constraints}

\begin{ThemeBoxA}
\themebullet \themekey{Related Methodologies.} \meth{physics-informed} \\
\themebullet \themekey{Addresses.} \challenge{covariate-shift}, \challenge{metrics}
\end{ThemeBoxA}
%\note{This section can be fleshed out further, it's a bit thin right now.}

For trustworthy results, we additionally demand that the \acrshort{ai}/\acrshort{ml} outputs at various stages of the pipeline satisfy null tests (e.g., B-modes in gravitational lensing or rho-statistics for \acrshort{psf} modeling, \citealp{10.1111/j.1365-2966.2010.16277.x}) or obey the laws of physics of the corresponding analysis component rather than merely report final results with high fidelity. The modularity of the pipelines and multi-scale nature of the phenomena asks for validations at every analysis stage. Crucially, our knowledge of physical relations (in the universe, in the atmosphere, in the instrument) permits a form of validation that is typically omitted or impossible in industry applications of AI/ML. This motivates research in areas such as invariant/equivariant representation learning and geometric learning, with possible interdisciplinary implications beyond the scope of \acrshort{desc} \citep{2025arXiv250902661F}. Furthermore, the use of symmetry-aware \acrfullpl{enn} could help with the extraction of more robustness features from the data and, in combination with domain adaptation, enable easier mitigation of covariate shifts~\citep{Pandya2025}. E.g., with \acrfullpl{pinn} one can, in principle, find solutions for explicitly specified differential equations if their optimization could be made more robust \citep{2024arXiv240201868R}. 

\subsection{Novelty Detection and Discovery}
\label{sec:discovery}

\begin{ThemeBoxA}
\themebullet \themekey{Related Methodologies.} \meth{anomaly-detection}, \meth{self-supervised}, \meth{active-learning} \\
\themebullet \themekey{Addresses.} \challenge{covariate-shift}, \challenge{data-sparsity}
\end{ThemeBoxA}
A central scientific promise of \acrshort{lsst} lies in its potential for unexpected discovery. Many now-fundamental astrophysical phenomena, such as strong lenses, fast radio bursts, and pulsars, as well as singular systems like the Bullet Cluster, were first identified as anomalies. With an anticipated catalog exceeding 20 billion galaxies and roughly 10 million alerts per night, however, detecting novel phenomena in LSST data represents an unprecedented challenge.

Generative models offer a powerful framework for unsupervised discovery by learning the ensemble properties of galaxy images, spectra, photometry, and time-domain behavior, and by enabling the detection of statistically anomalous signals without labeled training data~\citep{2023AJ....166...75L}. In time-series and high-energy astronomy, representation learning has already proven effective for discovery-oriented analyses~\citep{Dillmann2025,Song2025}. However, a common failure mode of generative models, where atypical signals appear highly typical \citep{2018arXiv181009136N}, will mean that true outliers may not be recognized, a loss if we seek to find them (e.g., rare SN types such as pair instability \acrshort{sne}) and a problem if they contaminate carefully selected samples used in high-precision cosmology (e.g.,unrecognized blends in shape catalogs, \citealp{Dawson16ublends}; or catastrophic outliers in \acrshort{photoz} estimates). 
More specifically, standard unsupervised methods often struggle in the dense and homogeneous latent spaces produced by deep representations~\citep[e.g.,][]{Baron2025,StarEmbed2025}. E.g., while \citet{AstronomalyDecals2024} successfully combined Zoobot (a foundation model discussed in \autoref{sec:foundation_models}) features with the \href{https://github.com/MichelleLochner/astronomaly}{\tt Astronomaly} framework~\citep{2021Lochner_Astronomaly} to identify new sources, \citet{ZoobotApplications2022} found that tailored anomaly-detection techniques were necessary even within the well-studied Galaxy Zoo dataset. This line of work culminated in \texttt{Astronomaly} \texttt{Protege}~\citep{Protege2025}, a general-purpose, active anomaly detection system optimized for exploration in deep latent spaces.

The emergence of deep learning and foundation models (\autoref{sec:foundation_models}) further elevates the importance of active learning, the tight integration of human expertise and machine-driven pattern recognition~\citep{Protege2025}. Self-supervised methods and foundation models promise the ability to generate rich, general-purpose representations, but expert oversight remains essential to interpret their outputs and assess scientific relevance. Automated systems may flag outliers or cluster data effectively, yet human judgment is required to determine which patterns constitute genuine discovery. As \acrshort{ai} systems evolve toward agentic operation (\autoref{sec:llm_agentic}), the collaboration between human and machine will become increasingly intertwined. Embedding active-learning capabilities directly within AI infrastructures will therefore be critical to enable rapid, scalable, and participatory scientific discovery—bridging expert analysis and citizen science within the LSST era.

%% file: sections/sec5_emerging.tex
\newpage
\section{Emerging Techniques}
\label{sec5:emerging_tech}
The priorities outlined in \autoref{sec4:aiml_research} establish the statistical and algorithmic foundations for \acrshort{ml}-enabled cosmology, but realizing these priorities at the scale and complexity of \acrshort{lsst} demands new approaches. Training distinct models for each science case is neither computationally sustainable nor conducive to the cross-probe consistency required for joint analyses. 

This section surveys two classes of emerging \acrshort{ai} techniques that may fundamentally change how \acrshort{desc} builds and maintains its analysis infrastructure: data \acrshortpl{fm} (\autoref{sec:foundation_models}) and \acrshort{llm}-based agentic systems (\autoref{sec:llm_agentic}). By building general-purpose and reusable models from large unlabeled datasets, data \acrshortpl{fm} offer the potential for a shared backbone \acrshort{ml} infrastructure that can be applied across science cases, greatly reducing time to science for analyses involving \acrshort{ml}. \acrshortpl{llm} and \acrfullpl{mas} target a different bottleneck: the coordination of complex workflows, synthesis of documentation and literature, and accessibility of the DESC software infrastructure to new collaboration members. Together, these techniques provide a framework for scaling the \acrshort{ml} efforts in DESC from individual analyses to reproducible, collaboration-wide pipelines.

\subsection{Data Foundation Models}
\label{sec:foundation_models}
\acrshortpl{fm}, \acrshort{ai} systems trained on massive datasets to perform a broad spectrum of tasks \citep{Bommasani2021FoundationModels}, have not only revolutionized \acrshort{ai} research but are also rapidly reshaping modern life. Vision \acrshortpl{fm} are now enabling breakthroughs in robotics \citep{RobotApplication2024}, medical diagnostics \citep{MedicalApplication2024}, and remote sensing \citep{RemoteSensing2024}. Beyond these applications, their adoption in scientific disciplines such as genetics \citep{Evo2025} and heliophysics \citep{Surya2025} has highlighted their powerful predictive capabilities and their capacity to uncover fundamental relationships in complex data. The burgeoning field of \acrshortpl{fm} presents a significant opportunity to enable and accelerate cutting-edge astrophysics.

Zoobot \citep{Zoobot2022} can be considered the first vision \acrshort{fm} in optical astronomy. Having been trained on labels from the Galaxy Zoo citizen science project \citep{GalaxyZoo2008, GalaxyZoo2011, GalaxyZoo2013}, it has demonstrated versatility on a range of downstream tasks. These include broad morphological classification problems critical for studies of galaxy evolution, such as identifying merging galaxies \citep[e.g.,][]{MergersHSC2023,MergerChallenge2024, MergerSims2025}, and anomaly detection for finding rare phenomena like strong lenses \citep{walmsley25, lines25}. Formally released in \citet{ZoobotRelease2023}, Zoobot has been adapted to imaging data from multiple surveys, including \acrshort{desi} \citep{DECALS2019}, \textit{Euclid} \citep{ZoobotEuclid2024}, \acrshort{hst} \citep{ZoobotHST2023}, and \acrfull{jwst} simulations \citep{ZoobotJWST2024}, among others \citep{ZoobotGAMA2024, ZoobotUNIONS2025, ZoobotHSC2025}.

\Acrshortpl{fm} have also been designed for astronomical time-series datasets, spurred by the need for automated photometric classification of Galactic and extragalactic transients in the Rubin era. Some examples of these models include Astromer \citep{Astromer2023, Astromer2025}, AstroCo \citep{Astroco2025}, ATAT \citep{ATAT2024}, FALCO \citep{FALCO2025}, and RoMaE \citep{ROMAE2025}. While the primary requirement for these models is high classification accuracy, they also enable the discovery of new classes of transients through anomaly detection, and, in some cases, provide lightcurve interpolation for further downstream analysis.

Despite early successes in modality-specific \acrshortpl{fm}, many astrophysical questions can only be answered by fusing different data types. This is a task for which traditional methods, which rely on reducing data to summary statistics, are quickly being outpaced by powerful multimodal \acrshortpl{fm}. Though the ideal neural architectures and training objectives for these multi-modal models are areas of active research, models have now been developed for galaxies \citep{Astroclip2024}, \acrshort{sne} \citep{2024Zhang_Maven}, variable and non-variable stars \citep{2024Leung_stellarfm,AstroM2025}, and cosmological simulations \citep{MOSAIC2025}. AION-1 \citep{parker2025aion} represents a significant step toward survey-scale multimodal \acrshortpl{fm}: trained on over 200 million observations of both stars and galaxies from five major surveys, it integrates images, spectra, and scalar measurements into a billion-parameter model. AION-1 demonstrates strong performance in low-data regimes, enables zero-shot detection of rare objects such as strong gravitational lenses, and produces survey-invariant representations that facilitate knowledge transfer across telescopes. 

\subsubsection{Foundation Models for DESC Science}
\paragraph{Pre-training and reusability} 
\Acrshortpl{fm} offer a significant advantage over existing techniques by reducing heterogeneous data types into a unified and simplified numerical representation known as a latent space, representation or feature space. Instead of reprocessing a full dataset for each analysis, a single, powerful \acrshort{fm} can generate rich data representations once. These representations can then be used directly or rapidly finetuned for numerous specific science cases, resulting in significant savings of computing resources. 

This shared representational basis also changes how \acrshort{desc} can approach cosmological inference. Traditional analyses reduce complex image data to limited summary statistics (e.g., moments, colors, or flux ratios), inevitably discarding information. Deep learning enables direct inference from raw observations, but training bespoke networks for each task across the petabyte-scale archive of \acrshort{lsst} is infeasible. A general-purpose \acrshort{fm}, trained once at scale, can thus act as a reusable ``backbone'' for all DESC pipelines, propagating consistent representations across tasks. 

For time-domain astronomy, the representations produced by \acrshortpl{fm} are especially powerful. Because models learn features directly from the data, they are not constrained by the a priori assumptions of human-engineered features. This makes them ideal for identifying new or unexpected classes of objects via anomaly and novelty detection. Furthermore, these representations could be highly effective for standard tasks such as early transient classification, which is critical for triggering spectroscopic follow-up.

For simulation-based inference (\autoref{sec4:sbi}), \acrshort{fm} latent spaces can act as highly flexible encoders, providing a more powerful data compression than traditional statistical summaries. In terms of data handling, multimodal models can naturally accommodate missing data when fusing datasets. Moreover, the ability to pre-train on vast unlabeled datasets enables \acrshortpl{fm} to address the representational bias (dataset shift) often encountered between the training and test sets in supervised learning.

\paragraph{Opportunities offered by multimodality}
 In astronomy, multi-wavelength models learn a shared latent space from multiple observational wavelengths (e.g., optical and infrared) of the same object. Multimodal models extend this concept by integrating fundamentally different data types into a shared latent space, often through self-supervised learning techniques. While the specific AI architectures for combining heterogeneous datasets remain in active development, all approaches fundamentally treat different data types as complementary views of the same underlying astrophysical system. The resulting multimodal representations provide a computationally efficient and powerful framework for a broad spectrum of scientific analyses.

In the time domain, joint modeling of photometric light curves and spectra provides a direct path to improving supernova cosmology and our understanding of explosion physics. A multimodal \acrshort{fm} can perform cross-modal imputation, predicting the spectral properties of SNe Ia from irregular photometric sequences alone \citep[as has been demonstrated on synthetic data;][]{2025Shen_DitSNe,2025Shen_MMVAE}, thereby recovering physically meaningful features (line velocities or continuum temperatures) that are otherwise inaccessible from broadband imaging data. These inferred spectra can serve as additional standardization parameters for SNe Ia, potentially reducing residuals in the Hubble diagram by incorporating information linked to, e.g., progenitor diversity \citep{2025Son_AgeBias} or host metallicity \citep{2013Childress_Metallicity}.

Beyond improving standardization, the same cross-modal embeddings enable proactive discovery. By comparing inferred to obtained spectra, outliers can be flagged for long-term monitoring. When combined with host-galaxy properties across the transient samples discovered by Rubin LSST, these models can provide a probabilistic mapping between host galaxy and transient properties, useful for exploring population-level correlations potentially linked to supernova physics and for obtaining sub-populations of highly-standardizable SNe Ia.

\paragraph{Challenges}
Despite rapid progress, technical and practical challenges must be addressed before multimodal \acrshortpl{fm} can be fully integrated into DESC pipelines. Propagating observational uncertainties to all studies conducted downstream of a DESC-wide \acrshort{fm} is critical, and the diversity of applications across spatial and temporal scales may not be well matched to a single architecture's inductive biases. Multiple application areas also require capturing the impact of selection effects in extant training samples, which may cause a model to over-weight well-observed populations and under-represent rare but scientifically informative objects. A related challenge is ensuring that learned embeddings disentangle instrumental systematics (e.g., tracking issues, bright sky backgrounds) from true astrophysical signal. Architectures that explicitly factorize instrumental and astrophysical contributions to observed data offer a promising avenue \citep[e.g.,][]{2025Audenaert_CausalFMs}, and further development of such approaches will be essential for meeting DESC's calibration requirements. 

\subsubsection{Training Objectives}
Training objectives determine whether \acrshortpl{fm} learn astrophysically meaningful structure or merely reproduce observational correlations. For DESC, these objectives must explicitly promote representations that encode physical invariants (e.g., morphology–redshift relations, color–temperature gradients) while remaining robust to the observational systematics and covariate shifts defined as calibratable in the Science Requirements Document. \Acrfull{ssl} offers the most practical and scalable route toward this goal, as it enables the extraction of representations from vast unlabeled datasets without compromising the requirement for bias quantification and calibration in DESC.

\paragraph{Self-supervised learning}
\acrshort{ssl} encompasses a range of approaches: reconstructive methods, such as autoencoders \citep{Autoencoders1993, Autoencoders2006}, learn to compress data into a low-dimensional representation and then reconstruct the original input; contrastive learning \citep{Contrastive2020, NonContrastive2021} trains models to produce invariant representations for augmented versions of the same data point (e.g., zoomed or rotated); and predictive methods, often implemented via transformer architectures \citep{Transformers2017}, learn by predicting masked or omitted sections of data based on their surrounding context. The principal advantage of \acrshort{ssl} is its scalability, allowing models to be trained on vast datasets without costly manual annotation. 

\paragraph{Generative approaches}
Beyond these paradigms, generative and diffusion-based models are also being adapted for self-supervised representation learning in the astronomical domain. Although originally designed for data synthesis, diffusion objectives \citep{2022Yang_Diffusion} can act as powerful denoisers and uncertainty estimators,  aligning with the DESC requirement that calibratable systematic errors remain subdominant to statistical uncertainties. How to manage a good generative model in the context of galaxy image synthesis has been investigated for instance using this denoising capability of score-based diffusion models \citep{Campagne_2025}. Hybrid diffusion autoencoders \citep{2021Preechakul_DiffAEs} combine reconstructive and generative losses, yielding latent spaces that capture astrophysical variation while marginalizing over observational noise. Methods such as these will be critical for DESC to achieve unbiased shear and photometric-redshift inference with Rubin data.

\paragraph{Unsupervised learning considerations}
As highlighted in \autoref{sec:discovery}, the vast LSST dataset will hold immense potential for scientific discoveries. Given this sheer scale, \acrshortpl{fm} will be critical for creating powerful representations that enable anomaly detection, unsupervised source classification, and similarity searches. However, a significant challenge remains: \acrshortpl{fm} are typically designed and optimized for supervised tasks, while their use for unsupervised applications is often an afterthought. Recent work by \citet{ZoobotApplications2022} and \citet{Protege2025} demonstrates this gap. They found that traditional anomaly detection methods fail when applied to the deep latent features learned by both supervised and self-supervised methods. This indicates a clear need for new research: new unsupervised methods compatible with these features must be developed \citep[such as \href{https://github.com/MichelleLochner/astronomaly}{\texttt{Astronomaly} \texttt{Protege}};][]{Protege2025} and \acrshortpl{fm} must be optimized specifically for unsupervised discovery.

\subsubsection{Architectural Innovations}
Realizing the capabilities of \acrshortpl{fm} for DESC science requires architectural and training advancements. Astronomical data presents unique barriers to large-scale training, including wide dynamic ranges from faint to bright objects, Poisson noise properties, multi-wavelength observations, and irregular sampling. These characteristics, combined with the science priorities of DESC, necessitate architectures that not only achieve optimal representational power on LSST data but also permit robust uncertainty propagation.

\paragraph{Attention}
Efficient attention mechanisms, which allow models to weigh the importance of different parts of the input data, have evolved significantly beyond standard transformers, offering new architectures that can scale to LSST-level data volumes. Methods like Flash Attention \citep{FlashAttention2022} use techniques such as tiling and kernel fusion to reduce the memory footprint by 10-–100$\times$. This significant reduction for the first time makes practical application of attention mechanisms at the scale of individual \acrfullpl{ccd} or rafts. 

\paragraph{Hierarchical Approaches}
Astronomical data is inherently hierarchical, with structures ranging from individual galaxies to massive clusters. Sparse attention models like Longformer \citep{Longformer2020} and BigBird \citep{BigBird2020} are well-suited to this, as their architecture directly mirrors this physical structure. They use local attention (e.g., sliding windows) to model interactions between nearby objects, while global tokens aggregate information about the entire system. These designs are critical for DESC, as they enable the joint modeling of small- and large-scale correlations essential for controlling systematics in shear, clustering, and supernova analyses. This hierarchical approach is well-developed in vision models, with hierarchical vision transformers \citep[e.g., Swin Transformer V2,][]{Swin2022} being especially promising for DESC imaging. Unlike the token patterns in sparse models, these architectures use multi-resolution attention windows that mirror the multi-scale nature of cosmological information, enabling pixel-level \acrshort{psf} modeling up to large-scale galaxy clustering. This design could enable the unification of weak lensing and large-scale structure analyses using a shared image encoder. Such a model would facilitate end-to-end uncertainty propagation across spatial scales, directly addressing the DESC requirement for cross-probe consistency in systematic control.

\paragraph{Mixture-of-Experts}
The principle of a shared, unifying backbone extends to \acrfull{moe} architectures, which offer natural alignment with DESC objectives. Rather than training independent networks for each object class or redshift regime, sparse MoE layers -- such as the Switch Transformer \citep{Switch2022} and Mixtral \citep{Mixtral2024} -- can dynamically route inputs through specialized sub-networks while preserving a common latent backbone. This paradigm mirrors the DESC software model itself: probe-specific inference modules built atop a common analysis infrastructure. E.g., expert sub-networks could specialize in quiescent versus star-forming galaxies or early- versus late-time transients, while shared latent features will ensure cross-consistency in calibration and selection functions across working groups.

\paragraph{Data Fusion}
A final architectural requirement is multimodal data fusion, driven by the operational need to combine LSST data with a continuous stream of ancillary datasets (e.g., \textit{Roman}, \acrshort{desi}, \acrshort{4most}/\acrshort{tides}) as well as managing evolving observing conditions. This challenge can be addressed with early-fusion models, which tokenize heterogeneous data types into a common representation space for joint training \citep[e.g., 4M, Chameleon;][]{4M2024,Chameleon2024}; AION-1 \citep{parker2025aion} exemplifies this approach in astronomy, using modality-specific tokenizers to convert images, spectra, and scalar measurements into discrete tokens before unified transformer-based masked modeling. Alternatively, late-fusion approaches merge specialized encoders post hoc \citep{2023Pereira_LateFusion}. More generalized architectures, like the Perceiver family \citep{2021Jaegle_Perceiver, 2021Jaegle_PerceiverIO}, are particularly advantageous. By learning a single, compressed latent array, they are explicitly designed to flexibly accommodate new data modalities. This provides a clear operational path for DESC to update and expand its shared analysis space continuously. For a comprehensive review of fusion strategies and their trade-offs in astronomical applications, see \citet{SHAO2026104103}.

Together, these architectural innovations suggest a coherent design philosophy for DESC \acrshortpl{fm}: hierarchical representations that preserve cosmological structure, modular routing across scientific workflows, and learned compression layers capable of cross-probe alignment. Each of these desiderata drive toward reproducible, uncertainty-aware analyses within a unified DESC software framework.

\subsubsection{Evaluation}
In contrast to traditional single-purpose \acrshort{ml} models trained end-to-end for a specific science objective, \acrshortpl{fm} derive their value from serving a broad range of downstream use cases. This generality introduces new evaluation challenges: it is critical to assess performance across the full spectrum of tasks to which the \acrshort{fm} will be applied, ensuring that it can be successfully adapted to each. In addition, because \acrshortpl{fm} are trained on large and difficult-to-characterize datasets, they include implicit biases that must be corrected for downstream scientific applications. It is therefore important to verify the correctness of adaptation and calibration procedures for each task. Establishing a common framework of benchmarks to evaluate this adaptability must be a research priority, particularly given that such validation is not standard practice for many industry-developed \acrshortpl{fm}.

\paragraph{Benchmarks}
The development or deployment of \acrshortpl{fm} within the DESC ecosystem should be accompanied by a comprehensive suite of benchmarks designed to evaluate predictive performance across a representative range of science tasks. Particular attention should be paid to robustness under survey systematics, sensitivity to biases inherited from the \acrshort{fm}'s training data, and the ability of adaptation procedures to yield well-calibrated uncertainties for each downstream application.

\paragraph{Interpretability}
Mechanistic interpretability will provide a complementary route to validation. Techniques such as attention visualization, activation clustering, or sparse dictionary learning can be adapted to determine whether internal model representations recover known astrophysical relations including the color–magnitude diagram, the fundamental plane, or the Tully–Fisher relation. Developing astronomy-specific interpretability tools would enable DESC to quantify whether \acrshortpl{fm} encode physically meaningful correlations or merely reproduce empirical correlations.

\paragraph{Distribution shift}
Evaluation under distribution shift is also essential for robustness. DESC models must maintain reliability under temporal drift across survey years, spatial variation in observing conditions, and transfer to external datasets such as \textit{Euclid} or \textit{Roman}. Dedicated stress tests should replace the assumption of \acrfull{iid} validation, using importance-weighted calibration errors and worst-group accuracy measures to reveal biases that emerge only under covariate change. These tests will be crucial for ensuring that DESC \acrshortpl{fm} remain stable as LSST transitions from early to full survey operations.

\paragraph{Long-term impact}
Finally, the deployment of large-scale models must consider sustainability and community governance. Training and fine-tuning \acrshortpl{fm} require substantial computational resources, underscoring the need for shared development, standardized documentation, and reproducible training pipelines. Strategic coordination within DESC and with external collaborations will ensure that these models serve as transparent, responsible, and scientifically verifiable assets for the next generation of cosmological analysis.

\subsection{Large Language Models \& Agentic AI}
\label{sec:llm_agentic}

\Acrfullpl{llm} have demonstrated an astonishing capacity for performing cognitive tasks and knowledge work (e.g. synthesizing information, generating text, writing and explaining code) that has triggered an \acrshort{ai} revolution over the past few years. When equipped with tools, \acrshort{llm}-powered agentic systems go further: they can develop and execute code, perform exploratory data analysis, or orchestrate multi-step workflows. As of late 2025, complex tasks, especially in software engineering, can increasingly be delegated to such systems and completed with high accuracy. Although integration into scientific workflows remains nascent and faces genuine obstacles, the pace of progress and the potentially transformative impact of the ability to delegate significant fractions of the research workflow to \acrshort{ai} suggests \acrshort{desc} should plan for how to effectively integrate these tools and guide the evolution of best practices, in line with the guiding principles in \autoref{sec:exec_summary}. The goal of integrating these tools should not be full automation or dehumanization, but empowerment: elevating the level at which researchers can engage within the scientific process, enabling them to tackle more ambitious projects while concentrating on fundamental questions, critical interpretation, and creative reasoning. \\
Below, we review the existing capabilities and limitations in the adoption of \acrshortpl{llm} for science, and examine potential applications for \acrshort{desc}, ranging from mature (e.g. documentation assistance) to aspirational (e.g. AI co-scientists). Overall, this subsection aims to answer the following question: what would successful integration of \acrshortpl{llm} and agentic systems within \acrshort{desc} look like?

\subsubsection{From LLMs to Agentic AI}

\paragraph{LLMs as knowledge tools}
At baseline, \acrshortpl{llm} are sophisticated systems for natural language processing. They process, synthesize, and generate text, though they remain subject to hallucinations (i.e.\ false statements expressed with confidence), out-of-date knowledge, and limited problem-solving capabilities. Some of these limitations can be mitigated to engineer reliable production-ready systems using \acrfull{rag}, a technique which allows LLMs to ground their responses using context-specific documentation \citep{lewis2020rag, fan2024survey}. OpenScholar \citep{asai2024openscholar}, for example, uses \acrshort{rag} over 45 million open-access papers to synthesize citation-backed responses, substantially outperforming general-purpose models while avoiding hallucinated citations. Over the last year, the proficiency of \acrshortpl{llm} at problem solving has also significantly improved with ``reasoning models'' (e.g.\ OpenAI's o1/o3, DeepSeek-R1 \citep{deepseekR1}) which allocate additional compute at inference time to perform deliberate, multi-step problem-solving (a paradigm sometimes called test-time compute scaling). These models show marked improvements on tasks requiring extended chains of reasoning. On physics tasks specifically, TPBench \citep{chung2025tpbench} (a benchmark of original theoretical physics problems ranging in difficulty from undergraduate to research-level) finds that newer reasoning models substantially outperform earlier systems, though research-level problems remain largely unsolved. This trajectory suggests that \acrshort{llm} capabilities in scientific reasoning will continue to improve, even as fundamental limitations persist. It is important to note that \acrshortpl{llm} alone can inform and assist, but they cannot act: they process information rather than executing tasks.

\paragraph{Agentic AI: from knowledge to action}
Agentic systems represent a conceptual leap beyond individual \acrshortpl{llm}. Where \acrshortpl{llm} generate text, agents take actions: writing and executing code, calling \acrshortpl{api}, modifying files, and interacting with other computational infrastructure. Where reasoning models are used, these agents can also dynamically revise their actions based on feedback and provide interpretable explanations for their decisions. These capabilities make it possible to delegate complex and loosely-defined tasks to agents much as one might delegate them to a colleague. However, not all tasks are equivalent, and a useful mental model is to ask: what could one hand off to a person of varying competency? At the ``intern'' level, current agentic systems can reliably fix small bugs, run analyses with different parameters, and write simple tests to verify their outputs or those of another code. At the ``junior'' level, they can implement well-specified features or debug failing pipelines. ``Senior''-level capabilities (building entirely new software systems, weighing design trade-offs in solving a specific problem, and validating scientific methodology) remain largely out of reach at present, though the complexity and execution horizon of these systems continues to increase. Tools such as Claude Code\footnote{\url{https://code.claude.com/}}, Cursor\footnote{\url{https://cursor.com}}, and Devin\footnote{\url{https://devin.ai/}} demonstrate agents completing real software engineering tasks: calling \acrshortpl{api}, executing shell commands, maintaining context across dozens of steps, recovering from errors, and iterating toward solutions. On SWE-bench \citep{jimenez2024swebench}, a benchmark of real GitHub issues from open-source Python repositories, leading agents now resolve over 70\% of issues in the human-verified subset; on the harder SWE-bench Pro \citep{aleithan2025swebenchpro}, which targets enterprise-level problems, the best systems solve roughly 25\%. 

\paragraph{Towards Scientific Agentic Systems} Beyond software engineering, agentic systems are increasingly being applied to scientific research. A first generation of agents has focused on literature search, going beyond static \acrshort{rag} systems by using tools and iterative refinement. PaperQA2 \citep{skarlinski2024paperqa2}, developed by FutureHouse, exemplifies this approach: rather than simply retrieving from a fixed index, it uses search tools to find relevant papers, traverses citation graphs to discover related work, and iteratively refines its own queries based on retrieved text. This agentic approach matches or exceeds PhD-level researchers on literature retrieval benchmarks. More ambitious systems extend beyond literature search to integrate data analysis and hypothesis generation. Google DeepMind's AI co-scientist \citep{gottweis2025coscientist}, a multi-agent system built on Gemini, uses a ``generate, debate, and evolve'' approach to produce novel hypotheses; it has identified drug-repurposing candidates for liver fibrosis that were later validated in laboratory experiments. FutureHouse's Robin \citep{robin2025}, an open-source multi-agent system, autonomously proposed ripasudil as a treatment for dry age-related macular degeneration and validated the hypothesis through wet-lab experiments. Edison Scientific's Kosmos \citep{kosmos2025} orchestrates parallel data-analysis and literature-search agents over 12-hour runs, producing detailed scientific reports that independent evaluators found 79\% accurate. Sakana AI's AI Scientist v2 \citep{yamada2025aiscientistv2} uses agentic tree search to autonomously generate hypotheses and run experiments, and demonstrated its ability to run end-to-end machine learning research projects. This list is by no means exhaustive, and these systems remain by and large proofs-of-concept, with little independent data on their utility and reliability for real-world complex scientific research. Nonetheless, these examples illustrate both the interest and rapid advancement in agentic systems for scientific research.

\paragraph{LLM and agentic AI in astronomy}
Domain-tuned \acrshortpl{llm} such as AstroSage-Llama-3.1-8B \citep{dehaan2025astrosage} demonstrate that compact, astronomy-specific models can match larger general-purpose systems at lower cost. In parallel, \acrshortpl{llm} are enabling large-scale knowledge synthesis: Pathfinder \citep{iyer2024pathfinder} applies semantic retrieval and citation-aware summarization across ${\sim}350{,}000$ astrophysical papers in the \acrfull{ads}, allowing users to move from keyword-centric searches to concept-level exploration. ChatGaia\footnote{\url{https://chatgpt.com/g/g-aYZOjK5zy-chatgaia}} demonstrates the potential for natural-language interfaces to complex astronomical databases: it translates user queries into \acrshort{adql} for the Gaia Archive, making ${\sim}2$ billion stellar sources accessible without requiring query-language expertise. Several agentic proofs-of-concept have also emerged. CMBAgent \citep{laverick2024mas, xu2025cmbagent} coordinates \acrshort{rag} with local code execution to run \acrshort{mcmc} pipelines for cosmological parameter estimation from \acrfull{act} data. Mephisto \citep{mephisto} iteratively refines stellar population parameters by orchestrating multi-band galaxy observations with the \acrshort{cigale} \acrshort{sed}-fitting code \citep{cigale}. Denario \citep{villaescusanavarro2025denarioprojectdeepknowledge} extends this paradigm further, spanning a full research lifecycle from idea generation through data selection, modeling, and manuscript drafting. These systems remain demonstrations rather than production tools, with scientific validation of each system's outputs still performed manually.

As a concrete example of the state of the art, and one directly relevant to \acrshort{desc}, the 2025 NeurIPS Weak Lensing Uncertainty Challenge\footnote{\url{https://www.codabench.org/competitions/8934/}} (a competition to infer cosmological parameters from simulated convergence maps while quantifying uncertainty) was won by a team using the CMBAgent system \citep{CMBAGENT}, beating a team of domain experts in weak lensing and ML who placed second without AI assistance. The CMBAgent-assisted team, though not specialists in weak lensing inference, overtook the experts within weeks and held the lead through the end of the competition. This result illustrates how \acrshort{llm}-based agentic systems can accelerate scientific research and lower the barrier to complex analysis techniques.

\paragraph{Fundamental limitations}
While early successes have shown the promise of \acrshortpl{llm} and agentic systems for science, the current generation of these systems face fundamental constraints that limit their applicability to science. As emphasized by \citet{ilievski2025aligning}, human scientific reasoning relies on abstraction, causal inference, and conceptual transfer, whereas \acrshort{ai} systems depend on statistical interpolation. This leads to what has been termed ``jagged intelligence'': systems that solve Olympiad-level problems while failing at kindergarten logic, with capabilities forming an uneven landscape rather than a coherent skill set. Hallucination (generating confident but incorrect outputs) also remains a fundamental problem for scientific applications of \acrshortpl{llm}. \citet{kalai2025hallucinate} provide a theoretical explanation: current training objectives and evaluation benchmarks structurally incentivize guessing over expressing uncertainty. When a model says ``I don't know,'' benchmarks penalize the response in the same way as the model producing an incorrect output, so models learn to guess confidently even when uncertain. Hallucinations, as a result, are a direct consequence of how these systems are trained and evaluated. Hallucinations span a broad taxonomy, from factual errors about well-documented knowledge to fabricated citations to arbitrary confabulations \citep{ji2023hallucination}, and they can be subtle enough to evade casual review. For science, the danger is acute: a hallucinated statistical result or misremembered prior work could propagate through analysis undetected if careful validation frameworks are not imposed on model output. Beyond hallucination, these systems face limitations particularly acute for scientific discovery. They are trained on existing literature and may fail precisely where discovery happens (at the frontier where established statistical patterns break down). They also lack reliable uncertainty quantification: unlike a careful scientist who flags when they have extended beyond their expertise, \acrshortpl{llm} produce outputs with uniform confidence regardless of whether they are interpolating within training data or extrapolating beyond it. Finally, complete reproducibility of \acrshort{llm} outputs remains challenging: outputs are stochastic, sensitive to prompt phrasing, and subject to version drift as underlying models change (challenges that conflict with science's demand for verifiable, repeatable results). For all these reasons, human oversight remains essential for science-critical output.

\subsubsection{Potential applications for DESC}

The following sections organize potential application areas of \acrshortpl{llm} by the outcomes they serve for \acrshort{desc}, rather than by the underlying technology. Each area sits at a different level of technological maturity, and the path toward realizing each application differs accordingly.

\paragraph{Knowledge work and research support}
This is the most mature application area, and the one where \acrshort{desc} members can immediately benefit. Success in this area would mean that new members can query \acrshort{desc} documentation in natural language and receive accurate, relevant answers rather than navigating distributed technical documentation and obscure Slack channels. It would mean that literature search moves from keywords to concepts, helping researchers find relevant work faster and discover connections across subfields. Writing assistance for drafts, summaries, and documentation would be readily available, though always with human review. As summarization capabilities improve, collaboration maintenance (meeting summaries, cross-working-group communication) could also become increasingly efficient.\\
The methodology is relatively mature for these tasks. For literature work, tools like OpenScholar, PaperQA2, and Pathfinder (discussed above) already enable concept-level exploration and synthesis across large scientific corpora. General-purpose \acrshortpl{llm} can summarize papers, extract key results, bridge jargon gaps across subfields, and assist with writing tasks. Domain-tuned models like AstroSage (discussed above) have shown that compact, astronomy-specific \acrshortpl{llm} can match larger general-purpose systems on astrophysical question-answering at lower cost. \acrshort{rag} techniques allow these systems to ground their responses in specific documentation, reducing hallucinations and enabling citation of sources.\\
The path forward for \acrshort{desc} involves building a \acrshort{rag} system over collaboration documentation, tutorials, and pipeline code; establishing human review guidelines for any generated content; and starting with low-stakes applications (e.g., onboarding Q\&A) before moving to higher-stakes use cases. Looking further ahead, a retrieval-augmented knowledge graph that unifies documentation, code, and literature could provide a continuously evolving record of methodological provenance, making it easier to trace how analysis choices change across the collaboration and how they connect to the broader literature.\\
The main challenges are hallucination (confident wrong answers erode trust, so citation of sources is essential), the maintenance burden of keeping \acrshort{rag} indices current as documentation evolves, and establishing clear guidelines and training procedures for how to use, review, and acknowledge generated content.

\paragraph{Natural language interaction with data}
Natural-language interfaces can dramatically reduce the barrier to accessing complex astronomical databases and archives. Success in this application area would mean that \acrshort{desc} members can query \acrshort{lsst} catalogs, simulation products, and image archives dynamically using plain language rather than specialized query languages or custom hard-coded scripts. A researcher could ask ``show me all galaxies with strong tidal features in the deep drilling fields'' and retrieve relevant candidates without writing \acrshort{sql} or navigating file systems. \\
At the time of writing, this methodology is now emerging. ChatGaia\footnote{\url{https://chatgpt.com/g/g-aYZOjK5zy-chatgaia}} demonstrates that conversational interfaces to catalog data are feasible: it translates natural-language queries into \acrshort{adql} for the Gaia Archive, making ${\sim}2$ billion stellar sources accessible without query-language expertise. For imaging data, AION-Search \citep{koblischke2025aionsearch} enables semantic search across 140 million galaxy images by using vision-language models to generate captions and align image embeddings with text queries, allowing researchers to search for morphological features or rare phenomena using free-form descriptions rather than predefined categories. \\
The path forward for \acrshort{desc} involves building natural-language interfaces to key data products: LSST catalogs, difference imaging outputs, and simulation archives. Such interfaces could be grounded in schema documentation and example queries to reduce the likelihood of hallucinated or malformed outputs. Techniques for constrained generation offer another avenue for guaranteeing that model outputs adhere to a pre-defined schema, and research in this area is ongoing \citep{2025Mundler_ConstrainedDecoding}. Integration into analysis notebooks would allow seamless transitions between exploratory queries and downstream analysis. \\
The main challenges to achieving this application area are accuracy (incorrect query translations could return misleading results), coverage (not all queries map cleanly to database operations), and validation (users need techniques for verifying that the system understood their intent). For image search, an additional challenge is that rare phenomena are precisely where training data is sparse, making retrieval of scientifically-interesting objects a non-trivial task.

\paragraph{Software engineering acceleration}
This area spans two levels of maturity: interactive coding assistance (mature) and autonomous operation (emerging). Success would mean that \acrshort{desc} developers spend less time on routine implementation and more time on high-level design and scientific validation. Interactive tools would help members implement features, fix bugs, and write tests more efficiently. Autonomous agents would handle low-risk maintenance tasks (updating tutorials when APIs change and package dependencies when softwares do, generating test coverage, fixing CI failures) and open pull requests for human review. Ultimately, the implementation of new analysis pipelines could also be delegated to agents, with researchers specifying the requirements and validation tests in natural language. \\
The methodology for interactive use is already mature. Tools such as Claude Code, Cursor, and GitHub Copilot are widely deployed and demonstrate real productivity gains on well-specified tasks. Autonomous operation is less mature: agents can be triggered by \acrshort{ci} failures to diagnose issues and propose fixes, but reliability remains inconsistent for complex, long-horizon problems due to limiting problem-solving abilities and finite context windows. \\
The path forward for \acrshort{desc} involves broader adoption of interactive coding assistants for pipeline development, along with collaboration-specific contexts tailored to \acrshort{desc} workflows (e.g., \acrshort{lsst} data structures, pipeline conventions, DESC software guidelines). For autonomous agents, the prudent approach is to pilot these systems first on low-risk tasks such as test generation and docstring updates before integrating them into \acrshort{ci} workflows. \\
The main challenges are code correctness (agents can introduce subtle bugs, so review and validation remains essential), security (generated code may have vulnerabilities), and institutional knowledge (members may not understand code they did not write). Observability is also an important point: tracking which agent and version contributed to each change, and ensuring edits stay limited to the relevant code context rather than allowing regular rewrites of entire scripts, will more easily enable debugging and auditing.

\paragraph{Analysis support}
This area is less mature and requires additional validation infrastructure to be put in place before deployment should be considered. Success would mean that \acrshort{ai} assistants, integrated directly into analysis notebooks, help researchers move faster through the exploratory phase of analysis: generating visualization code from natural-language requests, suggesting statistical tests, and helping iterate on plots and diagnostics. More ambitiously, such assistants could help construct null tests and systematics checks, lowering the barrier to rigorous validation while freeing the user to explore additional aspects of the data. A junior researcher could more easily implement the suite of tests that an experienced analyst would know to run in a given scientific context, making scientific best practices more accessible and standardizable beyond individual working groups and across the full collaboration. Beyond individual analysis, agents could proactively monitor nightly data streams and flag anomalies, generate human-readable \acrshort{qa} summaries, or diagnose pipeline failures. Human-in-the-loop frameworks, where active-learning strategies solicit expert input only for the most ambiguous cases \citep{settles2009active, christiano2017feedback}, could further optimize the use of limited expert time for monitoring these systems. \\
The methodology for \acrshort{llm}-enabled analysis support is nascent. While the building blocks exist (code execution, \acrshort{rag}, multimodal understanding), integrated systems for scientific analysis support are largely experimental at present. Projects like Jupyter AI\footnote{\url{https://jupyter-ai.readthedocs.io/}} provide infrastructure for integrating \acrshortpl{llm} directly into notebook environments, enabling conversational assistance and code generation within analysis workflows, but robust evaluation frameworks for analysis tasks remain underdeveloped. \\
The path forward for \acrshort{desc} requires building evaluation benchmarks before deployment: curated test cases where correct answers are known, so that agent outputs can be validated. Piloting on low-stakes analysis tasks (e.g., generating diagnostic plots, summarizing pipeline logs) would build experience before moving to higher-stakes applications. Most importantly, developing clear guidelines for human oversight is critical to ensuring that researchers remain accountable for scientific conclusions: \acrshort{ai} can accelerate the work, but the responsibility for validating results and interpreting their meaning must stay with humans. \\
The main challenges are automation bias (over-trusting fluent outputs that may contain subtle errors), the difficulty of validating agent reasoning on novel analyses, and the risk that diagnostics look valid but miss real issues. For science-critical workflows, the bar for reliability is high, and current systems do not yet meet it. There is also an educational tension: critically evaluating \acrshort{ai}-generated analysis requires understanding the underlying methods, yet that understanding has traditionally come from wrestling with implementation details oneself. If \acrshort{ai} removes the friction of implementation, how do junior researchers develop the judgment needed to recognize when the \acrshort{ai} is wrong? Training the next generation to use these tools effectively, without becoming dependent on them, is an open challenge. 

\paragraph{Toward an AI co-scientist}
This is the most ambitious and speculative application area. Success would mean an agent capable of PhD-student-level work: running analyses, interpreting results, drafting manuscript sections, and iterating on feedback with sufficient robustness that a PI could trust the output as they would a competent junior colleague. Such an agent might surface connections across \acrshort{desc} working groups that humans miss, generate hypotheses grounded in both literature and data patterns, or systematically explore parameter spaces or analysis choices that would be tedious for humans to cover. The measure of success in this area will be in its ability to amplify existing scientific efforts: researchers would report that these tools help them think more deeply and explore more broadly, not merely execute faster. \\
The methodology remains largely aspirational. The proofs-of-concept discussed earlier (CMBAgent, Mephisto, Denario, AI co-scientist, Robin) demonstrate that end-to-end workflow orchestration is possible, but none has demonstrated PhD-student-level robustness on real scientific problems. Scientific validation of their outputs remains entirely manual. Evaluation frameworks are beginning to emerge: \citet{ye2025replicationbench} introduce ReplicationBench, an astrophysics-specific benchmark testing whether agents can faithfully reproduce published analyses; Gravity-Bench \citep{koblischke2025gravitybench} evaluates agents on physics discovery tasks, testing their ability to plan experiments and infer physical laws from simulated gravitational systems; while ScienceBoard \citep{scienceboard2025} and DiscoveryWorld \citep{discoveryworld} assess compositional scientific reasoning more broadly. \\
The path forward is necessarily incremental. \acrshort{desc} should build on the higher-maturity applications first (documentation, coding assistance, data access), developing institutional experience and trust before attempting more autonomous scientific work. \acrshort{desc}-specific evaluation benchmarks, perhaps based on reproducing known analyses or detecting injected errors, would help measure readiness. Human oversight must remain central at every stage. \\
The challenges here are the deepest. Reaching PhD-student-level robustness requires solving the fundamental limitations described earlier: agents must know when they are outside their competence, reason about physics rather than pattern-match on surface features, and produce outputs that are reproducible with full provenance. The bar for scientific discovery is extraordinarily high. A single undetected error in an analysis pipeline could propagate into published results and impede progress instead of enabling it. Finally, proprietary \acrshortpl{llm} are trained by companies to be agreeable and affirming; a model valuable for enhancing scientific discovery should be proactive in challenging the user's assumptions, critiquing its own outputs, and pursuing independent (but relevant) lines of inquiry. Whether current architectures can be made reliable enough for this level of autonomy, or whether fundamentally new approaches specific to the physical sciences are needed, remains an open question.

\subsubsection{Implementation Considerations}

Beyond the application-specific concerns discussed above, three cross-cutting considerations will shape how \acrshort{desc} integrates these tools: evaluation, governance, and infrastructure.

\paragraph{Evaluation and benchmarking}
Benchmarking is essential for improving \acrshort{ai} systems, and meaningful benchmarks can only be created by domain experts who can accurately evaluate the quality of a provided solution. \acrshort{desc} is well-positioned to contribute here. Benchmarks for \acrshort{desc} applications should measure: (1) factual accuracy on domain literature and documentation; (2) code validity and runtime safety; (3) reproducibility under random seed variation; and (4) robustness to distribution shift and systematic differences (e.g., early vs.\ late \acrshort{lsst} years, cross-survey transfers). Existing efforts like ReplicationBench, Gravity-Bench, and ScienceBoard (discussed above) provide templates for these efforts, but \acrshort{desc}-specific benchmarks (perhaps based on reproducing published analyses, detecting injected systematics, or validating pipeline outputs) would directly measure readiness for collaboration-wide use. Establishing such benchmarks is itself a scientific contribution: it encodes expert knowledge about what problems are worth solving and provides a foundation for systematic improvement of \acrshort{ai}.

\paragraph{Governance and reproducibility}
Integrating \acrshortpl{llm} into scientific workflows raises unresolved questions about reproducibility, provenance, and attribution. Each automated workflow should maintain full provenance metadata: prompts, model versions, retrieved sources, generated code, and execution logs. Observability (logging all agent actions) will enable debugging and audit \citep{zhou2025autonomous}, and increase credibility in designed agentic systems. Sandboxed execution environments enforce deterministic behavior and ensure that agent-generated code runs in secure, auditable contexts. Reproducibility, however, remains difficult. With the pace of technical improvement rapidly increasing, ensuring that an analysis performed today can be replicated exactly next year is an open problem. Attribution also lacks established norms: how should \acrshort{ai} tools be credited in publications? Should they be listed as co-authors or as software tools, and how should their contributions be presented? These questions require collaboration-wide discussion and policy development, but are extremely topical as papers presenting physics research fully conducted by \acrshortpl{llm} begin to appear \citep[e.g.,][]{2026Schwartz_Claude}.

\paragraph{Infrastructure and sustainability}
The physical deployment of \acrshortpl{llm} within \acrshort{desc} must respect both scale and sustainability. Large models are expensive to run, and the cost-capability trade-off matters for a collaboration operating over \acrshort{lsst}'s decade-long baseline. Compact open-weight models, fine-tuned on domain-specific data, offer a path to balancing capability with efficiency while avoiding vendor lock-in---AstroSage exemplifies this approach. A dedicated \acrshort{rag} layer connecting \acrshort{desc} documentation, simulation archives, and survey data products could form the backbone of an agentic knowledge infrastructure. But even with efficient models, infrastructure requires ongoing maintenance: \acrshort{rag} indices must be updated as documentation evolves, prompts need revision as models change, and \acrshort{llm} \acrshortpl{api} themselves drift over time. Planning for this maintenance burden from the outset is essential.

\bigskip
\noindent \acrshortpl{llm} and agentic systems will not replace scientific judgment, but they may extend researchers' reach (handling routine tasks, surfacing relevant literature, accelerating code development) so that human effort can concentrate where it matters most. Finding the right interaction patterns between scientists and these systems remains an open challenge. \acrshort{desc} should engage now, building evaluation infrastructure and piloting lower-risk applications, to be ready as capabilities mature.

%% file: sections/sec6_infrastructure.tex
\newpage
\section{Infrastructure Requirements to Support AI/ML in DESC}
\label{sec6:infra_requirements}

Infrastructure is the shared technology needed to realize the methods, models, and scientific opportunities outlined in the preceding sections. It may be deployed in a distributed mode with individuals or small teams using local or institutional resources, or in a coordinated mode within a shared platform or environment. This latter mode is most relevant for the largest-scale \acrshort{ai}/\acrshort{ml}-enabled model training and inference workflows at the scale of the entire Rubin-\acrshort{lsst} data set, including via core infrastructure through \acrshort{doe}-funded high performance computing facilities in the US. The following subsections review the infrastructure elements most relevant to the future of AI/ML in \acrshort{desc}.

\subsection{Software}

Software is foundational to modern cosmology, especially for \acrshort{desc}, where scientific insight increasingly depends on sophisticated computational workflows. As \acrshort{ai} and \acrshort{ml} mature into core scientific technologies, software itself becomes a form of research infrastructure. In this context, two tightly connected priorities emerge: first, the development and long-term stewardship of a robust AI software stack that enables model development and experimentation; second, the strategic integration of AI methods into the DESC scientific pipelines, where they will ultimately shape how we reduce data, extract measurements, and perform inference.

\subsubsection{The AI Software Stack} 

DESC has long demonstrated leadership in scientific software development: from collaboration-wide development guidelines\footnote{\url{https://lsstdesc.org/assets/pdf/docs/DESC_Coding_Guidelines_latest.pdf}}
 and software-oriented publication policies, to a culture of reproducibility and sustained support for collaboration-wide software stacks. As AI becomes a first-class component of scientific analysis, extending that same discipline and strategic thinking to the AI software ecosystem is increasingly important. The goal is to define a modern, durable, and interoperable stack built on best practices for reproducibility and maintainability. This accelerates research while preserving the transparency that has always characterized DESC software.
 
To meet this goal, we recommend converging on a small set of shared practices and services that make ML development reproducible, portable to the \acrfull{cc-in2p3} and \acrfull{nersc}, and sustainable over the 10-year duration of the LSST survey. Key components include:

\begin{itemize}
\item \textbf{Frameworks for model development}, likely PyTorch for large models and JAX for differentiable physics.
\item \textbf{Experiment tracking}, capturing code/data versions, configurations, metrics, and compute environments to ensure reproducibility.
\item \textbf{Model and artifact registries} to version and archive trained models, datasets, and reports, mirroring survey data-release practices.
\item \textbf{Standardized export formats} such as \acrfull{onnx} so models integrate cleanly with Rubin/DESC pipelines and \acrfull{hpc} environments.
\item \textbf{\Acrfull{ci} and \acrfull{cd} for models} to test and validate training configurations, exported artifacts, and deployment environments.
\item \textbf{Observability} and drift monitoring so \acrshort{ml} components used in production behave predictably and transparently.
\end{itemize}

These elements are not overhead; they are the operational foundations that allow AI to become reliable scientific machinery rather than episodic experimentation. They also reduce long-term maintenance burden by enforcing shared conventions, minimizing bespoke tooling, and allowing learned models to be audited, reused, and trusted throughout the \acrshort{lsst} decade.

In addition to these considerations for ML model development, DESC will increasingly rely on \acrshortpl{llm} as flexible interfaces to data, documentation, and tooling. LLMs are unusual compared to conventional models in that they are supplied by a rapidly changing ecosystem of commercial providers, open models, and self-hosted deployments. Versions change frequently, and some use cases may require on-premises or institutionally-hosted models---e.g., at NERSC, CC-IN2P3, the \acrfull{csd3}, or similar facilities---via serving stacks such as vLLM for data-governance or cost reasons. To remain agile in this landscape, DESC should treat LLMs as interchangeable backends behind a stable internal \acrshort{api}. Such an abstraction layer enables swapping models without rewriting pipelines and moving workloads between commercial services and collaboration-owned \acrshort{gpu} resources as needs evolve. Beyond the LLMs themselves, agentic frameworks (libraries that orchestrate multi-step LLM workflows, tool calls, and human-in-the-loop interactions) are even more volatile, with new options appearing and disappearing on timescales of months. Frameworks such as LangGraph exemplify the current direction, representing agents as graphs of tools, memory, and control logic, and providing execution and tracing engines around them. 

Framework sustainability and openness should play an important role in guiding tooling choices. Solutions with strong governance and broad adoption (e.g., PyTorch under the Linux Foundation, ONNX, MLflow or self-hostable experiment-tracking systems) provide long-term stability and avoid dependence on proprietary silos. Finally, coordination with computing partners is essential to ensure portability of the software stack and deployment on more HPC-aligned facilities.

\subsubsection{Integration of AI/ML within Analysis Pipelines}

As AI components mature, DESC pipelines may evolve from purely sequential “raw-to-reduced” workflows to systems where learned models are first-class, production-grade elements. The emphasis is on embedding AI in ways that enhance measurement fidelity, calibration control, and inference efficiency, while preserving transparency, reproducibility, and smooth integration with existing practices.

\paragraph{Data reduction pipelines} AI/ML is already present in DESC workflows (e.g., \acrshort{photoz} estimation within \acrshort{rail}). Over time, more components are likely to incorporate learned models at defined stages such as deblending, \acrshort{psf} estimation, artifact rejection, and photometric calibration, tightly coupled to Rubin/DESC data structures and HPC execution. In parallel, DESC has interest in end-to-end approaches that optimize models directly against observational data and simulators, potentially replacing brittle stage boundaries while maintaining provenance through experiment tracking and model registries.

\paragraph{Foundation Models as Services} With the advent of the foundation model paradigm, we envision the possibility of providing “X-as-a-service” (photometry, astrometry, cross-matching, classification, anomaly detection) behind stable \acrshortpl{api}, so models can evolve without churn in downstream code.

\paragraph{AI/ML-based cosmological inference} AI-based methods are shifting inference from sampler-centric workflows toward models that learn mappings from data to posteriors or summaries. Foundation-style surrogates can amortize computation and reduce \acrshort{mcmc}-heavy workloads, while simulation-based inference trains directly on forward models and field-level methods operate on minimally reduced data. Active-learning loops may trigger simulations on the fly to refine surrogates and concentrate the HPC budget where it most reduces uncertainty. For pipeline use, these elements remain tied to provenance systems and model registries.

\paragraph{Shared infrastructure for emulators and surrogate models} Consistency across emulator efforts can lower reuse costs. A lightweight scaffold could include common interfaces (inputs, units, cosmology/observational context, stochastic controls, uncertainty outputs), a minimal dataset schema for training/evaluation, embedded metadata for provenance and validity domains, and containerized artifacts with dual exports (native framework and ONNX where feasible). A small validation suite (accuracy, coverage, calibration under shift, throughput) running in CI on facility images would help with portability to DESC computing facilities, and hooks for active learning can keep surrogates co-evolving with forward models.

\paragraph{LLMs and agents} Large language models and specialized agents can contribute at multiple levels of DESC work: in notebook environments as integrated assistants that accelerate iteration on analysis code, diagnostics, and documentation/query synthesis; in data discovery and curation by searching heterogeneous archives and DESC repositories, proposing cross-matches, and flagging anomalies; and at facility scale by assembling templated workflows, submitting authenticated jobs, and recording outcomes into experiment tracking and model registries. Adoption pairs capability with governance: standardized interfaces for serving models, authenticated access, audit logs of prompts and actions, and human-in-the-loop checkpoints for any decision with scientific impact.

Likely, investment in general-purpose AI tooling across industry, open-source, and public efforts will continue to exceed resources available for cosmology-specific software. When such tools meet DESC’s scientific and operational needs, adopting them can leverage broader community advances while allowing collaboration effort to focus on domain-specific modeling and validation. Finally, AI will also influence our approach to engaging with computing in the future. Computational infrastructure will increasingly be harnessed with the assistance of \acrshortpl{llm} and agentic frameworks. Depending on how this transition proceeds, it may enable a larger community of researchers within DESC to engage directly with large-scale scientific computing.

\subsection{Computing}

\subsubsection{Workflows and Scales}

Estimating the full scale of resources needed to support the range of \acrshort{desc} \acrshort{ai}/\acrshort{ml} use cases is beyond the scope of this white paper, and any estimates will evolve with time as new methodologies and science applications are developed. Major resource categories include \acrshort{gpu} and \acrshort{cpu} time, short- and long-term storage, and network bandwidth both between and within analysis facilities. AI-oriented workloads will range from small-scale \acrfull{rd} to training and serving \acrshortpl{fm}, serving tokens for agents/\acrshortpl{llm}, running on-the-fly simulations for active learning loops for \acrshort{acr:sbi}, up to potentially running simulations on the fly as part of explicit inference loops. All of these will bring their own requirements for computing cycles, data locality and throughput, interactivity, and orchestration.

At the low end of requirements, resource-augmented instances of the commercial cloud-based \acrfull{rsp} would provide an accessible route for individuals and small teams to scale up AI/ML workflows that require integration with the Rubin data and software stack. For cost efficiency, these resources would likely need to be managed through either a batch-processing system or through an elastic data-science workflow system. Allocating GPU-based interactive servers (virtual or physical) in the cloud-based RSP context is unlikely to be viable at scale for many users, given the typical idle time in interactive sessions and workflows. Larger AI workflows could also be accommodated through individual or DESC-wide allocations of CPU, GPU, and working storage at \acrshort{nersc} or through proposal-driven allocations under the 10\% of computing time reserved for Rubin science users at the Rubin \acrfull{usdf} at \acrfull{slac}.

On the high end, significant computing resources  may be needed for new simulations to train simulation-based inference approaches to large-scale cosmology analyses. This includes not only the computing needed for simulation but also the storage to save and share large simulation outputs. Another major driver of resource needs would be training of data-oriented FMs at the scale of the full Rubin data set. Again, an accurate estimate of the resources needed for this would require further study of an appropriate reference architecture and training strategy. A rough guide to the scale can be found from the (CPU-based) compute sizing model for Rubin--\acrshort{lsst} Data Release Production\footnote{\url{https://dmtn-135.lsst.io}} since it is operating on roughly the same scale of data as a full-scale LSST-based FM would train on. The operations compute model estimates a need for about 50M core-hours in year 1 of the survey, rising linearly year over year (assuming annual data releases) to roughly 10x this amount in year 10. Storage needs are estimated to increase from 30PB in year 1 to over 800PB in year 10 although not all of the associated data products would necessarily be needed for FM training.

In addition to ``offline'' computing needs, a number of time-critical AI/ML-driven DESC workflows will be driven by the nightly alert stream, including ML-driven algorithms for classification, anomaly detection, and intelligent follow-up observations. Many of these will be implemented within the broker and marshal systems that receive the LSST alert stream; depending on their level of resource-intensiveness, broker/marshal computing capacity may need to be further augmented, and/or integrated with elastic or preemptive compute allocations within research computing facilities or in commercial cloud.

\subsubsection{Computing Resource Providers}
In the US, DESC computing needs are primarily supported by \acrshort{doe} through its network of computing facilities under the \acrfull{ascr} program within the DOE Office of Science. This includes NERSC as the primary user facility for DESC science analysis, \acrfull{esnet} for advanced wide-area networking and data movement, and the \acrshort{alcf} and \acrshort{olcf} for larger-scale computational work. ASCR is also developing the \acrshort{hpdf}, led by Jefferson Lab in partnership with Lawrence Berkeley National Laboratory. DOE is also advancing the development of an \acrfull{iri} to support flexible, powerful, and accessible implementation of scientific workflows across all these facilities.

Anticipating an increasingly prominent role for AI in the scientific exploitation of data created by Rubin and other DOE-funded facilities across all disciplines, DOE launched the Genesis Mission in November 2025\footnote{\url{https://www.whitehouse.gov/presidential-actions/2025/11/launching-the-genesis-mission/}}, a national effort to accelerate AI-driven scientific discovery across its 17 national laboratories. As infrastructure for this mission, the US Congress has funded the creation of \acrshort{amsc}\footnote{\url{https://science.osti.gov/-/media/grants/pdf/lab-announcements/2025/LAB-25-3555.pdf}} to develop and deploy the next generation of AI-oriented capabilities on the foundation of the DOE computing platforms noted above. AmSC development is underway now, with funding distributed across an AmSC infrastructure component, a core AI model-development consortium (ModCon), pilot funding for discipline-specific data curation and science benchmarking activities, and seed funding for discipline-specific AI model teams. As a cornerstone of the Genesis Mission, computing and AI model development infrastructure within the AmSC represent a significant opportunity to address ambitious DESC AI/ML goals.

The joint \acrshort{nsf}--DOE nature of Rubin Observatory opens the possibility of leveraging significant NSF-supported computing facility resources, especially if pursued in coordination with other Science Collaborations working in areas typically supported by NSF. These resources include the \acrfull{lccf} entering production at the Texas Advanced Computing Center. DESC members also participate in both astrophysics-oriented National Artificial Intelligence Research Institutes, funded jointly by NSF and the Simons Foundation (\acrshort{skai}, led by Northwestern University; and CosmicAI, led by the University of Texas at Austin), and have access to their associated computing resource allocations. Considering interests in Rubin--\textit{Roman}--\textit{Euclid} joint analysis, \acrfull{nasa}-funded computing resources may also be a viable option.

More broadly, many DESC members have access to significant campus-level computing at their institutions. Members outside the US have access to their own networks of national resources, including national and regional initiatives prioritizing computing for AI in science. Through the in-kind contribution program that supports LSST data rights for scientists outside of the US and Chile, a network of \acrshortpl{idac} is being deployed, with some sites bringing additional \acrshort{cpu} and \acrshort{gpu} capabilities that DESC would be well positioned to make use of. The UK will host an IDAC, sized to satisfy the resource needs of 20\% of the global LSST community during survey operations. The UK IDAC will be connected to UK national research computing facilities, both traditional (simulation and modeling) supercomputing services and the coming generation of AI-focused Digital Research infrastructure currently being specified and prototyped through the \acrshort{ascend} process\footnote{\acrlong{ascend}; see \url{https://engagementhub.ukri.org/ukri-infrastructure/ascend-process/}}.

Hyperscale commercial enterprises may also offer a viable path for certain novel and ambitious Rubin-LSST AI/ML applications, provided that their resources can be engaged through partnership or at significant discount. Potential partners include major cloud providers such as Google, Amazon, and Microsoft; major AI players such as OpenAI, Anthropic, and (again) Google; and \acrshort{gpu} manufacturers such as NVIDIA and AMD. Additional effort would be required to develop private-sector partnerships that are mutually beneficial and compatible with the proprietary and non-commercial requirements of the Rubin--LSST data policy\footnote{\url{http://ls.st/RDO-013}}. On the positive side, timescales in industry are typically much shorter than in academia: work with an engaged hyperscale commercial partner could potentially deliver large-scale results quickly.

\subsection{Data}

The primary data products relevant for \acrshort{desc} \acrshort{ai}/\acrshort{ml} work fall into several categories:
\begin{enumerate}
\item \acrshort{lsst} data products delivered by Rubin Observatory
\item Derived data products produced through DESC collaboration efforts
\item Data from other major surveys that enhance and expand DESC AI/ML science
\item Simulation data
\item Model weights and biases from AI models trained on the above
\end{enumerate}

Rubin-LSST data products are organized into three categories distinguished by the timescale of their delivery. The most immediate data are the world-public alert packets that will be distributed within minutes of shutter-close, including difference-image detections of transient and variable objects along with associated postage stamps and (for repeat detections) a 1-year time series. ``Prompt products'' will be released to the LSST data-rights community after 24 hours (for catalogs) and 80 hours (for full focal plane images). On a longer cadence, uniform reprocessing and coaddition across all epochs will deliver annual data releases of catalogs and images. The alert stream will be distributed via the network of LSST Community Brokers, while the prompt products and annual data releases will be accessible via the \acrshort{rsp} and also available for bulk download to DESC via the Rubin \acrshort{usdf} at \acrshort{slac}. A workflow based on the Rucio data management software has been implemented to mirror LSST data to \acrshort{nersc} from the USDF, and could be employed for staging data at \acrshort{alcf} and \acrshort{olcf} as well.

Other major data sets of interest for cross-match, co-analysis, and multi-modal FM training in conjunction with Rubin data include: space-based surveys such as \textit{Roman}, \textit{Euclid}, the \acrfull{spherex}, and the \acrfull{wise}; spectroscopic surveys such as the \acrfull{sdss}, \acrshort{desi}, and \acrshort{4most}; precursor imaging and time-domain surveys such as \acrshort{des}, the \acrfull{decals}, \acrfull{pan-starrs}, and \acrfull{ztf}; and \acrshort{cmb} data from facilities such as Planck, \acrshort{act}, the \acrfull{spt}, and \acrfull{so}.

Given the multi-petabyte size of the LSST data (and of simulations and other survey data sets on similar scales), both network transfer and disk storage will be limiting factors. DESC would likely benefit from strategic coordination with other LSST science collaborations, \acrshort{lincc} (see \autoref{sec7:broader_coordination}), and \acrshortpl{idac} regarding which LSST data products are mirrored where, for how long, at what quality of service, in conjunction with which \acrshort{cpu} and \acrshort{gpu} allocations, and for which analysis purposes.

The current Rubin Data Management system was not primarily designed for large-scale AI/ML work. Hence the data will need to be fitted with additional data interfaces, \acrshortpl{api}, and standards that enable efficient use in this new context, such as the following examples:
\begin{itemize}
\item Adoption of tools like the Hyrax framework\footnote{\url{https://hyrax.readthedocs.io/en/stable/}} that provides modular components for a full AI/ML workflow tailored to astrophysical data.
\item Large-scale cross-match capabilities such as the \acrfull{hats}\footnote{\url{https://hats.readthedocs.io/en/latest/}} and Fink Xmatch\footnote{\url{https://fink-portal.org/xmatch}} that are critical to multi-modal dataset construction.
\item Performant and scalable services for streaming large batches of image cutouts into AI/ML training and inference workflows.
\item Data tokenization and embedding strategies that are well matched to AI/ML model architectures and downstream science tasks.
\item Active learning frameworks for alerts and images that maximize the value of limited human expert labeling time with respect to relevant modeling objectives.
\end{itemize}
In some cases, standards developed by the International Virtual Observatory Alliance\footnote{\url{https://www.ivoa.net}} may be fit for these purposes in DESC although they are typically conceived around classical astronomy use cases that do not map onto survey-scale AI/ML.

\subsection{Benchmarking and Reproducibility}

Challenges of reproducibility will only increase as \acrshort{ai}/\acrshort{ml}-based analyses become more common. Full computational reproducibility requires infrastructure for versioned retention of all input data, any pre-trained models used for inference, all analysis code and frameworks, and all software and environment dependencies. Solutions that allow for some or all of the above elements to be mutable over time fall short of true reproducibility but may be acceptable (or even preferable) to the extent that significant scientific conclusions remain replicable.

Additional challenges posed by the increasing adoption of AI/ML methods include:
\begin{itemize}
\item Defining the role and framework of blind analysis.
\item Maintaining independence of different experiments in the context of multi-modal FM-based analyses.
\item Defining training, validation, and test data sets when we ultimately want to use the full data set for cosmological inference.
\item Ensuring that any agentic software used or developed to generate \acrshort{desc} science records the provenance of the operations done within its task in ways that are compatible with scientific standards of reproducibility (explainable AI).
\end{itemize}

While the above are primarily questions of methodology rather than infrastructure, their solutions will have implications for infrastructure requirements. Some of the required infrastructure elements may include
\begin{itemize}
\item Persistent, accessible storage for testbed datasets, SBI training simulations, and pretrained model weights.
\item Hosted frameworks for deploying and running against science benchmarks.
\item Minified production environments that facilitate use of small-scale development instances to develop methods and benchmarks.
\item Standardized \acrshortpl{api} and architectures for reproducible large-scale model training and deployment.
\item Agentic frameworks for analysis reproduction \citep[e.g.,][]{ye2025replicationbench}
\item Comprehensive provenance tracking \& support for long-term co-archiving of data and analysis.
\end{itemize}

Our traditional thinking about reproducibility will be further challenged if \acrshortpl{llm} and AI agents continue to move us toward a more natural-language approach to dynamically and interactively extracting understanding from data. This trend can bring about a shift away from the traditionally sequential and siloed model of ``data, pipeline, results as papers'' and toward a more dynamic and interactive world characterized by prompts such as ``I want to regenerate the plot in Figure 1 of so-and-so's paper, except I want the magnitude cut at $g=22$ rather than $g=23$\dots'' Other disciplines will be experiencing similar transitions and DESC should look for opportunities both to lead by example and to benefit from broader trends and investment.

% Infrastructure text from 
The vast data scale of \acrshort{lsst} has necessitated a fundamental evolution in scientific methodology. This paradigm shift moves from local data processing on individual researchers' computers toward analytical tools designed to operate on shared, high-performance compute platforms. This centralization, combined with the anticipated widespread adoption of general-purpose foundation models, compels the establishment of a common implementation framework to ensure scientific analyses are reproducible, consistent, and interoperable.

While the \acrshort{rsp} serves as the primary portal for LSST data access, it was not designed for resource-intensive, large-scale AI applications. To bridge this gap, the Hyrax framework provides a  unified platform for exploring the latent space of foundation models. However, their operational deployment remains a significant undertaking that requires dedicated computational resources and a community-governed process for selection and validation.

The rapid pace of innovation in ML means that \acrshortpl{fm} cannot be considered static, long-term solutions. They must be constructed for a dynamic life cycle that includes systematic processes for review, evaluation, and replacement. This agile strategy is critical for all modalities, especially for time-series models deployed on real-time alert brokers, which must reliably process transient astronomical events to enable rapid discovery.

%% file: sections/sec7_coordination.tex
\newpage
\section{Opportunities for Broader AI/ML Coordination}
\label{sec7:broader_coordination}

\acrshort{desc} does not operate in isolation. The broader scientific impact of \acrshort{ai}/\acrshort{ml}-enabled analysis will depend critically on how well we coordinate with the rest of the Rubin ecosystem, other Stage-IV surveys, AI institutes, and large-scale compute providers. Many of these connections already exist in the form of shared personnel, joint projects, or informal collaborations. Here, we outline a non-exhaustive snapshot of this landscape and highlight opportunities to deepen and systematize these links. This section should be read as a living document: the specific institutes, infrastructures, and programs will evolve over the \acrshort{lsst} decade, but the underlying goal will remain positioning DESC as both a demanding scientific user and an influential driver of AI methodologies for fundamental science.

\paragraph{Coordination across the Rubin Community}
The Rubin LSST survey provides the most immediate opportunity for DESC coordination. \acrshort{lincc}, funded and supported by Schmidt Futures, prioritizes open-source and cross-project infrastructure for faint object detection, time-series data analysis, and photometric redshift estimation. Beyond producing intermediate Rubin data products that will be used in targeted ML pipelines, LINCC will serve a key role in maintaining a software ecosystem which has the potential to substantially accelerate the development of large-scale data foundation models and language models for science. DESC efforts in the initial years of the LSST survey can benefit LINCC Frameworks in optimizing its infrastructure toward these goals: for example, by stress-testing the throughput and accuracy of newly released pipelines and providing benchmark scientific datasets that LINCC can use to develop new detection and analysis algorithms.

In parallel with DESC, other LSST Science Collaborations are advancing AI methods for science, and coordinated work would accelerate progress. The \acrshort{issc} brings 150+ scientists from both academic and industry positions to shared discussions on large-scale data analysis with LSST. The ISSC could run independent audits of DESC \acrshort{photoz}, shear, and transient pipelines and partner with DESC to explore methodological advancements in AI/ML (\autoref{sec4:aiml_research}). The \acrshort{tvs} actively develops time-series analysis tools, and could partner with DESC in stress-testing broker infrastructure and classification benchmarks to evaluate cosmological readiness using photometric LSST samples. The \acrshort{galsc} collaboration could co-develop deblending and morphology benchmarks, in order to prevent biases in cosmological inference from mischaracterized shear and clustering constraints. The \acrfull{smwlv} analysis of crowded Milky Way fields will stress test deblending and photometric calibration, while the \acrfull{sssc} identification of solar-system objects and bogus alerts should inform the use in DESC of image products in large-scale scientific models. Across these collaborations, DESC should also encourage reproducible \acrshort{rsp} notebooks, tagged releases, and quarterly cross-collaboration readiness reviews in which independent teams reproduce results and report failure modes, potentially through the Rubin Community Forum\footnote{\url{https://community.lsst.org/}}.

\paragraph{Coordination of Stage IV Experiments}
Coordination between DESC (on behalf of the Rubin LSST) and other Stage-IV cosmological experiments will create additional opportunities for mitigating cross-survey systematics and optimizing the cosmological yield of LSST data.

\acrshort{desi} Data Release 2 is expected to contain $\sim$18.7M spectra across 4,000 deg$^2$ of overlap with the LSST footprint. This spectroscopic sample can be used as a primary calibrator for LSST redshift and lensing systematics. DESC should use DESI’s public releases to improve training sets for photometric-redshift models, require uncertainty coverage checks against those sets, and deploy clustering-redshift cross-correlations to validate $n(z)$ across tomographic bins. For weak lensing, DESC could cross-correlate LSST shear measurements with DESI density fields and test magnification and selection effects with controlled changes in DESI target completeness and fiber-assignment weights. 
The \acrshort{4most} \acrshort{tides} program \citep{tides}  will provide $\sim30,000$ spectroscopic transients. These data will enable real-time validation of classification algorithms and a sub-2\% measurement of the dark energy equation of state. The TiDES survey will also produce $>$200,000 spectroscopically-confirmed transient host galaxies, providing valuable contextual information for characterizing and marginalizing over environmental differences when curating standardizable SN~Ia samples. DESC could leverage these correlations to improve its survey simulations of the extragalactic time-domain sky in successors of the \acrshort{plasticc} and \acrshort{elasticc} challenges.
Coordination within the Roman Space Telescope presents possibly the most transformative opportunity for DESC. The OpenUniverse2024 simulations ($\sim$70 deg$^2$, $\sim$400 TB publicly available; \citealp{OpenUniverse2024}) enable immediate testing of how Roman's infrared imaging and superior resolution can be used for full multi-wavelength characterization of static sources in a joint data foundation model. Roman will also reveal blends invisible in Rubin data, allowing for validation of existing deblending/image segmentation methods for Rubin. Characterization with the high-redshift \acrshort{snia} population in Roman will also provide a laboratory for exploring any redshift-dependent systematics (e.g., changes in progenitor properties across cosmic time) that should be included in DESC cosmological analysis pipelines.
In addition, Schmidt Sciences announced in January 2026 the Eric and Wendy Schmidt Observatory System, a privately funded ``system-of-observatories'' designed for open-access time-domain and multi-messenger science complementary to LSST. The system comprises four facilities: the Argus Array \citep{2022Law_ArgusArray}, a $\sim$900-telescope optical array delivering $\sim$8,000~deg$^2$ instantaneous field of view with cadences down to $\sim$1~s; the Deep Synoptic Array \citep[DSA;][]{2019Hallinan_DSA2000}, a 1650$\times$6.15~m dish radio interferometer spanning 0.7--2~GHz with real-time imaging; the Large Fiber Array Spectroscopic Telescope \citep[LFAST;][]{2024Berkson_LFAST}, a scalable fiber-fed array of 0.76~m unit telescopes targeting ELT-class collecting area for photon-starved spectroscopy and rapid follow-up; and the Lazuli Space Observatory \citep{2026Roy_Lazuli}, a 3~m rapid-response optical--NIR facility (400--1700~nm) in lunar-resonant orbit with a wide-field imager and integral-field spectrograph capable of responding to targets of opportunity in $<$4 hours. With planned operations beginning as early as 2029 and a commitment to open data and shared analysis tools, this privately funded infrastructure could provide valuable cross-wavelength and high-cadence coverage for DESC time-domain and multi-messenger science, particularly for SN~Ia cosmology and transient follow-up. As private investment in astronomical infrastructure grows, DESC should monitor these developments for coordination opportunities.

\paragraph{Coordination with AI Institutes}
Two \acrshort{nsf}-Simons AI institutes have been launched as of September 2024, with funding and explicit scientific themes targeting LSST and cosmology. The \acrshort{skai} Institute between Northwestern, University of Illinois, and University of Chicago, is developing an \acrshort{fm} for transient science that can serve as a precursor to upcoming DESC models. CosmicAI at the University of Texas at Austin is developing \acrshort{llm}-powered AI ``copilots'' for research. These efforts create opportunities for DESC members to identify DESC pipelines most amenable to automation and provide CosmicAI with datasets for beta testing of their models. Across MIT, Harvard, Northeastern, and Tufts Universities, the \acrfull{iaifi} has also explored the use of generative models for field-level inference and multi-modal foundation models for transient science, which DESC can validate with synthetic Rubin datasets such as \acrshort{cosmodc2} 
\citep{2019Korytov_cosmodc2} and PLAsTiCC/ELAsTiCC \citep{PLAsTiCC1810.00001,2023AAS_ELAsTiCC}.  Further, the focus of DESC on AI integration at multiple stages of data processing will benefit the efforts of these institutes in incorporating realistic atmospheric effects and detector systematics directly into model architectures.

A feasible path toward training generalizable FMs for cosmology is to split the work. Models could be prototyped at individual universities or AI Institutes with support from LINCC Frameworks, and scaled through the pre-training of backbones on national and European supercomputers (pooled across \acrshort{doe} and \acrshort{eurohpc} facilities), since this will likely require hundreds of \acrshortpl{gpu} and training across multiple days. The models could then be fine‑tuned and calibrated near the data on DESC computing facilities such as\acrshort{nersc}, with brokers providing smaller fine-tuned heads as software filters for targeted streaming objectives.

A primary bottleneck to achieving this widespread scientific coordination is the development of robust, well-documented, and well-maintained software infrastructure. LINCC Frameworks, supported by the Schmidt Sciences, provides this support for the Rubin Observatory LSST, but this should be equally supported across all major observatories and international collaborations such as DESC in the coming years to enable the emerging technologies outlined in \autoref{sec5:emerging_tech}.

\paragraph{Coordination with European Networks}
\acrshort{eucaif} coordinates AI infrastructure and research across European institutions, and has produced white papers on infrastructure needs \citep[e.g.,][]{caron25}, and LLMs/FMs \citep{barman25}. DESC members at European institutions can, for instance, join EuCAIF WG4 (machine learning and artificial intelligence infrastructure) to contribute cosmology-specific challenges to EuCAIF's methods repository, or WG1 (foundation models \& discovery) to coordinate the development of FMs. These connections may facilitate successful applications for EuroHPC resources where DESC could test whether cutting-edge architectures will scale to LSST volumes. EuroHPC systems (Leonardo with 240 petaflops on NVIDIA A100 GPUs, \acrshort{lumi} with 380 petaflops on AMD MI250x GPUs, or the exascale \acrshort{jupiter} with NVIDIA GH200 superchips) enable training foundation models on billions of galaxy images, computationally infeasible on current NERSC allocations. These systems are already being deployed in astrophysics as a testbed for exascale and GPU-optimized implementations of simulation codes \citep[see e.g.][]{shukla25, lacopo26}. EuroHPC access follows a staged pathway from Benchmark (testing code scaling) to Development (algorithm validation) to Extreme Scale (production runs of up to 8M GPU hours). DESC could pursue Benchmark Access to validate early algorithms before committing to larger allocations. 

\paragraph{Collaborations with Industry}
Tech partnerships can provide expertise, computational resources, and opportunities to stress-test DESC methods at scale. NVIDIA's Academic Grant Program, along with complementary access through Google Cloud and Amazon Web Services, could allow DESC to rapidly prototype architectures and objective functions for foundation models at LSST scale.

Partnerships between DESC and \acrshort{llm} providers (e.g., Anthropic, OpenAI) should also be encouraged. Research credits would allow DESC to simultaneously explore the strengths and failure modes of the current generation of models. This compute could also be used to conduct systematic benchmarking of these models (through, e.g., HuggingFace) on targeted, science-specific use-cases. Any formal arrangement for \acrshort{llm} use across DESC would need to comply with LSST data rights policies (e.g., through private networking, complete audit trails, and explicit no-train/no-retain clauses). Such an arrangement would yield reproducible evaluation suites that could serve as a case study of language model readiness for science applications.

\paragraph{Rubin Alert Brokers}
The seven full-stream Rubin alert brokers are \acrshort{alerce} \citep{alerce}, \acrshort{ampel} \citep{ampel}, \acrshort{antares} \citep{antares}, Babamul \citep{babamul}, Fink \citep{fink}, Lasair \citep{lasair}, and Pitt-Google. These systems provide the primary filtering layer between Rubin streams and science-specific transient samples, turning raw alerts into ranked candidates, host associations, and early labels that will drive spectroscopic follow up and downstream analyses. Tight coordination with these brokers will give DESC direct leverage over the quantities that contribute to cosmological systematics: the completeness and purity of SN Ia samples, characterization of selection effects, and calibration of host-galaxy priors. Characterizing the selection functions of alert brokers as part of the analysis pipeline will help align transient discovery with the DESC requirements.

Brokers should maintain rigorous provenance tracking for all derived data features, host-galaxy associations, and classification/anomaly scores so that DESC can understand the selection effects of deployed algorithms. In return, collaboration with DESC can provide the alert brokers with benchmark datasets and the targeted science objectives used to validate their infrastructure and foster additional software development. Algorithms developed in the early years of the Rubin LSST can be ported upstream to broker environments after public release, providing additional metadata (e.g., embeddings from a data foundation model or concise, text-based descriptions of a subset of high-priority alerts) and allowing the broader scientific community and all Science Collaborations to benefit from DESC efforts without violating LSST data rights policies.

%% file: sections/sec8_risks_challenges.tex
\newpage 
\section{Risks, Challenges, and Mitigation Strategies for AI/ML in DESC}
\label{sec:aiml_risks}

The increasing reliance on \acrshort{ai}/\acrshort{ml} within \acrshort{desc} brings not only opportunities but also a set of technical, organizational, and societal risks that must be managed deliberately. The aim is not to discourage the use of AI, but to ensure that methods are deployed in ways that are scientifically robust, sustainable over the \acrshort{lsst} decade of operations, compatible with DESC’s standards for transparency and reproducibility and with the broader scientific and educational aims of DESC. Below we highlight key challenge areas together with concrete mitigation strategies.

\paragraph{Methodological Robustness and Interpretability}
AI/ML models trained directly on data are vulnerable to familiar statistical pitfalls: biased or incomplete training data, overfitting \citep{2023Huppenkothen,2023Smith}, and domain mismatch between simulations and real survey data \citep{2023Ciprijanovic,Pandya2025}. For methods that sit close to top-level cosmological inferences, there is a justified reluctance to adopt ``black-box'' results as reference constraints unless they can be thoroughly validated and stress-tested. This is amplified for neural summary statistics and learned emulators \citep{2022VillaescusaNavarro, 2023Huppenkothen, 2023Smith}, where it can be difficult to diagnose failure modes or to construct transparent null tests. There is also a subtle “ML-only” risk: some future analyses will be so data- and compute-intensive that no independent non-ML cross-check is feasible, making it even more important that AI-based pipelines be internally well understood and stress-tested.\\
From a DESC perspective, mitigation rests on \emph{explicit validation and interpretability practices}: (i) defining standardized simulation and challenge suites where AI and traditional methods are compared head-to-head in the regime where both are applicable; (ii) publishing diagnostic plots and ablation studies that isolate which data features drive the constraints; (iii) requiring explicit documentation of model training domains, assumptions, and known failure modes, and discouraging use outside those regimes; (iv) developing approximate surrogate models or interpretable summaries (e.g., response functions, feature attributions) that can be inspected by domain experts; and (v) encouraging redundancy where it matters most—for example, using different architectures, loss functions, or simulation pipelines to cross-check key inferences, even when all are “AI-based”. Any AI-based result used as a reference cosmological constraint should be accompanied by a documented validation program and, where possible, benchmarked against simpler baselines. We recognize that imposing these higher standards carries a non-negligible human effort cost, which in turn creates a natural opportunity for agentic AI systems to automate parts of the validation workflow and thereby reduce that burden.

\paragraph{Provenance, Reproducibility, and Integration into Pipelines}
As analyses become more complex, certifying full provenance (from raw data through simulations, training runs, and model selection) becomes harder but more important. If DESC cosmology results depend on opaque training pipelines, unversioned models, or undocumented hyperparameters, small bugs or biases can consume a non-negligible fraction of the “tension budget” in precision tests of \acrshort{lcdm}. While these issues are common to most long-term software infrastructures, the rapid pace of methodological advancement in AI/ML hinders provenance tracking, and the stochastic nature of most neural network training can prevent complete reproducibility. There is also the practical challenge of integrating AI components into mature pipelines that already rely on well-tested codes.\\
Mitigation here is largely infrastructural and procedural: (i) treat trained models as first-class data products, with versioning, metadata, and model cards describing training data, objectives, and known limitations; (ii) require that AI components be runnable from containerized environments and integrated into \acrshort{ci} pipelines with regression tests; (iii) maintain “shadow” implementations (simpler, slower, or more traditional pipelines) for cross-checks where feasible; and (iv) define clear deprecation and maintenance policies so that AI dependencies do not become unmaintainable over the survey lifetime.

\paragraph{Safe Usage, Data Rights, and External Services}
Widespread availability of commercial \acrshortpl{llm} and AI services lowers the barrier to experimentation but introduces new questions about data governance and safe usage. Uploading proprietary Rubin/DESC data or unpublished results to third-party services may raise data-rights concerns, and using off-the-shelf models without understanding their limitations can encourage application of techniques outside their domain of validity.\\
DESC can mitigate these issues by (i) establishing clear guidelines on what kinds of data and metadata may be used with external services, in coordination with Rubin, \acrfull{lsst-da}, and agency policies; (ii) prioritizing self-hosted or collaboration-controlled deployments (e.g., for LLMs and inference services) for sensitive workloads; and (iii) prioritizing services committed to long term availability of their AI models and transparent versioning of these models, for long term reproducibility.

\paragraph{Human Capital, Training, and Sustainability}
AI/ML tools (and, increasingly, agentic assistants) can accelerate research by automating repetitive tasks and lowering the barrier to entry for complex workflows. However, there is a real risk that early-career researchers learn to operate pipelines as “black boxes” or solely by prompt engineering, without acquiring a deep understanding of the underlying statistics and physics, and thus weakening long-term scientific literacy and even cognitive abilities~\citep{kosmnya25, Trotta2025AIThreat}. \\
DESC can turn this into an opportunity by (i) framing AI/ML training as an integral part of graduate and postdoctoral education, combining hands-on use of tools with explicit coverage of underlying concepts; (ii) pairing students with mentors who can help them “open the box” at least once (e.g., by reproducing a result from scratch or implementing a simplified version of a method); (iii) encouraging contributions to shared, well-documented libraries rather than bespoke one-off scripts, spreading maintenance across the collaboration and ensuring that successful methods become communal assets; and (iv) explicitly furthering a scientific culture that does not unduly favor efficiency and speed over in-depth engagement with modeling and creative thinking.  

\paragraph{Environmental Impact}
Large-scale adoption of AI/ML methods comes with significant environmental impact: data centers and computing farms are energy intensive, water hungry, and have generally a negative environmental impact in terms of land usage, noise and e-waste production. The first step toward an environmentally sustainable practice for DESC is to quantify the computing resources used in AI/ML development. This should cover not just the training, validation and deployment costs for methods presented in a research paper (as is common practice), but also track the computational expenditure for hyperparameter search, failed tests, aborted training runs and architectural dead-ends -- which often dwarf the compute needed for the ultimate implementation. \\
To this end, DESC needs to develop easy-to-use, adaptable and constantly reviewed guidelines regarding how to log and report transparently such computational costs, so as to monitor compute usage in DESC over the lifetime of LSST. Taking advantage of more energy-friendly implementations (both software and hardware) should also be encouraged, while training on the matter should be available to all DESC members. \\
Detailed evaluations of astronomy-specific activities in terms of their carbon footprints are scarce, and usually limited to research infrastructure \citep[see, e.g.,][]{knödlseder2025}. \cite{2020NatAs...4..843S} find that even before the rise of AI/ML, supercomputer usage already accounted for the majority of Australian astronomers' carbon footprint. However, compute-related carbon emissions are but one aspect of the multi-faceted environmental impact of AI/ML, which remains understudied. It would be desirable to address this gap by developing in-depth evaluations of the wider environmental cost of LSST-related AI/ML usage, as a way to support sustainability alongside scientific gains \citep{Bashir2024Climate}. 

\paragraph{Computing Resources and Infrastructure}
Realizing the full potential of AI/ML within DESC will place non-trivial demands on computing resources. Training large foundation models for images, catalogs, or time series, and running large-scale simulation-based inference, require sustained access to \acrshort{gpu} clusters at a scale beyond traditional analysis workloads. Similarly, if DESC wishes to host its own LLMs or other generative models for work involving sensitive or proprietary data, these services will need reliable, secure GPU backends and operational support over many years. Without careful planning, AI workloads risk competing destructively with other science uses for scarce accelerators, or fragmenting into ad hoc deployments that are hard to maintain.\\
Mitigation here is primarily strategic: (i) aligning major AI training campaigns with DESC’s existing resource-allocation processes and external partners (e.g., LSST-DA, national and international \acrshort{hpc} centers); (ii) prioritizing shared, reusable models and services over one-off experiments; (iii) investing in efficient training and inference schemes (e.g., parameter-efficient finetuning, mixed precision, model distillation); and (iv) treating any self-hosted LLM or foundation-model service as collaboration infrastructure, with clear policies on access, data governance, and long-term support.

Overall, the main risks associated with AI/ML in DESC are not existential but \emph{operational}: biases that are hard to diagnose, results that are difficult to reproduce, methods that are fragile under domain shift, and human capital that is either over- or under-reliant on automation. By treating AI components with the same methodological rigor as any other part of the cosmology pipeline requiring validation, documentation, governance, and training, DESC can reap the benefits of these tools while keeping these risks manageable.

%% file: sections/conclusion.tex
\newpage
\section{Summary and Conclusion}
\label{sec:conclusion}

The Vera C.\ Rubin Observatory \acrshort{lsst} will generate heterogeneous data at a scale and complexity that strain traditional analysis pipelines. \acrshort{desc}’s mission is to convert these data into robust constraints on the dark sector, which demands methods that are statistically powerful, scalable, and operationally reliable. \acrshort{ai}/\acrshort{ml}, from \acrshort{nde} for \acrshortpl{photoz} to \acrshort{acr:sbi} and generative models for field-level cosmology, have \textit{already} demonstrated that they can address key bottlenecks in this program. At the same time, their utility for precision cosmology hinges on trustworthy \acrshort{acr:uq}, explicit treatment of model misspecification and covariate shift, and fully reproducible integration into DESC workflows.

\autoref{sec3:use_case_for_aiml} and \autoref{sec4:aiml_research} demonstrate that DESC is at the forefront of cutting-edge machine learning applications in astronomy. Research into machine learning is now integral to the primary LSST cosmological probes—including strong and weak lensing, supernovae, galaxy clusters, and large-scale structure—as well as cross-cutting topics such as theory, photometric redshifts, simulations, and deblending. Across DESC working groups and the broader cosmology community, several critical themes and methodologies have crystallized:

\begin{itemize}
 \item \textbf{Simulation-Based Inference (SBI):} SBI has emerged as a powerful methodology, enabling analyses of a complexity that typically exceeds the capabilities of traditional forward modeling. This domain offers fertile ground for machine learning research, particularly in the development of emulators to accelerate pipeline components and in extending analyses beyond traditional point statistics. However, SBI remains sensitive to model misspecification, lossy summaries and inaccurate posterior approximations, problems which are particularly challenging to solve in a machine learning paradigm.
 \item \textbf{Bayesian Methodology and Uncertainty Quantification (UQ):} While Bayesian frameworks are ubiquitous in cosmology, machine learning is increasingly being explored to accelerate Bayesian inference on LSST-scale datasets that would otherwise be computationally intractable. Furthermore, the high precision required by cosmology requires accurate uncertainty estimates that go beyond common practice in machine learning. Building on the application of Bayesian neural networks and related methods, DESC is well positioned to drive fundamental developments in this area.
 \item \textbf{Validation and Benchmarking:} For cosmology, rigorous validation is paramount. Algorithms must not only be accurate and unbiased but also capable of correctly propagating uncertainty. Covariate shift, inevitable in many supervised learning contexts, must be mitigated through accurate simulations and techniques for domain adaptation. Benchmarking and validation are particularly important for algorithms used in products intended for broad usage, such as \acrshortpl{fm} and simulations. The \acrshort{rail} project (see \autoref{sec3:photo-z}), developed by DESC specifically to benchmark photometric redshift algorithms, serves as an excellent model for such validation frameworks.
 \item \textbf{Active Learning and Discovery:} Active learning has become an essential part of machine learning and will be crucial in managing LSST data. The \acrshort{resspect} project (see \autoref{sec3:td}), a collaborative initiative developing an active learning pipeline for transient classification, is an example of the comprehensive infrastructure required for effective active learning. Furthermore, human-in-the-loop workflows will be vital for anomaly detection and the identification of rare phenomena within the vast LSST dataset, facilitating novel discoveries that purely automated systems might overlook.
\end{itemize}

Realizing this potential requires DESC to treat AI/ML as primary components of the measurement pipeline. \autoref{sec6:infra_requirements}--\ref{sec:aiml_risks} of this paper outline the software, computing, and data infrastructure required to support AI/ML at scale. Sustainable integration of emerging tools requires a shared AI software stack, containerized and \acrshort{rsp}-compatible workflows, a DESC Data Registry for model and data products, and benchmark suites that tie model performance directly to cosmological and systematic-error budgets (see \autoref{sec6:infra_requirements}). These methods also present opportunities for broader coordination with Rubin operations, community brokers, external AI/ML institutes, and industry, which we outline in Section~\ref{sec7:broader_coordination}, but present risks ranging from model miscalibration and covariate shift to data rights compliance, environmental cost, and the erosion of human oversight (see \autoref{sec7:broader_coordination}).

On the basis of these insights, we have defined a series of recommendations (R1--R15) and opportunities (O1--O5) in the Executive Summary (\autoref{sec:exec_summary}), spanning methodological research, foundation models, LLMs and agentic AI, infrastructure and software, organizational coordination, human capital, and external partnerships. Implementing these recommendations would position DESC to use AI/ML for ambitious and disciplined science. LSST-era cosmology will be limited not by a lack of algorithms, but by our ability to validate, govern, and integrate them. By investing in that infrastructure now, DESC can shape how AI is used for precision cosmology and set a standard for its responsible deployment across the physical sciences.

%% file: desc-standard-ack.tex
We are thankful to DESC members Prakruth Adari, Keith Bechtol, Simon Birrer, Elisa Chisari, Andy Connolly, Cora Dvorkin, Eric Gawiser, Jimena Gonz\'alez, Katrin Heitmann, Xiangchong Li, Simona Mei, Irene Moskowitz, Peter Nugent, Natalia Porqueres, Mara Salvato, Anze Slosar, Crescenzo Tortora, and V.\ Ashley Villar for their contributions and feedback in the preparation of this manuscript.
The DESC acknowledges ongoing support from the Institut National de Physique Nucl\'eaire et de Physique des Particules (IN2P3) in France; the Science \& Technology Facilities Council (STFC) in the United Kingdom; and the Department of Energy (DOE), the National Science Foundation (NSF), and the LSST Discovery Alliance (LSST-DA) in the United States. DESC uses resources of the IN2P3 Computing Center (CC-IN2P3--Lyon/Villeurbanne - France) funded by the Centre National de la Recherche Scientifique; the National Energy Research Scientific Computing Center, a DOE Office of Science User Facility supported by the Office of Science of the U.S.\ Department of Energy under Contract No.\ DE-AC02-05CH11231; STFC DiRAC HPC Facilities, funded by UK BEIS National E-infrastructure capital grants; and the UK particle physics grid, supported by the GridPP Collaboration. This work was performed in part under DOE Contract DE-AC02-76SF00515.
% authors' acknowledgments
A.T.G.\ is supported by the National Science Foundation under Cooperative Agreement PHY-2019786 (The NSF AI Institute for Artificial Intelligence and Fundamental Interactions, \url{http://iaifi.org/}). 
A.M. is supported by the Australian Research Council Discovery Early Career Research Award (DE230100055). 
M.L.\ acknowledges support from the South African Radio Astronomy Observatory and the National Research Foundation (NRF) towards this research. Opinions expressed and conclusions arrived at, are those of the authors and are not necessarily to be attributed to the NRF. 
H.P.\ and S.T.\ have been supported by funding from the European Research Council (ERC) under the European Union's Horizon 2020 research and innovation programmes (grant agreement no.\ 101018897 CosmicExplorer). H.P.\ was additionally supported by the G\"{o}ran Gustafsson Foundation for Research in Natural Sciences and Medicine. 
C.D.L.\ is supported by the Science and Technology Facilities Council (STFC) [grant No. UKRI1172]. 
The work of Y.Z.\ is supported by NOIRLab, which is managed by the Association of Universities for Research in Astronomy (AURA) under a cooperative agreement with the U.S. National Science Foundation. 
J.d.V., I.S.-N., and L.T.S.C.\ are partially supported by the Spanish MICINN under grant PID2021-123012 and for the MAD4SPACE-CM TEC-2024/TEC-182 project funded by Comunidad de Madrid.
S.S.\ has received funding from the European Union’s Horizon 2022 research and innovation programme under the Marie Skłodowska-Curie grant agreement No 101105167 — FASTIDIoUS.
R.T.\ acknowledges co-funding from Next Generation EU, in the context of the National Recovery and Resilience Plan, Investment PE1 – Project FAIR ``Future Artificial Intelligence Research'', as well as Fondazione ICSC, Spoke 3 ``Astrophysics and Cosmos Observations'', Project ID CN00000013 ``Italian Research Center on High-Performance Computing, Big Data and Quantum Computing'', and partially supported by INFN INDARK grant. 
M.I.\ acknowledges support in part by the U.S. National Science Foundation under grant AST2327245, and also in part by the Department of Energy, office of Science, under Award Number DE-SC0022184. 
B.R.\ gratefully acknowledges the support of the NSF-Simons AI-Institute for the Sky (SkAI) via grants NSF AST-2421845 and Simons Foundation MPS-AI-00010513.
C.G.\ is funded by the MICINN project PID2022-141079NB-C32. IFAE is partially funded by the CERCA program of the Generalitat de Catalunya.
T.Z.\ is supported by Schmidt Sciences. 
J.Z.\ work is partially supported by Schmidt Futures, a philanthropic initiative founded by Eric and Wendy Schmidt as part of the Virtual Institute for Astrophysics (VIA).  
C.A. acknowledges support from DOE grant DE-SC009193.
A.G.\ is supported by an LSST-DA Catalyst Fellowship, made possible through the support of Grant 62192 from the John Templeton Foundation to LSST-DA. The opinions expressed in this publication are those of the author(s) and do not necessarily reflect the views of LSST-DA or the John Templeton Foundation.
T.T.\ acknowledges funding from the Swiss National Science Foundation under the Ambizione project PZ00P2\_193352. 
M.M.\ acknowledges support by the SNSF through return CH grant P5R5PT\_225598 and Ambizione grant PZ00P2\_223738. 
The work of A.A.P.M.\ was supported by the U.S. Department of Energy under contract number DE-AC02-76SF00515.
M.W.C.\ acknowledges support from the National Science Foundation with grant numbers PHY-2117997, PHY-2308862 and PHY-2409481. A.\'C.\, A.D.-W.\: This work was produced by FermiForward Discovery Group, LLC under Contract No.\ 89243024CSC000002 with the U.S. Department of Energy, Office of Science, Office of High Energy Physics. Publisher acknowledges the U.S.\ Government license to provide public access under the (\href{http://energy.gov/downloads/doe-public-access-plan}{DOE Public Access Plan}).
G.M.\ acknowledges support from Illinois Campus Research Board Award RB25035, NSF grant AST-2308174, and NASA grants 80NSSC24K0219 and 80NSSC25K7739. This work used Delta and DeltaAI at NCSA through allocations PHY240290, PHY250333, PHY250374, PHY250386, PHY250281 and PHY250308 from the Advanced Cyberinfrastructure Coordination Ecosystem: Services \& Support (ACCESS) program, which is supported by U.S. National Science Foundation grants \#2138259, \#2138286, \#2138307, \#2137603, and \#2138296. This work utilizes resources supported by the National Science Foundation's Major Research Instrumentation program, grant \#1725729, as well as the University of Illinois at Urbana-Champaign.
M.G.\ is supported by the European Union’s Horizon 2020 research and innovation programme under ERC Grant Agreement No.\ 101002652. 
Argonne National Laboratory’s work was supported under the US Department of Energy contract DE-AC02-06CH11357.
T.S.\ gratefully acknowledge the support of the NSF-Simons AI-Institute for the Sky (SkAI) via grants NSF AST-2421845 and Simons Foundation MPS-AI-00010513. T.S.\ was supported by NSF through grant AST-2510183 and by NASA through grants 22-ROMAN22-0055 and 22-ROMAN22-0013.
Y.-Y.M. is in part supported by the U.S. Department of Energy, Office of Science, Office of High Energy Physics, Experimental Research at the Cosmic Frontier program under Award Number DE-SC0009959.